\documentclass[10pt,
  prl,
  twocolumn,
  aps,
  amssymb,
  nofootinbib,
  longbibliography,
  showpacs,
  superscriptaddress,
  floatfix,
  colorlinks,
  linkcolor=blue,
  citecolor=blue,
  anchorcolor=blue
]{revtex4-2}

\usepackage{graphicx}
\usepackage{float}
\usepackage{dcolumn}
\usepackage{siunitx}
\usepackage{mathtools}

\usepackage{physics}
\usepackage{booktabs}
\usepackage{color}
\usepackage{comment}
\usepackage{dsfont}
\usepackage[export]{adjustbox} 
\usepackage{multirow}

\usepackage{lipsum}
\usepackage{amssymb}
\usepackage{amsmath}
\usepackage{subfigure}
\usepackage{natbib}
\usepackage{amsfonts}
\usepackage{bm}
\usepackage{mathrsfs}
\usepackage[toc,page,title,titletoc,header]{appendix}
\usepackage[colorlinks,linkcolor=blue,citecolor=blue,anchorcolor=blue]{hyperref}
\usepackage{dsfont,amsthm,amsbsy}
\usepackage{todonotes}
\usepackage{braket}
\usepackage{array}
\usepackage{url}
\usepackage[normalem]{ulem}

\newcommand{\SM}{\hyperref[sec:SM]{(Supp. Mat.)}}
\begin{document}
\title{Approximately-symmetric neural networks for quantum spin liquids}
\author{Dominik S. Kufel}
\thanks{These authors contributed equally to this work.}
\affiliation{Department of Physics, Harvard University, 17 Oxford St. MA 02138, USA}
\affiliation{Harvard Quantum Initiative, 60 Oxford St. MA 02138, USA}
\author{Jack Kemp}
\thanks{These authors contributed equally to this work.}
\affiliation{Department of Physics, Harvard University, 17 Oxford St. MA 02138, USA}
\affiliation{Harvard Quantum Initiative, 60 Oxford St. MA 02138, USA}
\author{DinhDuy Vu}
\affiliation{Department of Physics, Harvard University, 17 Oxford St. MA 02138, USA}
\affiliation{Harvard Quantum Initiative, 60 Oxford St. MA 02138, USA}
\author{Simon M. Linsel}
\affiliation{Department of Physics, Harvard University, 17 Oxford St. MA 02138, USA}
\affiliation{Faculty of Physics, Arnold Sommerfeld Centre for Theoretical Physics (ASC),\\Ludwig-Maximilians-Universit{\"a}t M{\"u}nchen, Theresienstr.~37, 80333 M{\"u}nchen, Germany}
\affiliation{Munich Center for Quantum Science and Technology (MCQST), Schellingstr. 4, 80799 M{\"u}nchen, Germany}
\author{Chris R. Laumann}
\affiliation{Department of Physics, Boston University, 590 Commonwealth Avenue, Boston, Massachusetts 02215, USA}
\author{Norman Y. Yao}
\affiliation{Department of Physics, Harvard University, 17 Oxford St. MA 02138, USA}
\affiliation{Harvard Quantum Initiative, 60 Oxford St. MA 02138, USA}

\date{30 July, 2025}

\begin{abstract}
We propose and analyze a family of \textit{approximately-symmetric} neural networks for quantum spin liquid problems. 
These tailored architectures are parameter-efficient, scalable, and significantly outperform existing symmetry-unaware neural network architectures.
Utilizing the mixed-field toric code and PXP Rydberg Hamiltonian models, we demonstrate that our approach is competitive with the state-of-the-art tensor network and quantum Monte Carlo methods.
Moreover, at the largest system sizes ($N = 480$ for toric code, $N=1584$ for Rydberg PXP),
our method allows us to explore Hamiltonians with sign problems beyond the reach of both quantum Monte Carlo and finite-size matrix-product states.
The network comprises an exactly symmetric block following a non-symmetric block, which we argue learns a transformation of the ground state analogous to quasiadiabatic continuation.
Our work paves the way toward investigating quantum spin liquid problems within interpretable neural network architectures.
\end{abstract}

\maketitle

Quantum spin liquids represent exotic phases of strongly-correlated matter exhibiting long-range entanglement and fractionalization~\cite{anderson1973resonating,kitaev2006anyons,savary2016quantum,verresen2021prediction}.
Their detection and characterization remain the subject of intense experimental interest in both quantum materials and  simulators~\cite{semeghini2021probing,googleTC,iqbal2024non,google2023non,broholm2020,xu2023realization,scheie2024proximate,zhang:2024}.
On the numerical front, the exploration of spin-liquid phases inevitably runs into the wall of an exponential Hilbert space. Quantum Monte Carlo (QMC) can efficiently explore this space, but suffers from the sign problem which limits its applicability. 

\begin{figure}[h]
\includegraphics[width=\columnwidth]{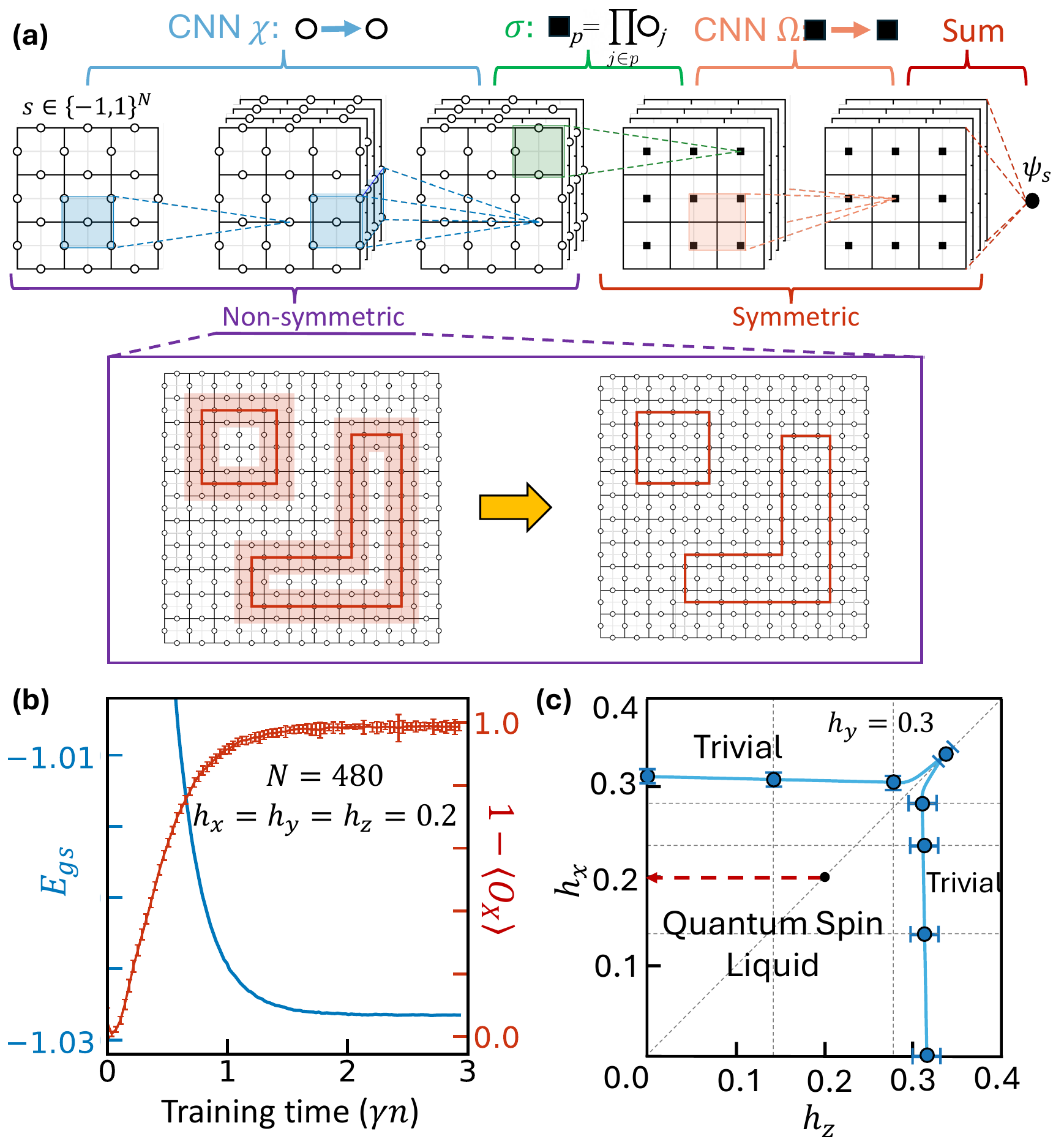}
\caption{%
(a) The approximately-symmetric NQS architecture for a mixed-field toric code model. The network computes the ground-state amplitude, $\psi_s = \Omega(\sigma(\chi(s)))$, given an input bit string $s$.
The circles (squares) represent edge (plaquette) variables. 
The convolutional neural networks (CNN) $\chi$ and $\Omega$ consist of between 1-16 layers and 2-16 channels (only 1 layer and 4 channels shown for each) and use normalized $\mathbb{C}$-sigmoid and $\mathbb{C}$-ELU non-linearities, respectively~\SM. 
The non-linear map $\sigma$ imposes invariance on all following layers. 
For training, $\chi$ is initialized to the identity, while $\Omega$ is randomly initialized. 
The non-symmetric layers ``unfatten" the loop symmetry operators of the quantum spin liquid (purple box) in the spirit of quasi-adiabatic continuation~\cite{hastingswen}. These symmetries are then enforced exactly in the following layers. 
(b) The convergence of the energy density (blue) and the non-local Bricmont-Fr\"olich-Fredenhagen-Marcu string order parameter (red) 
as a function of training time (step size times iteration number), in a regime of the mixed-field toric code model which suffers from the sign problem.
(c) Phase diagram of the toric code (Eq. (\ref{eq:tchamiltonian})) as a function of magnetic field strength, with $h_y=0.2$, imposing a sign problem. The phase-transition locations are extracted from finite-size extrapolation of the string order parameter~\SM.
The red arrow indicates the approximate-to-exact mapping carried out by the non-symmetric block of the network depicted in panel (a).
}
\label{fig:summary}
\end{figure}

Alternatively, variational methods, such as tensor networks \cite{SCHOLLWOCK201196,haegeman2011time,banuls2023tensor}  or variational Monte Carlo \cite{McMillan1965,Ceperley1977,Kent1999,Foulkes2001}, avoid the sign problem but instead restrict themselves to a small subspace, parameterized by physically-motivated variational ansatze.

A tremendous amount of recent attention has focused on a new  approach, which  utilizes neural networks as the variational ansatze (Fig.~\ref{fig:summary})~\cite{carleo2017solving}.
The interest in such neural quantum states (NQS), owes in part, to theoretical guarantees of their expressivity~\cite{hornik1989multilayer}, which is strictly greater than that of efficiently-contractible tensor networks \cite{sharir2022neural}.
Moreover, from a more pragmatic perspective, NQS  have achieved state-of-the-art ground state energies in certain archetypal models~\cite{sharir2020deep,sprague2023variational}, such as the two-dimensional transverse-field Ising model.

\begin{figure}
\includegraphics[width=1\columnwidth]{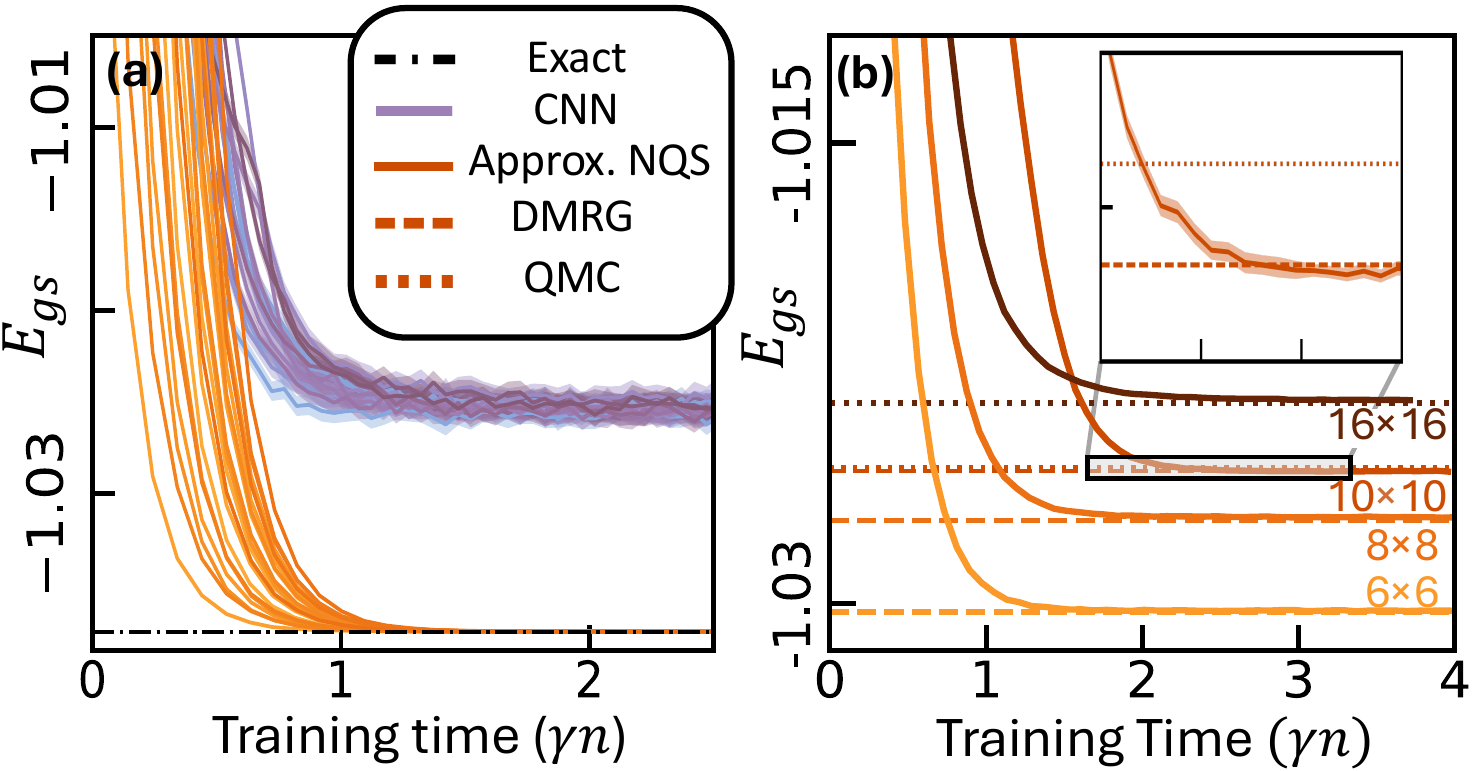} 
\caption{
Benchmarking the approximately-symmetric NQS applied to the sign-problem-free toric code with $(h_x,h_y,h_z)=(0.2,0.0,0.2)$.
(a) The convergence of the energy density 
as a function of training time (step size times iteration number) for a $4\times 4$ ($N=24$) lattice. 
Convolutional neural networks (CNNs) (solid purple) become stuck in local minima, while the approximately-symmetric NQS (solid brown) converges to the exact diagonalization result (dashed black).
Different shades correspond to different random initializations and network hyperparameters~\SM.
(b) At larger system sizes, where exact diagonalization is unavailable, we compare our approximately-symmetric NQS to state-of-the-art DMRG (dashed brown) and QMC (dotted brown) calculations. 
Due to memory constraints, we were unable to obtain converged DMRG results for the $16\times16$ lattice.
(Inset): Zoom-in comparing the NQS, DMRG and QMC energy densities for $L=10$. 
For further analysis under different perturbations and with different network hyperparameters, see \SM. NQS error bars are shown as shading. QMC uncertainty (not shown) is of order $\sim 10^{-4}$ in units of energy density.} 
\label{fig:architecture+energies}
\end{figure}

The simulation of more exotic and delicate quantum phases, such as spin liquids, remains challenging for both NQS and more traditional methods~\cite{patil2023quantum,verresen2021prediction,viteritti2022accuracy,valenti2022correlation}. 
Indeed, for neural quantum states, despite recent progress on the $J_1$-$J_2$  Heisenberg model \cite{choo2019two,roth2023high,chen2023autoregressive,beck2024phase,roth2104group}, the  long-range entangled nature of quantum spin liquids leads to inherently complicated optimization landscapes~\cite{zhang2022hamiltonian,duric2024spin}. 
This causes the training of generic network architectures to become trapped in local minima [Fig.~\ref{fig:architecture+energies}(a)].

One  strategy for simplifying the optimization landscape is to make use of symmetries.
Indeed, by imposing symmetries on the neural network via group equivariant methods~\cite{cohen2016group,roth2023high}, one can significantly reduce the number of optimization parameters without sacrificing expressivity. 
This strategy has been extensively employed for both lattice translation and point group symmetries~\cite{reh2023optimizing,zhang2022hamiltonian}.
In the context of quantum spin liquids, 
exploiting symmetry ought to yield even greater dividends as the ground states are invariant under an exponentially large emergent ``gauge'' group~\cite{hastingswen}. 
For certain models, for which these emergent symmetries are  exactly known, group-equivariant neural networks  have  been shown to yield significant improvements over more conventional methods such as restricted Boltzmann machines or multi-layered perceptrons \cite{luo2021gauge,luo2022gauge,luo2023gauge}.

Unfortunately, for generic spin liquids, it is only possible to specify the exact form of the emergent symmetry operators at particular points in phase space~\cite{exactsoluble}.
Away from these special regions, applying such operators will only leave the ground state approximately invariant.
This precludes their strict imposition on the neural network.

In this Letter, we demonstrate that \emph{approximately}-invariant neural networks can impose a soft inductive bias on the ground-state search while maintaining the flexibility to capture complex quantum states (e.g.~spin liquids) that are not exactly symmetric [Fig. \ref{fig:summary}(a)]. 

Our main results are threefold. 
First, to impose approximate symmetries on neural quantum states, we leverage techniques from the field of approximately group-equivariant networks \cite{finzi2021residual,wang2022approximately}.
We modify these constructions for quantum many-body problems, incorporating physical insights into the structure of the neural network. 
Next, we demonstrate the accuracy of our approach on a paradigmatic quantum spin liquid model: the $\mathbb{Z}_2$ toric code perturbed by a magnetic field. 
We show that the variational energies obtained: 
(i) outperform conventional NQS methods; 
(ii) converge to exact diagonalization results for small system sizes [Fig. \ref{fig:architecture+energies}(a)];
(iii) match state-of-the-art tensor network and quantum Monte Carlo results for larger system sizes [Fig. \ref{fig:architecture+energies}(b)];
and (iv) enable access to large system sizes ($N=480$) even when the Hamiltonian has a significant sign problem, beyond the reach of both QMC and finite-size matrix product state methods [Fig.~\ref{fig:summary}(b)].
Finally, we discuss how the approximate-symmetries framework facilitates NQS interpretability.  In particular, we argue that the neural network discovers a representation of the emergent ground-state symmetries of spin liquids in the spirit of the quasi-adiabatic continuation of Hastings and Wen [Fig. \ref{fig:summary}(a)]~\cite{hastingswen}.

\textit{Emergent symmetries and the toric code}---Consider an $L \times L$ square lattice with open boundary conditions and  $N=2L^2 - 2L$ qubits placed on its edges [Fig. \ref{fig:summary}(a)].
The mixed-field, $\mathbb{Z}_2$ toric-code model is given by the following Hamiltonian:
    \begin{equation}
        H=-\sum_v A_v - \sum_p B_p - \sum_i \left(h_x X_i + h_y Y_i + h_z Z_i\right),
        \label{eq:tchamiltonian}
    \end{equation}
where $X_i$, $Y_i$ and $Z_i$ are Pauli operators, the vertex operator $A_v=\prod_{j \in v} X_j$ acts on qubits neighboring a lattice vertex $v$, and the plaquette operator $B_p=\prod_{j \in p} Z_j$ acts on qubits around a square plaquette $p$. 
Let us begin by considering the case where $h_y=0$. 
In this regime, it is well-understood that the phase diagram hosts a gapped quantum spin liquid up to finite values of $h_x$ and $h_z$~\cite{wu2012phase,trebst2007breakdown,FradkinShenker1979}. 
Along the $h_z=0$ line, there is an exact local $\mathbb{Z}_2$ symmetry group, $G_{\textrm{TC}}=\mathbb{Z}_2^{\times N/2}$, generated by the $A_v$ operators. 
To wit, the ground state $|\psi\rangle$ is invariant under $G_{\textrm{TC}}$, $A_v |\psi\rangle = |\psi\rangle$.
We refer to elements of $G_{\textrm{TC}}$ as loop operators, $W_c$, since they have support on closed loops $c$ on the dual lattice. 

For $|h_z|>0$, exact symmetry under $G_{\textrm{TC}}$ no longer holds.
Nonetheless, as long as the system remains in the gapped spin liquid phase, it is possible to quasi-adiabatically continue the ground state back to the $h_z=0$ line, where $G_{\textrm{TC}}$ is again exact~\cite{hastingswen}. 
Crucially, the continuation is accomplished by a \emph{local} unitary, $U$, which implies that for $h_z \neq 0$ there are a set of ``fattened'' loop operators, $\tilde W_c = U W_c U^{\dagger}$, which remain symmetries of the ground state.
The fattened loop operators are supported on ribbons of finite width about the unperturbed contours $c$, up to exponentially small errors~[Fig. \ref{fig:summary}(a)].

\textit{Approximate symmetries in neural networks}---
In order to exploit the emergent symmetries to improve the NQS ansatz, we must first construct an approximately-invariant neural network. 
    %
    %
    For a system composed of $N$ qubits, each with state-space $\{ \left |\pm1 \right \rangle\}$, a many-body quantum state vector $|\psi \rangle$ can be decomposed into a  complete basis labeled by $2^N$ bit strings $s$: $|\psi \rangle = \sum_s \psi_s |s \rangle$,  where $\psi_s \in \mathbb{C}$ is a complex amplitude and ${|s\rangle}$ is the quantum state associated with the bit string (e.g.~$|s\rangle = |-1,1,1,-1,\cdots\rangle$).

    The key idea underlying NQS is to represent $\psi_s$ as a neural network that gives a complex scalar output for a particular bit string input, $s$. 
    %
 To compute the ground state of a Hamiltonian, $H$, one solves a variational energy-minimization problem with respect to the parameters, $\theta$, of the neural network: $\min_{\theta} \langle H \rangle = \min_\theta \frac{\langle \psi(\theta) | H | \psi(\theta) \rangle}{\langle \psi (\theta) | \psi (\theta) \rangle}$. 
 The energy $\langle H \rangle$ is typically evaluated via Monte Carlo Markov chain sampling \cite{carleo2017solving}, while the network parameters $\theta$ are  optimized  via either gradient descent or more complicated second order methods (e.g.~stochastic reconfiguration \cite{sorella1998green,amari1998natural,stokes2020quantum}). 

Let us now consider the problem of incorporating approximate symmetries into an NQS ansatz. 
Suppose that the ground state, $|\psi \rangle$, exhibits a particular group of symmetries $G$, such that $g|\psi \rangle = |\psi \rangle$ for all $g\in G$. 
We will assume that the basis $\{|s \rangle \}$ is chosen such that the group $G$ acts as a permutation on bit-string basis elements. 
In such a basis, the invariance of the state $|\psi \rangle$ under $G$ 
is ensured if two inputs of the network connected by symmetry yield the same output (i.e.~complex amplitude):  
 $\psi_{gs} = \psi_s$ for all $g \in G$~\cite{roth2023high,cohen2016group}.

For approximate symmetries, the strict invariance condition  above is relaxed to $\mathbb{E}_{g \in G}\mathbb{E}_{s} |\psi_{gs} - \psi_s | < \epsilon$, where the expectation value $\mathbb{E}$ is taken over all group elements and input bit strings~\cite{expectation}.
Given a fully invariant neural network with $\epsilon=0$,  one can  lift the strict constraints by adding an extra non-invariant layer to the network or by using a non-invariant skip-connection~\cite{wang2022approximately,finzi2021residual}.
In principle, for sufficiently large breaking $\epsilon$ of the fixed point symmetries, a neural network constructed in such a fashion can target any vector in the Hilbert space. In practice, the appropriate value of $\epsilon$ is learnt by the network itself, and is independent of network hyperparameters~\SM.


    %

\emph{Approximately-invariant neural quantum states for the toric code}---We propose a family of  approximately-symmetric neural networks utilizing the so-called ``combo" architecture~\cite{wang2022approximately}. 
While we focus on the mixed field toric code model [Eq. (\ref{eq:tchamiltonian})], our approach is applicable to a broad class of quantum spin liquid  problems. 
    Our proposed architecture is schematically depicted in Fig. \ref{fig:summary}(a) and structured as follows: we first impose the constraints $A_v |\psi \rangle = |\psi \rangle$ on the neural network and then weakly break these constraints by  transforming the input with a non-invariant layer.   
    More specifically,  the neural network is  defined by $\psi_s = \Omega(\sigma(\chi(s)))$. 
    Here, $s$ is the bit string input, $\chi$ is a non-invariant convolutional layer acting on the qubits and $\sigma$ is a $G_\mathrm{TC}$-invariant non-linearity which maps qubits to plaquettes.
    Finally, $\Omega$ is a further convolutional layer (consisting of square-shaped kernels) acting on the plaquettes themselves followed by a summation and exponentiation to calculate $\psi_s$ [Fig. \ref{fig:summary}(a)].  
    The non-invariant convolutional layer $\chi$ has a kernel centered at each link of the lattice, and explicitly breaks the $G_{\textrm{TC}}$ symmetry. 
    Meanwhile, the non-linear layer, $\sigma$, is constructed using the $G_\mathrm{TC}$-invariant operators, $B_p=\prod_{j \in p} Z_j$ (since $[B_p,A_v]=0$) and ensures the $G_\mathrm{TC}$-invariance of any further layers~\cite{luo2021gauge,favoni2022lattice}. 
    %
    We describe a number of other choices for approximately-symmetric architectures, along with their implementation details, in the ~\SM. 

Let us begin benchmarking the accuracy of the approximately-symmetric NQS architecture for the mixed-field toric-code Hamiltonian in the sign-problem-free case (with $h_x=h_z=0.2$ and $h_y=0$).
Starting with small $4\times4$ lattices ($N=24$ spins), we compute the ground state energy density, $E_\textrm{gs}$, using (i) our architecture; (ii) a conventional convolutional neural network (CNN) and (iii) exact diagonalization.
The energy as a function of training time (measured in units of stochastic reconfiguration stepsize $\gamma$ times iteration number $n$)  is depicted in Fig.~\ref{fig:architecture+energies}(a).
While the CNN becomes stuck in local minima for all runs (with different initializations and hyperparameters)~\cite{carleo2017solving,vicentini2022netket}, the approximately-symmetric architecture converges to the exact diagonalization energy up to a relative error, $\delta E = |E^{\textrm{NQS}}_{\textrm{gs}}-E^{\textrm{ED}}_{\textrm{gs}}|/E^{\textrm{ED}}_{\textrm{gs}} \sim 10^{-8}$ (see also Fig.~S2 in the supplemental material~\SM for further comparisons with full-lattice-symmetrized restricted Boltzmann machines and additional validation of the results).

We further find that the performance of the approximately-symmetric NQS is not limited to small values of the symmetry-violating $h_z$ field. 
Indeed, it is accurate to a relative error of $\sim 10^{-5}$ even outside of the topological phase at $h_z = 0.7$~\SM.
This suggests that the ansatz is more widely applicable than naively expected and, due to the inclusion of a non-invariant block in the architecture, enables us to identify the phases and phase transitions out of the spin-liquid state using a single \textit{approximately}-symmetric NQS architecture.

    \begin{figure} []
    
     \includegraphics[width=1\columnwidth]{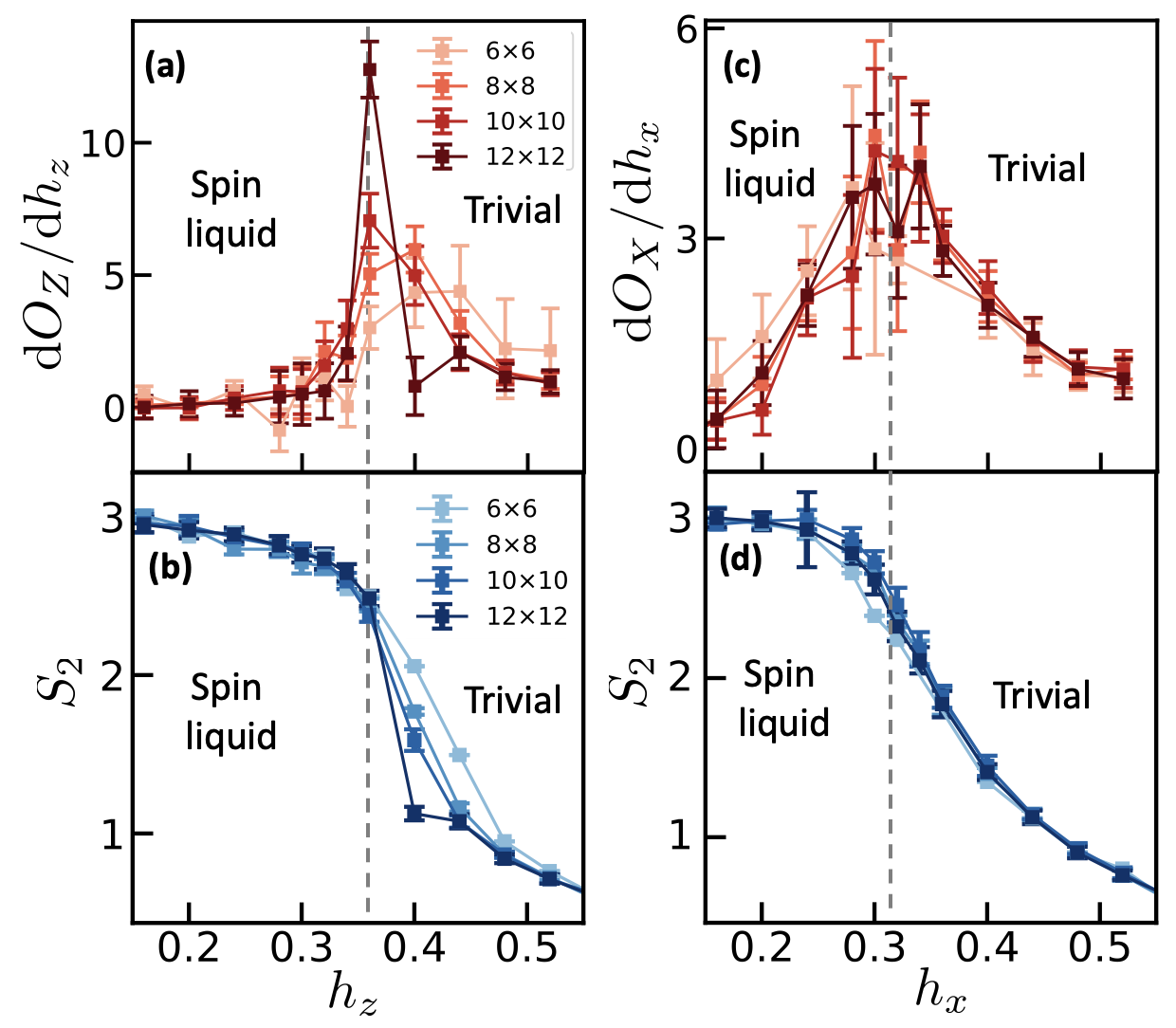}
    \caption{Detecting phase transitions in the toric code with a sign-problem at $h_y=0.3$ [Fig. \ref{fig:summary}(c)].
    (a,c) Depicts the derivative of the BFFM string order parameter $O_Z$ ($O_X$) with respect to the magnetic field, calculated for a square loop around the center of the lattice with perimeter $\ell=8$ ($\ell=4$).
    We use finite-size extrapolation of the position of the peaks in order to estimate the location of the phase transition (dashed gray line,  extrapolation in~\SM). 
    (b,d) The second Renyi entropy for a square, 4-qubit subsystem in the center of the lattice. 
    Panels (a) and (c) illustrate the horizontal cut (at $h_x=0$) through the phase diagram, while panels (b) and (d) illustrate the vertical cut (at $h_z=0$) through the phase diagram.}

    \label{fig:observables}
    \end{figure}

To demonstrate the scalability of the architecture, we perform an extensive set of numerical simulations on lattices up to $16\times16$ ($N=480$ spins) using three methods: state-of-the-art DMRG (density matrix renormalization group)~\cite{White1992,White1993,Schollwoeck2005,SCHOLLWOCK201196}, continuous-time QMC \cite{wu2012phase,linsel2024percolation,greitemann2018lecture}, and the approximately-symmetric NQS architecture. 
As illustrated in Fig.~\ref{fig:architecture+energies}(b), the approximately-symmetric NQS yields competitive energies at all system sizes [inset, Fig. \ref{fig:architecture+energies}(b)].

Crucially, this scalability extends to spin liquid models beyond the mixed-field toric code. 
As an example setting, which has received widespread recent attention, we utilize our approximately symmetric NQS to explore the so-called PXP model (on the ruby lattice), which describes the physics associated with Rydberg atom arrays in the blockade regime~\cite{verresen2021prediction,semeghini2021probing,sahay2022quantum,giudici2022dynamical,wang2025renormalized}.
Similar to the toric code setting, we find that our approximately-symmetric neural network significantly outperforms non-gauge-symmetry aware NQS and yields competitive ground state energies compared to finite DMRG methods (see Fig.~\ref{fig:PXP_results} in End Matter).
Moreover, we are able to reach  significantly larger system sizes (up to $N=1584$ Rydberg atoms), which will be crucial for benchmarking the next generation of Rydberg atom array experiments~\cite{manetsch2024tweezer}. 
Further application settings natural for our approximately symmetric NQS are discussed in Table I in the Supplementary Material.

Furthermore, to demonstrate the general utility of our architecture, we consider 
our approach is not limited to the toric code model but is applicable to a broad class of quantum spin liquid problems.

\textit{Toric code with a sign problem---}%
Many physical perturbations naturally lead to sign problems, including magnetic fields, frustrated long-range couplings, and antiferromagnetic Heisenberg interactions~\cite{savary2016quantum,Chaloupka2010,Jiang2012}.
In principle, variational methods such as DMRG can be used to study such models, even as QMC hits an exponential sampling barrier. 
However, as we have just seen in the sign-problem-free case, the memory requirements for DMRG on finite-size two-dimensional clusters can quickly become prohibitive. 
Accordingly, this is a regime in which the NQS approach should be uniquely well-suited.

We introduce a sign problem in the toric code by turning on $h_y = 0.3$.
We compute the ground state phase diagram as a function of $h_x$ and $h_z$ [Fig. \ref{fig:summary}(c)].
We utilize two diagnostics to identify the transition out of the spin liquid phase: 
    (i) the  string order parameter due to Bricmont, Fr\"olich, Fredenhagen, and  Marcu \cite{bricmont1983order,fredenhagen1983charged} 
    and (ii) the entanglement entropy. 

The string order parameter, $\mathcal{O}_Z$, diagnoses the confinement of the e-type excitations of the toric code.
Consider a closed square loop $C$ of side length $\ell$ on the primal lattice. 
The order parameter is defined in the limit $\ell \to \infty$,
\begin{equation}
\mathcal{O}_Z = \lim_{\ell\to \infty}\sqrt{| O_{Z}|}; \, \,O_{Z} = \frac{\langle \psi | \prod_{j \in \tilde{C}} Z_j | \psi \rangle}{\sqrt{\langle \psi | \prod_{j \in C} Z_j | \psi \rangle}}.\nonumber
\end{equation}
where $\tilde{C}$ is the open string corresponding to half of the square $C$.
One expects $\mathcal{O}_Z$ to be finite in the trivial, confining, phase, while it vanishes in the spin liquid phase because each end of the open string creates a deconfined excitation~\cite{fredenhagen1983charged,bricmont1983order,Gregor_2011}. 
An analogous order parameter, $\mathcal{O}_X$, diagnosing the confinement of m-type excitations, can be defined with strings on the dual lattice and Pauli $X$ operators.
Our second diagnostic is the entanglement entropy, $S_2(\rho) = -\log \mathrm{Tr}(\rho_{A}^2)$, where $A$ represents a particular subsystem. 
One expects the entanglement entropy to be larger deep in the spin liquid phase, since the trivial phase is smoothly connected to an unentangled product state.

As shown in Fig.~\ref{fig:observables}, both of these expectations are borne out by the NQS simulations (at $h_y = 0.3$). 
In particular, fixing $h_x=0$ and sweeping $h_z$, we observe a pronounced peak in the derivative of the string order parameter, $\mathrm{d} O_z/\mathrm{d} h_z$ [Fig.~\ref{fig:observables}(a)], as well as a sharp change in the entanglement entropy at the same field strength [Fig.~\ref{fig:observables}(b)].
Analogous results fixing $h_z=0$ and sweeping $h_x$ are depicted in Figs.~\ref{fig:observables}(c,d).
To estimate the location of the thermodynamic critical point [Fig.~\ref{fig:summary}(c)], we 
perform a power-law extrapolation in $1/L$
of the location of the observed finite-size cross-over~\SM.
After this finite-size extrapolation, the phase boundaries in [Fig.~\ref{fig:summary}(c)] match those obtained by ``infinite system-size'' infinite projected-entangled pair states (iPEPS) and perturbative continuous unitary transformation (pCUT) computations to about $5\%$ (Fig.~2 of~\cite{dusuel2011robustness,rescaling}).

\emph{Interpretability---}The approximately-symmetric NQS architecture facilitates partial interpretation of which physical features are learnt by different sections of the network.
In particular, the non-invariant $\chi$ block maps the model from an approximately-symmetric regime to an exactly-symmetric one, which the $G_{\textrm{TC}}$-invariant $\Omega$ block of the architecture can then learn efficiently. 
This mapping is analogous to the quasi-adiabatic continuation of Hastings and Wen~\cite{hastingswen}, in which the exact “emergent” symmetries of the model (i.e. the “fattened”  loops) are constructed by applying a finite-depth unitary circuit to the unperturbed toric code. 
We hypothesize that the dominant effect of the non-invariant  block of the network is to reverse this finite-depth unitary dressing.  

\begin{figure}[t]

 \includegraphics[width=1\columnwidth]{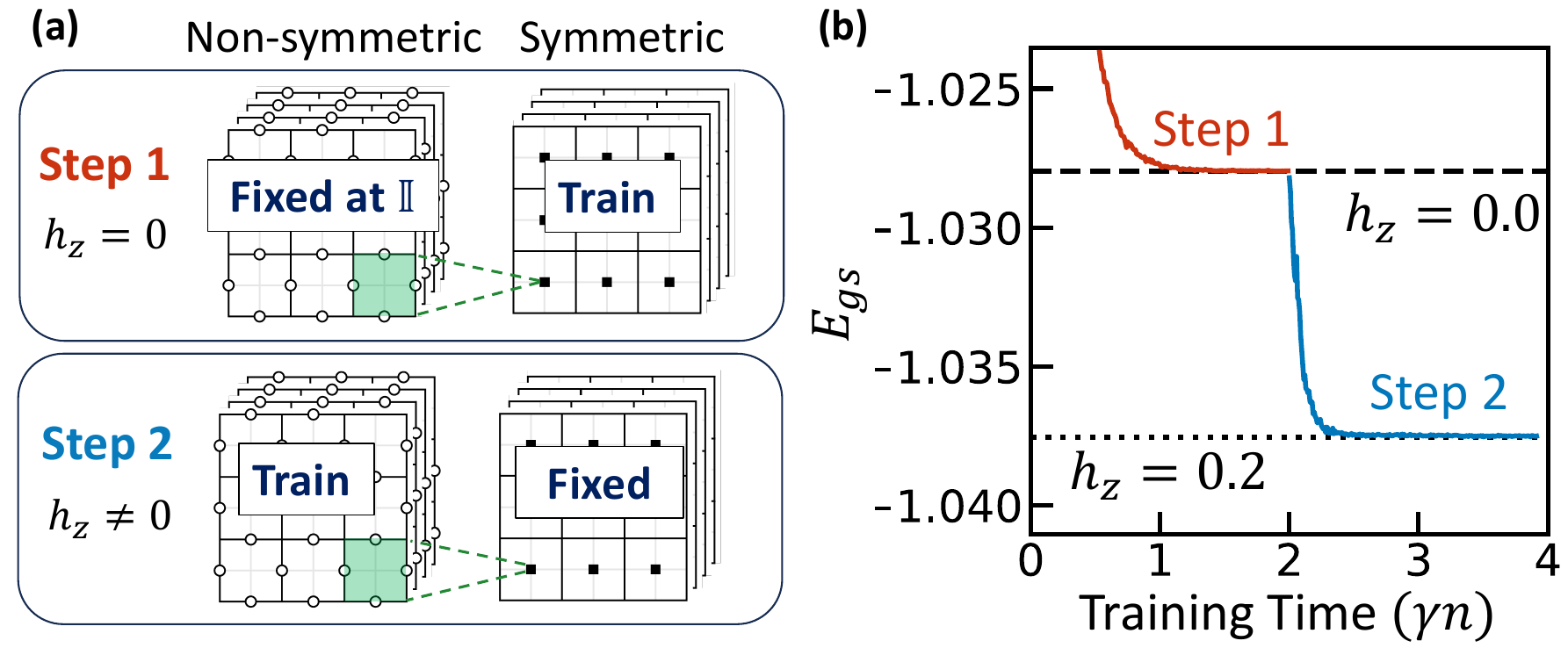}

\caption{(a) The training scheme used to demonstrate interpretability of the approximately-symmetric network. 
In Step 1, we find the ground state of the model at the exactly symmetric point $(h_x,h_y,h_z) = (0.2, 0.0, 0.0)$ by training only the symmetric part of the network. 
In Step 2, we turn to $(h_x,h_y,h_z) = (0.2, 0.0, 0.2)$, fix the symmetric weights within the network, and train the non-symmetric block.
(b) Energy convergence curve for this training scheme for a $4\times 4$ ($N=24$) lattice, compared with exact diagonalization (dashed lines).
}
\label{fig:interpretability}

\end{figure}

To test this hypothesis, we investigate the following two-step training scheme [Fig.~\ref{fig:interpretability}(a]):  First, we train the network in the fully symmetric regime at $(h_x,h_y,h_z)=(0.2,0.0,0.0)$, by optimizing only the parameters of the invariant $\Omega$ layer  and keeping the non-invariant $\chi$ layer fixed to the identity. 
Second, we fix the invariant $\Omega$ layer at these optimized parameters and turn on a field  ($h_z=0.2$) which breaks the exact invariance of the state under the group, $G_\textrm{TC}$; then, we find the ground state by only optimizing the parameters of the $\chi$ layer. 
Despite this restricted training procedure, we still obtain an excellent ground state energy [Fig.~\ref{fig:interpretability}(b)], strongly suggesting that  the  network is indeed effectively learning the ``fattened'' loops  [Fig. \ref{fig:summary}(a), red arrow]~\SM.

\textit{Outlook}---
    Our work opens the door to a number of intriguing directions. 
    First, straightforward extensions of the approximately-symmetric architecture may be used to study increasingly exotic abelian spin liquid models, ranging from experimentally-motivated microscopic models of $\mathds{Z}_2$ liquids \cite{semeghini2021probing}, to more general chiral and non-chiral $\mathds{Z}_p$ and $U(1)$ liquids. 
    By further adapting techniques from~\cite{diaconu2019learning,vieijra2021many}, models with non-abelian anyons and/or non-abelian continuous symmetries may also be accessible (see \SM for more discussion).

    Second, on the numerical front, there are multiple avenues for further improvement, including: 
    (i) tuning the depth of the architecture, 
    (ii) exploring alternative representations for the complex amplitudes required for models with sign problems,
    (iii) increasing parameter efficiency with pooling layers, or (iv) replacing the CNNs with more elaborate backbone architectures (e.g.~transformers)~\cite{sprague2023variational}. 
    Finally, although our focus here has been on finding ground states, the approximately-symmetric architecture can also be applied to simulate real-time dynamics in such systems.

\textit{Acknowledgements} We gratefully thank Andrea Pizzi, Johannes Feldmeier, Lode Pollet, Quynh Nguyen, Rui Wang, Wen-Tao Xu, Marc Machaczek, Di Luo, Phil Crowley, Yichen Huang, Marc Finzi, Max Welling, Shivaji Sondhi, Saumya Shivam, Ruben Verresen, Robert Huang, Leo Lo, Joaquin Rodriguez-Nieva,  Arthur Pesah for interesting discussions. We also thank Curtis McMullen for an inspiring group theory class. 
We acknowledge support from the NSF via the STAQ program and the QLCI program (grant no. OMA-2016245), and from the Wellcome Leap under the Q4Bio program.
D.K. acknowledges support from a Generation-Q AWS and HQI fellowship. 
S.M.L. acknowledges support from the European Research Council (ERC) under the European Union’s Horizon 2020 research and innovation program (Grant Agreement no 948141) from ERC Starting Grant SimUcQuam. 
N.Y.Y. acknowledges support from a Simons Investigator Award.  

The source code supporting this work is publicly available at \url{https://github.com/dom-kufel/Approximate-Symmetries-TC}.

\clearpage
\onecolumngrid
\appendix

\onecolumngrid
\newpage
\section*{End Matter}
\twocolumngrid
\textit{Spin liquid of the PXP model on the ruby lattice---}We present details associated with our study of the PXP model using approximately symmetric NQS. 
We begin by introducing the setting.
Consider a system of $N$ Rydberg atoms placed on an $L \times (2L+1)$ ruby lattice [$N=3L(2L+1)$] with open boundary conditions~\SM and described by the following Hamiltonian
\begin{equation}
    H = -\frac{\Omega}{2}\sum_i (b_i + b_i^\dagger) + \sum_{ij} V_{ij}n_i n_j - \delta \sum_i n_i
\end{equation}
where on-site Hilbert space is $\{ |g_i\rangle,|e_i\rangle \}$,  $n_i = |e_i \rangle \langle e_i|$ measures the $i$-th site atom occupation number, and $|e_i\rangle =b_i^{\dagger}|g_i \rangle$, $|g_i\rangle =b_i|e_i \rangle$. In the so-called the PXP limit of this Hamiltonian,  the interaction strength is assumed to be infinite for atoms within the blockade radius $R_b$ (i.e. $r_{ij} \le R_b=2a$ with $a$ being the shortest distance between two atoms), and zero elsewhere; this strictly prohibits the presence of two excited atoms within a Rydberg blockade radius.

For this  PXP model,  iDMRG numerics on cylinders have shown that the ground state phase diagram consists of three phases~\cite{verresen2021prediction}: a trivial phase connected to $\left | g \right \rangle^{\otimes N}$ for $\delta/\Omega \lesssim 1.4$, a $\mathbb{Z}_2$ quantum spin liquid phase (QSL) for $1.4 \lesssim \delta/\Omega \lesssim 2.0$, and a symmetry-broken valence-bond solid (VBS) phase for $\delta/\Omega \gtrsim 2.0$. The QSL and VBS phases arise from a local constraint permitting exactly one excited atom per lattice vertex when \( \delta/\Omega \gtrsim 1/4 \). There are exponentially many configurations that fulfill these local (Gauss' law) constraints. In the QSL phase, the ground state is approximately an equal superposition of all fully packed configurations, while in the VBS phase, the state spontaneously collapses into one such particular configuration.

To explore the PXP model using  approximately symmetric NQS, we tailor the Wilson loop operators (which act around each hexagon of the ruby lattice) such that they  map one Gauss'-law-satisfying configuration to another~\cite{verresen2021prediction}. 
Since this symmetry is only \textit{approximate} for generic values of \( \delta \) (even within the PXP limit), our architecture includes a gauge-invariant block alongside a non-symmetric block.
Additional details of the implementation are provided in the supplemental material~\SM. 
We compare the performance of this architecture to both RBMs and finite-cluster DMRG.
In all cases, we focus on \( \delta = 1.6 \), which corresponds to a point within the QSL phase.

\begin{figure}[ht]
\vspace{5mm}
\includegraphics[width=0.99\columnwidth]{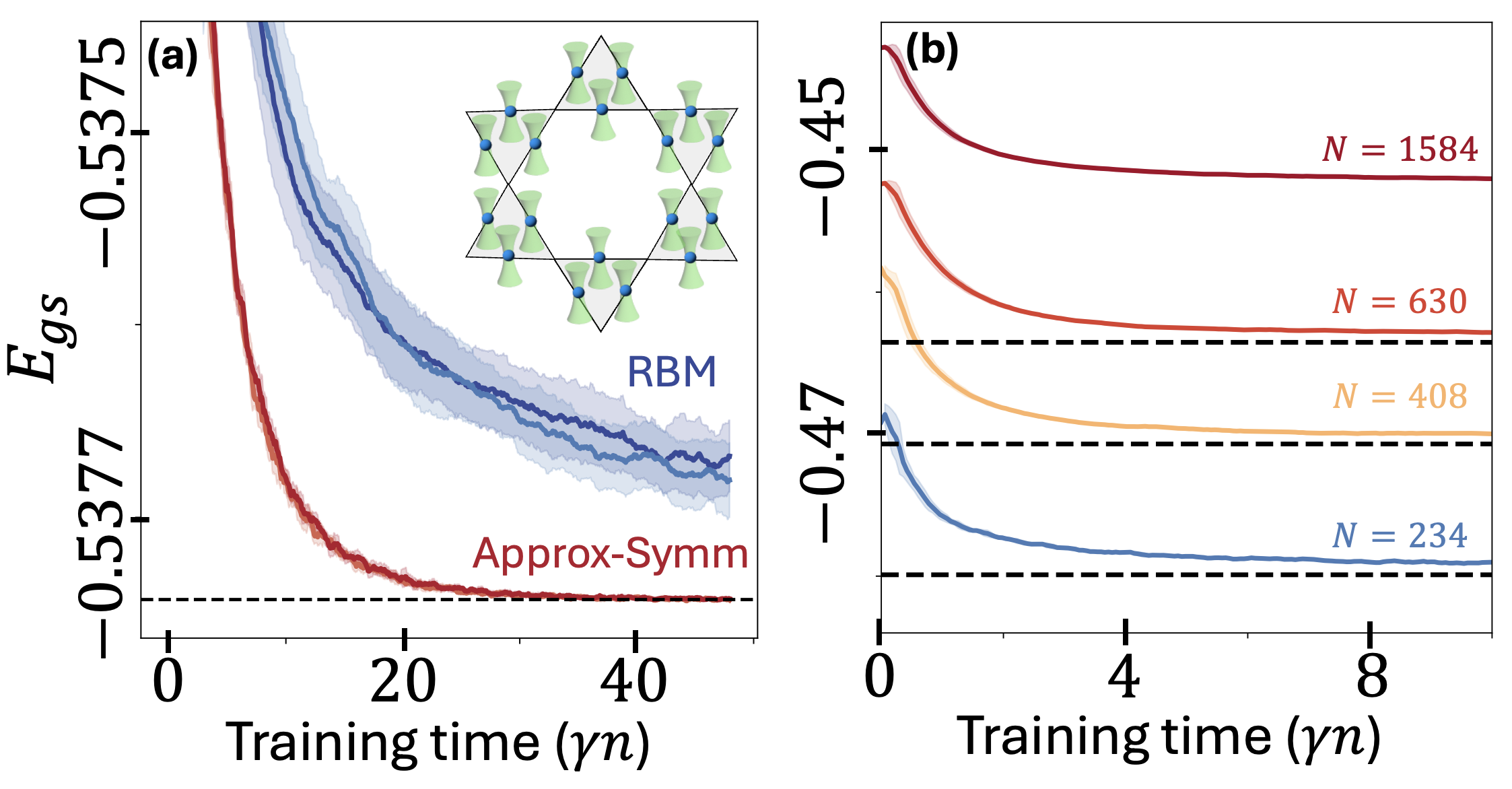} 
    \caption{%
    (a) Convergence of the energy density \( E_{\mathrm{GS}} = \langle H \rangle / N \) for approximately symmetric neural networks (red curves), compared to symmetrized RBMs (blue curves) and exact diagonalization (dashed black) on a system of \( N = 30 \) Rydberg atoms. Inset: schematic depiction  of Rydberg atoms trapped in optical tweezers on a ruby lattice~\cite{semeghini2021probing}. (b) Energy density convergence for larger system sizes \( N = 234, 408, 630, 1584 \) using approximately symmetric neural quantum states (solid curves), compared to finite-cluster DMRG (dashed black curves) with bond dimension \( \chi = 1024 \) and $100$ sweeps.
    }
    \label{fig:PXP_results}
\end{figure}

As depicted in Fig.~\ref{fig:PXP_results}(a), we find that, much as in the mixed-field toric code case, our approximately symmetric NQS yields a nearly three order of magnitude improvement in terms of relative energies (dashed line obtained from exact diagonalization) as compared to an RBM architecture.
In Fig.~\ref{fig:PXP_results}(b), we present a comparison of the ground-state energy densities obtained via our approximately symmetric NQS with finite DMRG results. 
For those system sizes accessible to DMRG ($L=6,8,10$), we observe excellent agreement between NQS and DMRG, with a relative energy difference  within $\epsilon_{rel} = 10^{-3}$--$10^{-4}$. 
In the same figure, we also show the NQS numerics for $L=16$ (equivalent to $N=1584$ Rydberg atoms) which goes beyond the system sizes that have been investigated using either finite DMRG or QMC~\cite{wang2025renormalized}.
Finally, we note that our approach for studying the PXP limit of the Rydberg Hamiltonian, can also be directly applied to the full long-range \( 1/r^6 \) interacting model~\cite{semeghini2021probing}, where the existence of a spin liquid ground state remains under debate \cite{verresen2021prediction,wang2025renormalized}.

\clearpage
\clearpage\onecolumngrid
\section*{Supplementary Material \label{sec:SM}} 

\setcounter{figure}{0}
\setcounter{table}{0}
\setcounter{equation}{0}
\setcounter{section}{0}

\renewcommand{\thefigure}{S\arabic{figure}}
\renewcommand{\thetable}{S\arabic{table}}
\renewcommand{\theequation}{S\arabic{equation}}
\renewcommand{\thesection}{S\arabic{section}}

\makeatletter
\renewcommand{\theHfigure}{S\arabic{figure}}
\renewcommand{\theHtable}{S\arabic{table}}
\renewcommand{\theHequation}{S\arabic{equation}}
\renewcommand{\theHsection}{S\arabic{section}}
\makeatother

\tableofcontents
\section{Neural quantum states and symmetries of many-body systems}
In the main text, we discussed the concept of approximately-symmetric networks and illustrated an explicit example of an approximately-symmetric NQS. Here we discuss the general methodology for constructing approximately-invariant NQS architectures. We start by discussing different ways of imposing many-body system symmetries on neural networks, followed by highlighting connections between our approach and group-equivariant neural network research in the ML community. We detail an explicit example of a different approximately-invariant architecture from the ``combo'' architecture in the main text, based on an alternate approach in the ML literature known as ``residual pathway priors'' (RPP).

\subsection{Overview of different ways of imposing NQS symmetries}
Symmetries of many body states ($g | \psi \rangle = |\psi \rangle \ \forall g \in G$) might be imposed on a neural network in multiple ways. If the symmetries of the states are also symmetries of the Hamiltonian, then the approach perhaps most familiar to a physicist is that of mimicking imposing symmetries in the exact diagonalization (ED). As discussed in the main text, in NQS one decomposes an arbitrary state in a certain basis $|\psi \rangle = \sum_s \psi_s |s \rangle$ where $\{|s \rangle \}$ is typically chosen to be an eigenbasis of the $Z_j$ operators. Then a simple way of imposing Hamiltonian symmetries is to turn to the shared eigenbasis of the Hamiltonian and symmetry group $G$. This approach works for an \textit{arbitrary} group $G$ and has been demonstrated in NQS for e.g., Heisenberg $SU(2)$ symmetric model by \cite{vieijra2020restricted}. Within such approach one rotates the basis while recalculating matrix elements of the Hamiltonian (in a new basis) and picks a symmetry sector (particular representation of the symmetry group) e.g., $J=0$ for $SU(2)$ by selecting a \textit{subset} of input bit strings. This approach plays well together with the standard Monte Carlo Markov Chain (MCMC) sampling of bit string configurations in NQS: one initializes the MCMC chain in the symmetry-consistent configuration and then applies a symmetry-preserving update rule. For instance for a $U(1)$ symmetry (e.g., of a quantum $XY$ model) one would only sample configurations with a total spin $0$ by initializing the chain accordingly and later using total spin conserving rule for its updates (e.g., one which only exchanges individual spins within the configuration). It should be noted that within this approach the architecture of the neural network does not need to be constrained in any way, yet requires many less parameters to achieve the same relative error in ground state energy (see Fig. 2 in Ref. \cite{vieijra2020restricted}).  

Alternatively, another commonly used approach to imposing symmetries is ``post-symmetrization" \cite{choo2019two,reh2023optimizing} where one averages the output of the unconstrained NQS over the symmetry group i.e. in the simplest form $\tilde \psi_s = \frac{1}{|G|} \sum_{g \in G} \psi_{g s} \chi_g $ where $\chi_g$ are characters of a chosen representation of the symmetry group $G$ (where $G$ is abelian). This approach suffers from an extra computational overhead proportional to the size of the group, and thus becomes infeasible for gauge groups or continuous symmetry groups. Its success also heavily depends on the specific choice of the post-symmetrization method: see Ref. \cite{reh2023optimizing} for details.  

Finally in the autoregressive NQS, another method is instead to simply enforce that the probability of sampling configurations violating the state symmetry constraints vanishes \cite{chen2023antn,luo2022gauge,hibat2020recurrent}.

We note that, for all the above methods, it is not immediately clear how to generalize them to the \textit{approximate} symmetries context. We therefore turn to a class of methods which put constraints on the neural-network architecture itself, and allow flexible inclusion of approximate symmetries: group-equivariant neural networks.

\subsection{(Approximately) group-equivariant neural networks}

\paragraph{Group-equivariance and group-invariance} Group-equivariant networks \cite{cohen2016group} are neural network architectures which by construction are \textit{equivariant} under the action of a particular symmetry group. Group equivariance is one of the essential building blocks for the success of the accurate protein structure prediction with AlphaFold 2 architecture \cite{jumper2021highly}. So far, in the NQS context, group-equivariant networks have mostly been applied within the context of imposing lattice symmetries \cite{roth2023high}. Within this approach, one restricts the NQS to (by construction) fulfill a certain constraint: in our case $\psi_{gs} = \psi_s$ i.e. that of group-invariance of the output. Here we assume that the group action maps bit strings to bit strings. (Extending the applicability of the method beyond bit string to bit string mapping is perhaps possible by proceeding along lines of Ref. \cite{vieijra2021many} where one imposes $SU(2)$ + lattice symmetries within the constrained neural network architecture utilizing idea of an equivariant Clebsch-Gordan nets \cite{kondor2018clebsch}). 

The usual way of imposing group-invariance on a neural network is to ensure \textit{group-equivariance} of each of its layers, $\Xi: V \rightarrow W$, and non-linearities. Group-equivariance of each layer means that $\Xi(\rho_{in}(g) x) = \rho_{out}(g) \Xi(x) \ \ \forall x \in V, g \in G$ where $V$ is an $N$-dimensional input vector space, $W$ is an $n$-dimensional output space, and $\rho_{in}: G \rightarrow GL(N, \mathbb{C})$ and $\rho_{out}: G \rightarrow GL(n, \mathbb{C})$ are the input and output representations of the symmetry group $G$. In other words, for a $G$-equivariant neural network, transforming the input by a certain group element corresponds to transforming features by the same group element (though perhaps expressed in a different representation). The difference between ``equivariance" and ``invariance" is intuitively illustrated in the \href{https://github.com/dom-kufel/g_equiv_networks/blob/main/vectorfield_gcnn.gif}{animation} \cite{S_websiteequivariance}. As one transforms (rotates) an input (image in the left panel) to the \textit{$G$-equivariant} neural network, features extracted by the neural network transform accordingly (i.e. rotate; see middle panel). In a \textit{$G$-invariant} neural network (see right panel), as one transforms the input (left panel), feature fields remain unchanged. In other words, rotational \textit{equivariance} might be thought of as an \textit{invariance} in a co-transforming (co-rotating) frame. The fact that generic neural network architectures are not $G$-equivariant might be illustrated on the case of the convolutional neural networks (CNNs) for the case of e.g., $SO(2)$ rotation group symmetries. CNNs are by construction translationally equivariant but are not rotationally equivariant. For any equivariant network, a final layer with a scalar output can be used to promote the equivarance to invariance, because it can be chosen to transform under the trivial representation of the symmetry group.

\paragraph{Approximate invariance}
Critically, generalizing the above approach to approximate symmetries is straightforward: instead of demanding $\psi_{gs} = \psi_s \ \ \forall g \in G, \ s\in V$ we demand $\mathbb{E}_{g \in G}$ $\mathbb{E}_{s \sim |\psi(s)|^2}|| \psi_{gs} - \psi_s || < \epsilon$ for some $\epsilon > 0$ as discussed in the main text (and where we average over group elements in $G$ and bit strings $s \in V$). We emphasize that $ \epsilon$ is learnt by the network itself, perhaps with a help of initialization at the fully symmetric point $\epsilon = 0$.

For completeness, we mention that our approach to imposing approximate symmetries within NQS might be perhaps also extended to the framework of Ref. \cite{choo2018symmetries}. Therein one evaluates NQS only on ``canonical'' bit strings which ensures equivariance of the output (for any Abelian group). A ``canonical'' bit string is a fixed representative of each equivalence class under the group action. 
Although it is unclear how efficient evaluation of such canonicalization would be for general Abelian (gauge) groups, we point out an interesting connection to the recent ML literature: \cite{kaba2023equivariance}. 
Therein a related canonicalization function is learnt efficiently by an equivariant neural network. 
It is thus potentially feasible that the above approximate-symmetries framework might be carried on to this context as well (akin to \cite{kaba2023symmetry}) - we leave this direction for future work.

\subsection{General construction for approximately-symmetric NQS}

\paragraph{Exactly-symmetric architecture} 
Here we present a general approach for imposing group equivariance in quantum many-body physics problems, provided that group action maps bit strings to bit strings i.e. $g s \in V \ \forall g \in G, \ s \in V$. Following \cite{finzi2021practical}, we construct group-equivariant / group-invariant layers of the network by using equivariant multi-layered perceptrons (EMLP). There, equivariance is achieved by appropriately restricting weights of the multi-layered perceptron architecture. In special cases, this approach reduces to other group-equivariant frameworks such as G-convolutional \cite{cohen2016group}, G-steerable \cite{cohen2016steerable} or deep set \cite{zaheer2017deep} architectures. We note that G-convolutional networks cannot be directly applied to problems possessing gauge symmetry since their evaluation cost scales with the size of the group. G-steerable convolutions, on the other hand, require decomposition of the group of symmetries $G$ onto semidirect product of group of translations and some other group \cite{weiler2019general} - which does not seem to be feasible in our case and would require recomputation of irreducible representations for every new group. 

Within the EMLP framework, we consider linear layers with dimensions $O(N)$, which are constructed by appropriately restricting weights of the otherwise fully-connected layer. Weight restriction is obtained from an efficient algorithm ~\cite{finzi2021practical}, which is feasible since the cost of imposing equivariance couples only to the size of the generating set of the group---at worst $O(\textrm{poly}(N))$ for the groups we consider. 

Due to the non-regular hidden-layer representations, one needs to ensure that the activation functions $\sigma(x)$ are also equivariant (i.e. that $\sigma(\rho_{in}(g) x) = \rho_{out}(g) \sigma(x)$. Instead of traditionally used gated/norm \cite{weiler20183d} non-linearities, we utilize other gauge-equivariant non-linearities derived from a physical model in mind i.e. we construct the non-linearity as $\sigma(x)=x h(x)$ where $h(x)$ is a gauge-invariant function of the input to a layer (an example of $h(x)$ for a $\mathbb{Z}_2$ lattice gauge theory is described in the main text). Finally, in the last layer one applies a gauge-invariant non-linearity constructed in a similar fashion by $\tilde{\sigma}(x)= h(x)$. Note that for large symmetry groups (such as gauge groups discussed in the main text), most general linear equivariant layers fulfilling the gauge constraint might be trivial, e.g., for the $\mathds{Z}_2$ gauge group from the main text, if one considers $N$-dimensional hidden layer vector space, a linear equivariant layer would simply be proportional to the identity.

Finally, we remark that gauge-invariant architectures discussed above might also in principle be used to find selected excited states of the fully symmetric model. For example, for a purely $h_x$-perturbed toric code, the lowest lying excitations would correspond to the lowest energy states within the $e$-anyon excitation sectors. In practice, this might be achieved by applying a local basis transformation for the Hamiltonian $H \mapsto U H U^{\dagger}$ where e.g., $U=Z_l$ would introduce e-anyon excitation to vertices $v,v'$ belonging to the boundary of the link $l$. This method of finding an excited state might be further combined with the orthogonalization-based approach used in Refs. \cite{choo2018symmetries,valenti2022correlation}. 

\begin{figure}
\includegraphics[width=1\columnwidth]{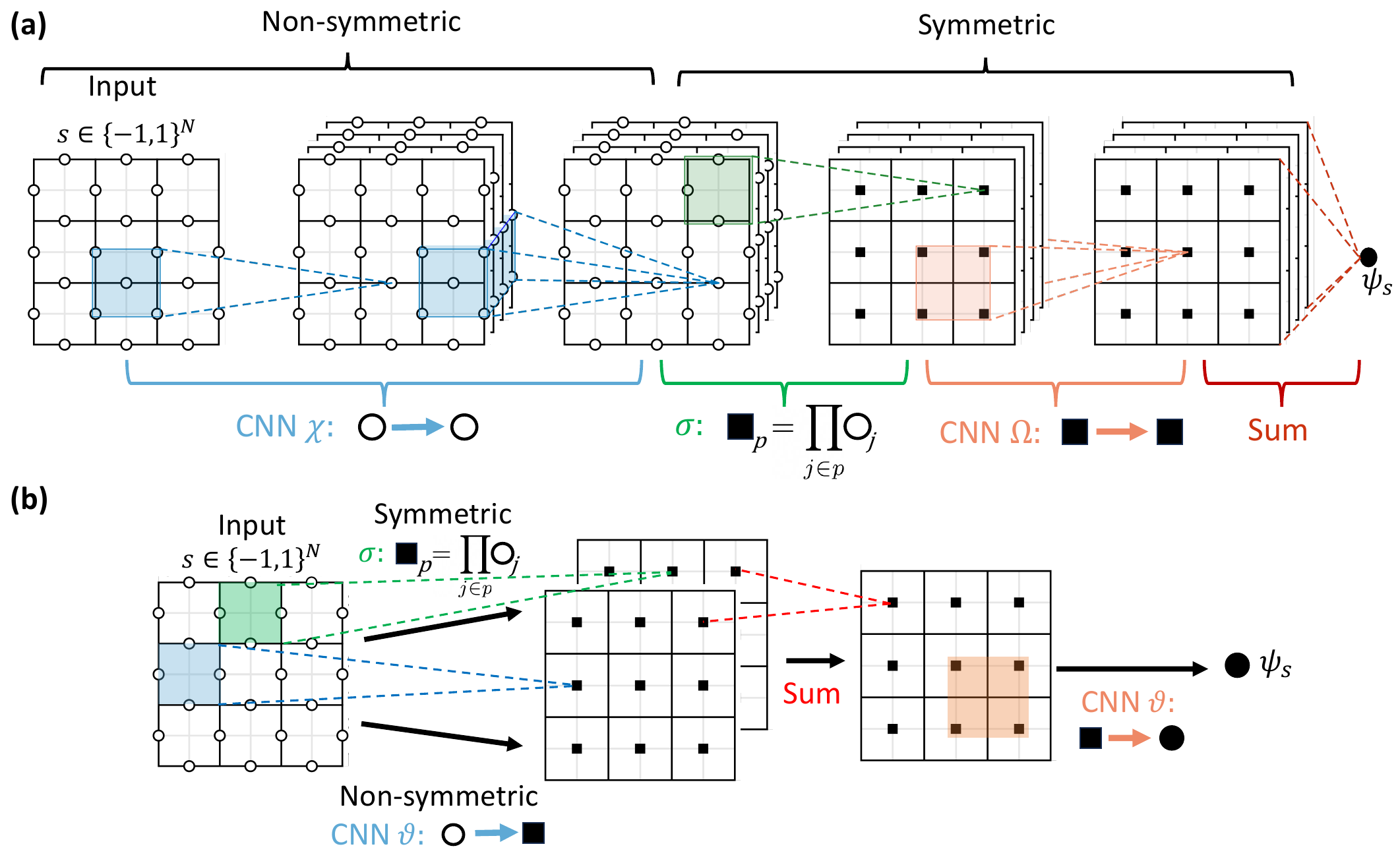}
\caption{(a) Replotted "Combo" approximately-symmetric architecture from the main text. (b) For comparison, a sketch of an alternative ``RPP" approximately-symmetric architecture. In practice each layer and ``branch" (symmetric/non-symmetric) of the network will have multiple channels, but only one has been depicted for simplicity. Initialization of all layers is according to a random Gaussian distribution.} 
\label{fig:rpp_architecture}
\end{figure}

\paragraph{Approximately-symmetric architecture}
Let us now discuss how to generalize to the case when the symmetries are only approximate. The basic idea behind any such generalization is to add non-invariant blocks to an otherwise symmetric architecture. We demonstrate this approach with two examples: the ``Combo" architecture in the spirit of \cite{wang2022approximately}, discussed in the main text, and the residual pathway prior ``RPP" architecture in the spirit of \cite{finzi2021residual}. The RPP architecture is constructed to mimic the ResNet \cite{he2016deep} skip-connection architecture. Apart from a gauge-invariant pathway, it also contains a non-invariant skip-connection which allows incorporation of non-symmetric features. In practice, for $G=\mathbb{Z}_2^{\times N/2} $, we construct the skip connection with a non-$G$-equivariant convolutional layer $\Omega: E \rightarrow P$ which maps set of edges to plaquettes. Finally the two pathways (equivariant and non-equivariant) are summed together and post-processed with another convolutional layer acting on plaquettes (as for the ``Combo" architecture) -- see Fig. \ref{fig:rpp_architecture}(b) for ``RPP" as compared with ``Combo" in Fig. \ref{fig:rpp_architecture}(a). 

We have implemented and tested both architectures, and find both yield comparable performance (see details below). However, the ``Combo'' architecture has the advantage that we have some physical intuition about the action the non-invariant layer (see the interpretability sections of the main text and also below for more details), which is why we chose to focus on that architecture.

\vspace{-8mm}
\subsection{Potential use cases for approximately-symmetric models}
\vspace{-2mm}
We emphasize that there exists many target systems with multiple open physics questions where our approximately-symmetric neural network may be applied. These can be addressed using our approximately symmetric architecture, with minor extensions, for both ground state and real-time evolution studies. For clarity we summarize them in the Table \ref{tab:applications}.

\begin{table}[h!]
\centering
\begin{tabular}{|p{4.5cm}|p{6cm}|p{4.5cm}|}
\hline

\centering{\textbf{Extension}}               & \centering \textbf{Motivation}            &                       \textbf{Framework modification}                         \\ \hline
\textbf{$\mathds{Z}_2$ Spin Liquids}           & Realizable in quantum simulators via engineering \cite{semeghini2021probing,verresen2021prediction}; resulting ground states (and their dynamical preparation) away from exactly symmetric fixed point.                                           & No modification \\ \hline
\textbf{Chiral $\mathds{Z}_d$ toric code}      &  $\mathds{Z}_d$ has an exact symmetry limit and for $d >2$ permits chiral (exact local symmetry-breaking) perturbations; allows study of fractional quantum Hall physics  \cite{weerda2024fractional}.                                   & Increase onsite dimension $d$, modify group nonlinearity \cite{luo2021gauge}: $x_1 x_2 x_3 x_4 \rightarrow x_1 x_2 x^{*}_3 x^{*}_4$ (around each plaquette) \\ \hline
\textbf{$U(1)$ lattice gauge theory} & Approximately realizable in quantum simulators via engineering \cite{feldmeier2024quantum}; study fractionalization and confinement dynamics in $U(1)$ gauge theories                                                                & Similar to $\mathds{Z}_d$; promote input variables from binary to continuous \cite{luo2022gauge}. \\ 
\hline
\textbf{Kitaev double}             & Study phases and dynamics of models with discrete non-abelian symmetries and non-abelian anyons~\cite{iqbal2024non}.                                           & Similar to $\mathds{Z}_d$ with special care taken for multiplication ordering \cite{luo2021gauge}. \\ 
\hline
\textbf{Double semion}             & Study of sign-problem-full~\cite{hastings2016quantum} twisted gauge theories featuring semionic excitations~\cite{levin2005string}.                                                                                                           & Modify product non-linearity from group invariance to group equivariance~\cite{choo2018symmetries,kaba2023equivariance}.   \\ \hline
\textbf{Non-abelian local continuous symmetries}              & Study of realizations of non-Abelian lattice gauge theories such as QCD in quantum simulators \cite{wiese2014towards}.                                                                                                                                                                                          & Generalizing framework to non-Abelian continuous symmetries~\cite{diaconu2019learning}.  \\ \hline

\textbf{Approximate global symmetries} & Study of weakly positionally disordered systems e.g., with approximate translation symmetry. & Using group equivariant neural networks with relaxed group convolutions~\cite{wang2022approximately}. \\ \hline

\end{tabular}
\caption{Partial list of models where approximately symmetric neural networks may be helpful and the extensions that they would require.}
\label{tab:applications}
\end{table}

\section{Approximately-symmetric NQS toric code performance study}
In this section, we first detail the specific hyperparameters used within our simulations, as well as the Monte Carlo Markov chain update rule we use. Second, we discuss the performance and characteristics of the model for a toric code in $h_x$ and $h_z$ fields. We claim that increasing the depth of the network allows us to systematically improve the accuracy of the simulations. Third, we then discuss the performance of the model under a variety of perturbations, both with and without a sign problem, and identify its key architectural components. Fourth, we track the main issues for further improving the accuracy of the network predictions in different regimes, culminating in suggestions for future architecture improvements with increased parameter-efficiency and enhanced performance. Finally, we provide more details on observable evaluations and extracting phase transitions with NQS. 

\subsection{Simulation parameters}
For the simulations in the main text, we consider neural networks with real bit string $s \in \{-1,1\}^N$ inputs, (i) complex parameters and outputs if the Hamiltonian has a sign problem ($h_y \neq 0$) and (ii) real parameters, and scalar positive outputs if the Hamiltonian is sign-problem-free ($h_y=0$) (as then underlying Hamiltonian is stoquastic so by Perron-Frobenius theorem the coefficients of the eigenvector can be chosen real and non-negative in the chosen basis \cite{pillai2005perron}). 

We evaluate the energy $\langle H \rangle$ through MCMC sampling. It is performed by noting that $\langle H \rangle = \langle \psi | H | \psi \rangle / \langle \psi | \psi \rangle = \sum_s p(s) E_{loc,s}$ where $p(s)=|\psi(s)|^2$ and $E_{loc,s}= \sum_{\mathbf{s}'} \frac{H_{s s'}}{\psi_{s}} \psi_s'$ where the latter is evaluated exactly for each bit string $s$ (due to only $\mathcal{O}(poly(N))$ non-zero matrix elements of the Hamiltonian). Then for $N_{\textrm{samples}}$ samples one approximates $\langle H \rangle \approx \frac{1}{N_{\textrm{samples}}}\sum_{i=1}^{N_{\textrm{samples}}} E_{loc,s_i}$ where samples $\{ s_i \}$ are found by applying a Metropolis-Hastings algorithm. 

We use a custom sampling rule, closely related to the one used in \cite{marcmachaczekthesis}, which involves flipping either a single spin per update step, or all the spins surrounding a single vertex. The intuition for this update rule stems from the fact that in the ground state of the toric code, a single spin-flip creates (two) excitations, whereas a vertex-flip does not. Thus close to the toric code fixed point, a single-spin flip update would generically be expected to take a high-amplitude state to a low amplitude state, in contrast to a vertex-flip. Nevertheless, single-spin flips are still required for ergodicity. Given $N_v$ vertices and $N$ spins, on each update we choose to flip a vertex with probability $p=N_v/N$ or else flip a spin. For open boundary conditions, we neglect any vertex with spins on the boundary in this procedure (for periodic boundary conditions, a more complicated procedure is required, see \cite{marcmachaczekthesis}). We find that this vertex-spin update rule empirically yields a significant improvement in MCMC sample acceptance probabilities compared to only using single spin updates.

In order to benefit from GPU parallelization, as well as improving the ergodicity of exploring possibly multi-modal probability landscapes, we draw samples from $\mathcal{O}(10^3)$ independent MCMC chains processed in parallel. 
We apply $N_{\textrm{subsample}}$ updates between each collected sample in order to reduce the autocorrelations within each chain.
Furthermore, we discard the first $N_\textrm{burn-in}$ samples while the chain thermalizes past its initial transient. 
We evaluate $\langle H \rangle$ using $C_L$ samples per chain, so that the total number of samples is $N_{\textrm{samples}} = C_L N_{\textrm{chains}}$. 
The total number of updates which must be applied is thus $\left(N_{\textrm{burn-in}} + C_L \right) N_{\textrm{subsample}} N_{\textrm{chains}} $.

We initialize the neural network parameters such that the neural network is fully-symmetric before training. This is achieved by initializing the weights of the non-invariant kernels to $\mathds{1}$ and utilizing the $\textrm{sigmoid}$ non-linearity in the non-invariant layers, given for real inputs as $\phi_{\textrm{sigmoid}}(x)=\frac{\tanh(x/2)}{\tanh(1/2)}=\frac{2+2e}{e-1} \left(\textrm{sigmoid}(x)-1/2\right)$ and $\textrm{sigmoid}(x) = \frac{1}{1+e^{-x}}$ (shift and rescaling of the sigmoid to ensure that $\phi_{\textrm{sigmoid}}(x=\pm 1)=\pm 1$ for identity initialization of the non-invariant block). 
For invariant layers we use ELU non-linearity \cite{clevert2015fast}. For complex neural network parameters, we separately pass real and imaginary parts of the input to the non-linearity $\mathbb{C}-\textrm{sigmoid}$ (non-invariant layer) and $\mathbb{C}-\textrm{ELU}$ (invariant layer) e.g., $\phi_\mathbb{C}(x)=\phi(Re[x])+i \phi(Im[x])$ for either non-linearity $\phi$. 

We optimize the network parameters using stochastic reconfiguration \cite{sorella1998green} with diagonal shift regularization. This might be thought as an imaginary time evolution for the states \cite{stokes2020quantum} and amounts to 1st order time evolution on a TDVP manifold, yielding linear equations of motion for the vectorized parameters $\bm{\theta}_t$: $\mathbf{S} \dot{\bm{\theta}}_t = - \gamma \mathbf{F}$ where $\mathbf{F}$ is a force vector, $\gamma$ is a learning rate, and $\mathbf{S}$ is a quantum geometric tensor (see e.g., \cite{carleo2017solving} for details). We solve this equation by performing a singular value decomposition on the $\mathbf{S}$ matrix, which has complexity of $\mathcal{O}(N_{\textrm{parameters}}^3 + N_{\textrm{parameters}}^2 N_{\textrm{samples}})$ \cite{chen2023efficient}. We choose it over the conjugate gradients solver with complexity $\mathcal{O} (N_{\textrm{parameters}} \kappa N_{\textrm{samples}})$ (where $\kappa$ is a condition number of the matrix $\mathbf{S}$), because we empirically observed that for the family of ground states of the system and architecture under consideration $\kappa$ is very large ($\mathbf{S}$ eigenvalues span $15$ to $30$ orders of magnitude) which in practice yields rather unpredictable and slower runtimes. Furthermore, in order to stabilize the simulations we add a small shift to the diagonal of the $\mathbf{S}$ matrix i.e., $\mathbf{S} \mapsto \mathbf{S} + d \mathds{1}$. 

Parameters used for running simulations with the architecture presented in Fig. (1)a of the main text for $16 \times 16$ are listed in the Table \ref{tab:parameters} below. 
All simulations are run in \textit{NetKet} \cite{vicentini2022netket} that benefits from \textit{JAX}\cite{jax2018github} auto-differentiation and just-in-time compilation. We follow the convention in \textit{NetKet 3.16} such that in code neural network outputs represent a \textit{logarithm} of an amplitude $\psi_s$ to facilitate calculations. All neural networks are coded in \textit{FLAX} \cite{flax2020github}. Density Matrix Renormalization Group (DMRG) simulations were performed in \textit{iTensor} \cite{itensor,itensor-r0.3} and exact diagonalization in \textit{Dynamite} \cite{gregjulia}. Energy densities as used in the figures are defined as $E_{gs}=\frac{\langle H \rangle}{N+1}$ i.e. energy per stabilizer. Training time in all figures refers to the $\gamma n$ (where $n$ is iteration number, and $\gamma$, as above, is a learning rate). 

\begin{table}[h]
\centering
\begin{tabular}{|c|c|c|}
\hline
\textbf{Parameter Type} & \textbf{Parameter} & \textbf{Value} \\ 
\hline
\multirow{3}{*}{Optimization parameters}   
                           & Learning rate       & 7e-3                     \\ \cline{2-3}
                           & Diagonal shift $d$   & 5e-5                           \\ \hline
\multirow{9}{*}{Architecture parameters}    
                           & NIB: Channels \& Depth         &   $[1, 2, 4]$               \\ \cline{2-3}
                           &  NIB: Initialization & Identity $\mathds{1}$                      \\ \cline{2-3}
                           & NIB: Kernel Size  &    3                \\ \cline{2-3}
                           & NIB: Non-linearity  &     $\mathbb{C}-\textrm{sigmoid}$               \\ \cline{2-3}
                           & IB: Channels \& Depth & $[4, 4, 4]$   \\ \cline{2-3}
                           &  IB: Initialization &  Random Gaussian $\mu=0,\sigma=0.02$ \\ \cline{2-3}
                           &  IB: Kernel Size  &  15 \\ \cline{2-3}
                           &  IB: Non-linearity & $\mathbb{C}-\textrm{ELU}$ \\ \cline{2-3} 
                           &  Total $N_{\textrm{parameters}}$  & 11324   \\ \hline
\multirow{4}{*}{Sampling characteristics} 
                           & $N_{\textrm{chains}}$ & $1024$ \\ \cline{2-3}
                           & $N_{\textrm{burn-in}}$ (per chain) & $8$ \\ \cline{2-3}
                           & $N_{\textrm{subsample}}$ & $480$ \\ \cline{2-3}
                           & $N_{\textrm{samples}}$  & $8192$ \\ 
\hline
\end{tabular}
\caption{Summary of neural network parameters for $16\times16$ runs. NIB=Non-invariant block, IB=Invariant-block; channels \& depth specifies the number of channels in each layer within a block.}
\label{tab:parameters}
\end{table}

\vspace{-8mm}
\subsection{Comparisons with baseline neural networks for a $4 \times 4$ system}
\begin{figure}
\includegraphics[width=0.35\columnwidth]{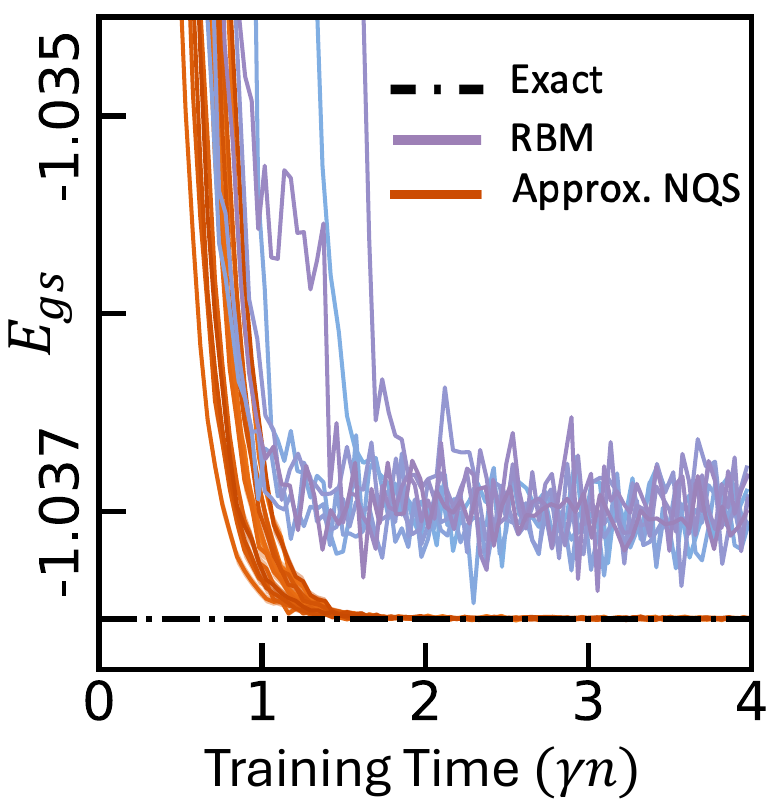}
\caption{Energy convergence curve for approximately-symmetric networks and RBMs as compared with exact diagonalization on $N=24$ system in $(h_x,h_y,h_z)=(0.2,0.0,0.2)$ fields.}
\label{fig:rbmcomparison}
\end{figure}

We provide more details on comparing our approach with baseline neural network architectures which incorporate lattice symmetries but no local gauge symmetries. In the main text this was performed for CNNs, which by construction incorporate translation symmetries and contain $4-10$ layers with multiple filters per layer. We note that the main difference between CNNs and our architecture is the presence of the group-invariant block and both networks have a comparable runtime. As mentioned in the main text (cf. Fig. 2(a) therein), we found that unlike our approximately-symmetric network approach, CNNs are trapped in local minima, thus conclusively showing the importance of imposing approximate gauge symmetries. Additionally, in Fig. \ref{fig:rbmcomparison}, we show that symmetric restricted Boltzmann machines (RBMs), which incorporate \textit{all} lattice symmetries rather than only translations \cite{carleo2017solving}, perform better than CNNs. However, they still get trapped in local minima with errors several orders of magnitude larger than those achieved by our architecture. Furthermore, we expect the performance gap between symmetric RBMs (or their deeper variants, such as GCNNs) and approximate symmetry-based approaches to grow with system size, as the local symmetry space scales exponentially ($\mathcal{O}(2^{N/2})$) compared to the linear scaling of lattice symmetries ($\mathcal{O}(N)$).

\subsection{Performance in $h_x$ and $h_z$ fields for a $4 \times 4$ system}

\begin{figure}
\includegraphics[width=1\columnwidth]{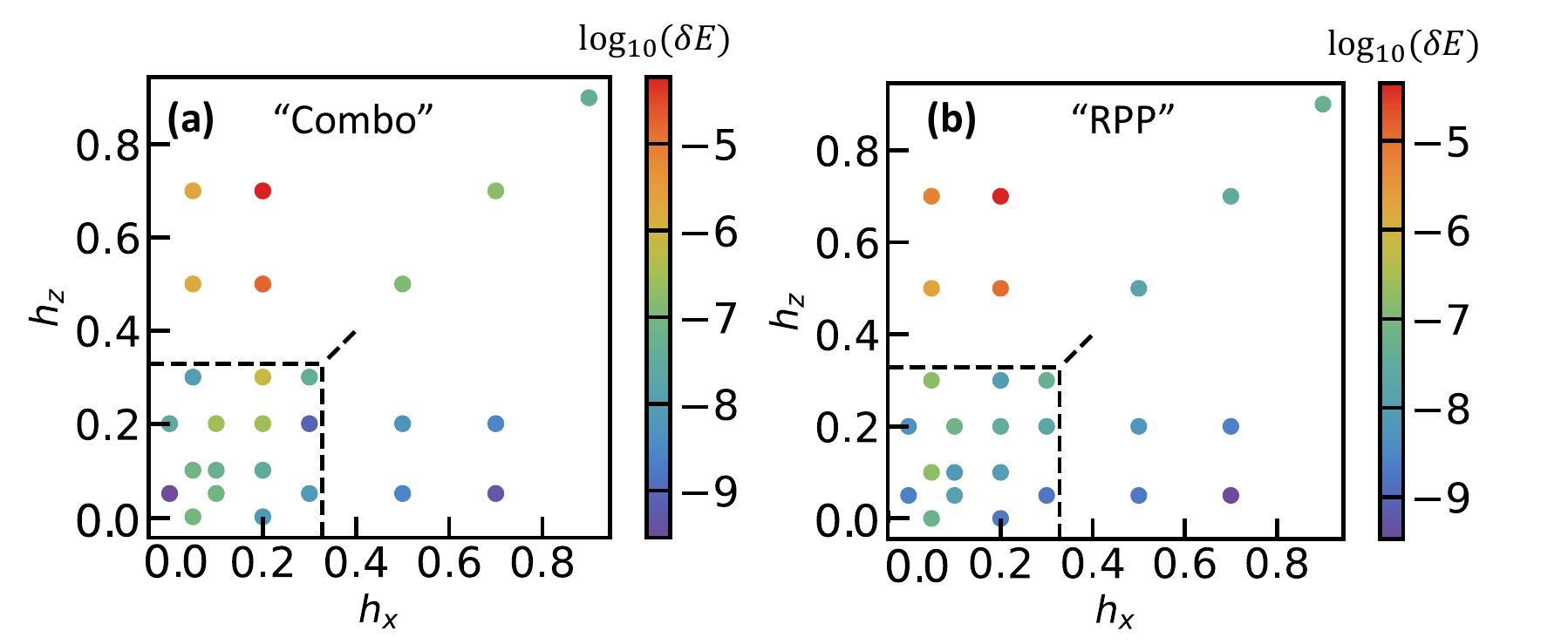}
\caption{Performance of (a) "Combo" and (b) "RPP" neural network architectures at different magnetic field strengths for system size $4 \times 4$ and $h_y=0$.} 
\label{fig:performance_architectures}
\end{figure}

For a $4 \times 4$ system and the ``Combo" and ``RPP" architectures, we plot the best relative error for different values of fields $(h_x,h_z)$, optimized over hyperparameters after multiple simulation runs on a cluster (Fig. \ref{fig:performance_architectures}). 

As expected from the built-in inductive bias, the neural network has the best performance for the small $h_z$ fields (where the approximate symmetries of the model are close to being exact). Although the network's performance drops slightly as the value of the $h_x$ field is increased (owing to the increasing complexity of the wavefunction associated with the higher correlation length), the drop along the $h_z$ field is much more pronounced. As mentioned in the main text, the accuracy does not sharply diminish at the edge of the topological phase, but instead slowly decreases with the $h_z$ field, making the ansatz applicable even when $h_z$ is roughly comparable to the strength of the toric-code Hamiltonian coupling. Further, the energies obtained from our architecture (although not directly comparable due to different boundary conditions and system sizes) are much lower than these obtained in Ref. \cite{valenti2022correlation} when studied on the perturbed toric code model. Finally, we do not observe any significant differences in performance between the ``RPP" and ``Combo" architectures. We note that for $4\times4$ systems, having a kernel size for the invariant part of the network covering the entire system ($O(N)$) is crucial for achieving the reported accuracies (in contrast to the non-invariant network kernel size which can be $O(1)$ without any significant loss in accuracy). 

\subsection{Performance under different perturbations for $4 \times 4$}
We demonstrate that our ansatz is stable against a variety of different Hamiltonian perturbations: including ones introducing a sign problem. We broadly divide perturbations of the toric code Hamiltonian into different classes (see Table \ref{tab:perturbations}). For sign-problem-full perturbations, the Hamiltonian will not be stoquastic, and therefore the Frobenius theorem does not apply, permitting a non-trivial ground-state sign structure. For sign-problem-full perturbations we therefore allow for complex network outputs by using complex parameters.

\begin{table}[h]
\centering
\begin{tabular}{|c|c|c|}
\hline
\textbf{Perturbation type} & \textbf{Perturbation} & \textbf{$\epsilon_{rel}$} \\ \hline
\multirow{1}{*}{Sign-problem-free, invariant}   
                           & XX       & 1e-9                     \\ \hline
\multirow{3}{*}{Sign-problem-free, non-invariant}    
                           & ZZ       & 6e-8                           \\ \cline{2-3}
                           & XX+YY+ZZ       & 2e-5                      \\   \cline{2-3}
                           & XX+YY       & 1e-5                      \\  \hline
\multirow{1}{*}{Sign-problem-full, invariant}    & YYYY on plaquettes  & 1e-5        \\ \hline
\multirow{3}{*}{Sign-problem-full, non-invariant}    & Y  &  1e-4       \\ \cline{2-3} 
                           & YY       & 1e-3                           \\  \cline{2-3}
                           & YYYY on vertices      & 4e-6                      \\  \hline
\end{tabular}
\caption{Relative energies obtained within Combo architecture for different perturbations on N=24 toric code. All perturbations are fixed to a value of $0.2$ in magnitude and in addition we add a small $h_x=0.05$ field to all of them.}
\label{tab:perturbations}
\end{table}

We note that the performance of our ansatz has accuracy below $10^{-3}$ for all perturbations tested. As generically expected \cite{szabo2020neural}, perturbations introducing sign-problem are more difficult for the NQS. This is most likely due to difficulty in propagation of complex phases through the network \cite{jing2017tunable}. We notice that achieving the performance we have reached for the \textit{sign-problem-full} case hinges upon including correlations between different channels of the non-invariant architecture as depicted in Fig. 1(a) of the main text. For instance, architecture with a "bottleneck" where one sums over all channels before passing the output to the Wilson loop non-linearity, performs one to two orders of magnitude worse in accuracy as compared with the presented results.

Although we have not further tailored our ansatz for the sign-problem-full case, there are several ways of doing this e.g., by (i) experimenting with different complex phase representation e.g., by two decoupled networks with real parameters (one representing amplitude and the other the phase) \cite{szabo2020neural,astrakhantsev2021broken} and performing their sequential training  \cite{szabo2020neural,astrakhantsev2021broken}, (ii) modifying form of complex non-linearities within the complex parameters architectures \cite{jing2017tunable}. 

\subsection{Performance in $(h_x,h_y,h_z)$ magnetic fields for a $10 \times 10$ system}
\begin{figure}
\includegraphics[width=1\columnwidth]{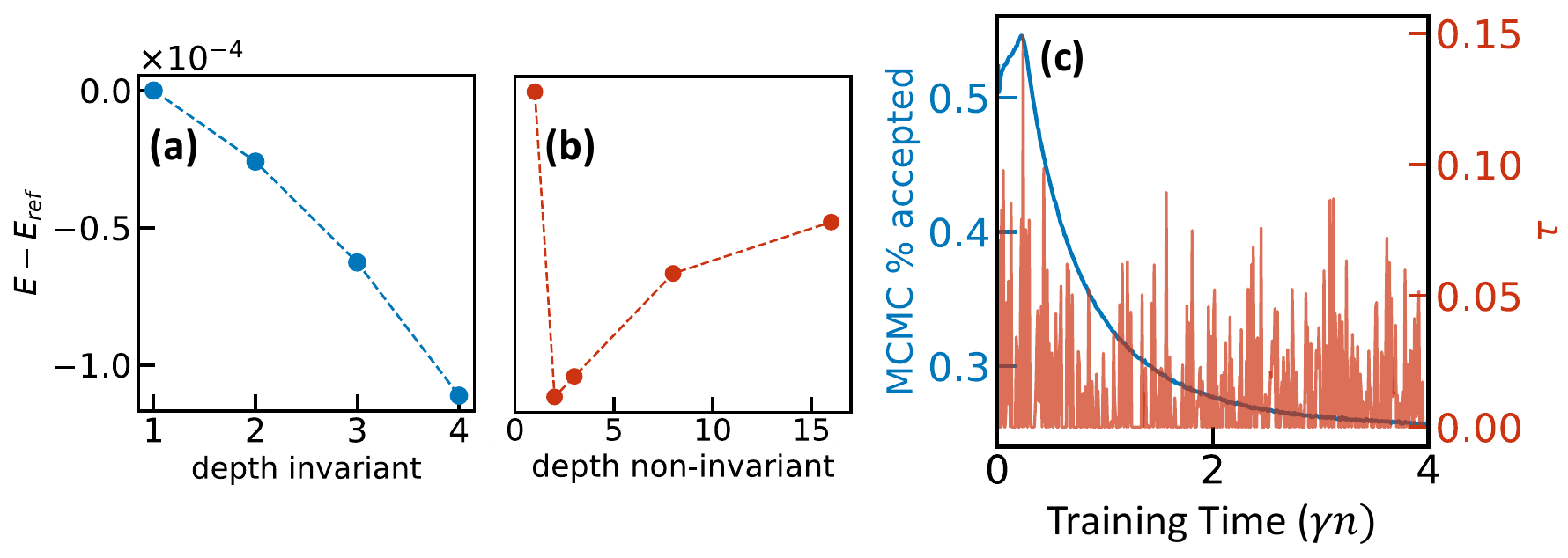}
\caption{ Scaling of the relative variational energies with the number of layers (depth) of the neural network architecture on a $10 \times 10$ model for (a) the invariant and (b) the non-invariant block under $(h_x,h_y,h_z)=(0.2,0.0,0.2)$ field. The accuracy of the ansatz can be systematically improved with the depth of the invariant block (but not the non-invariant one).  Reference energy corresponds to the depth$=1$ value. (c) Sampling statistics for a $(h_x,h_y,h_z)=(0.2,0.2,0.2)$ on $10 \times 10$ system. $\tau$ refers to an energy autocorrelation time. We see that the acceptance rate remains high throughout, while the autocorrelation time does not significantly increase.} 
\label{fig:depthscaling}
\end{figure}

\paragraph{Approximately-symmetric neural network analysis and improvement}
We proceed to investigating what limits the performance of the approximately-symmetric neural network. We first study the performance of the purely gauge-invariant architecture (constructed by omitting non-invariant block in the architecture in Fig. (2a) in the main text). We performed additional validations to characterize the converged ground states—such as evaluating order parameters for the topological phase—and found that the V-score \cite{wu2024variational} was of the same order of magnitude as the relative error compared to state-of-the-art DMRG/QMC results, consistent with expectations. Additionally, we note that the relative energies achieved for $(h_x,h_y,h_z)=(0.2,0.0,0.0)$ fields with a fully gauge-invariant architecture are comparable to these for $(h_x,h_y,h_z)=(0.2,0.0,0.2)$ with an approximately-symmetric one. We further observe a systematic scaling with the number of parameters and particularly depth of the network. We suggest therefore that the accuracy of the approximately-symmetric network behavior is not limited by the non-invariant block of the network but the invariant one. We investigate this hypothesis by investigating scaling of the relative energy error with the number of channels and depth of the symmetric and non-symmetric block of the network. In agreement with the purely gauge-invariant network results, we observe that by increasing the depth of the invariant part of the network, one can systematically improve the achievable energies (Fig. \ref{fig:depthscaling}(a)). No similar scaling was observed with the depth of the non-invariant block (Fig. \ref{fig:depthscaling}(b)) or the number of channels in the non-invariant or invariant part blocks (not shown). This conclusion holds for a range of other values of magnetic fields and system sizes. 

We therefore suggest that the accuracy of our approximately-symmetric architecture might be improved by increasing the depth of the invariant part of the architecture. This might be potentially achieved by trying to further stabilize the training of deep neural network architectures (e.g., by introducing batch or layer normalization, see also \cite{goodfellow2016deep}) and by increasing the efficiency of utilizing neural network parameters e.g., through pooling layers and decreasing invariant kernel sizes (simulations were run for $O(N)$-sized invariant kernels, but we did not observe any noticeable drop of accuracy in network performance with the reduction of the kernel size). Finally, one could try applying some of the transfer learning techniques \cite{zen2020transfer} and optimizing ansatz with minSR for shorter runtimes \cite{chen2023efficient}. We leave these suggested improvements for future work. 

\paragraph{Sampling statistics}
We provide more data on MCMC statistics for simulations on a $10 \times 10$  system under an $h_y$ field to demonstrate that, for the system under consideration and architecture chosen, one does not run into any significant sampling issues. Sampling for the sign-problem full toric code with an $h_y \neq 0$ field should be the most difficult: e.g., for stoquastic Hamiltonians it can be proven \cite{bravyi2022simulate,bravyi2023rapidly} that MCMC mixing time for sampling their ground states scales polynomially with the number of qubits. To demonstrate that there are no issues with sampling, we show that (Fig. \ref{fig:depthscaling}(c)): (i) the acceptance probabilities of the MCMC chain remain large ($ > 25 \%$) throughout the  optimization; (ii) the energy autocorrelation time $\tau$ is small (where $\tau=0$ is ideal and one would like to avoid $\tau \gg 1 $). Furthermore, we verified that the split-$\hat{R}$ characteristics \cite{vehtari2021rank} remains $<1.01$ throughout the simulation (where $\hat{R}$ is ideally $1$ and $\hat{R}<1.01$ tolerance is heuristically recommended \cite{vehtari2021rank}). 

\subsection{Extracting phase transitions for a mixed field toric code}
We provide more details about evaluating observables and extracting the phase diagram of the mixed field toric code under $h_y=0.3$ field (Fig. 1(c) main text). As mentioned in the main text, to find the phase transition we evaluate BFFM $X$ and $Z$ observables ($O_X$ and $O_Z$) and second Renyi entropy. Assuming that observables of interests have only a polynomial number of non-zero matrix elements in each column of the matrix, they can be efficiently evaluated in the NQS by MCMC sampling \cite{vicentini2022netket}. This is indeed the case for the $O_Z$ ($O_X$) order parameters (defined in the main text). Formally these observables become order parameters ($\mathcal{O}_Z$ and $\mathcal{O}_X$) only as the loop perimeter $m\rightarrow \infty$. In practice however, relatively small value of $m$ allows one to observe phase transitions. The intuition behind the $\mathcal{O}_Z$ ($\mathcal{O}_X$) order parameters is the following: close to $h_x=h_y=h_z=0$ (unperturbed toric code limit) one would expect that the numerator is vanishing (open half-loop string creates two excitations at its ends which have a vanishing expectation value in a ground state) and denominator finite (closed Wilson loops are near exact symmetries of the ground state); away from $h_x=h_y=h_z=0$ point, closed Wilson loop decay exponentially with the loop perimeter $m$, but this decay is compensated by a similar decay of the half-loops, leaving only the  contributions from the endpoints yielding vanishing expectation value of the string order parameter \cite{xu2024critical}. 

On the other hand, the second Renyi entropy can be efficiently evaluated using Monte Carlo method as the expectation value of a $\textrm{SWAP}$ operator on two copies of the state \cite{hastings2010measuring}. 

\begin{figure}
\includegraphics[width=0.4\columnwidth]{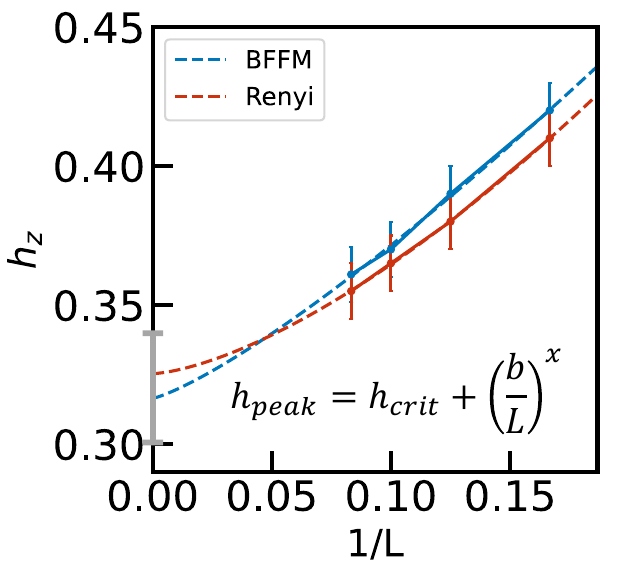}
\caption{Finite-size extrapolation of BFFM and Renyi observables for an $h_x=0$ cut through the phase diagram. $b$ and $x$ are fit parameters. Uncertainty in the $1/L \rightarrow 0$ intercept corresponds to the gray errorbar.} 
\label{fig:fss}
\end{figure}

For a fixed $h_z$ cut through the phase diagram we find no significant shift of the peak in the derivative of both BFFM and second Renyi entropy with increasing $L$ which we associate with the location of the phase transition. For a fixed $h_x$ cut, on the other hand, we find stronger finite-size effects. (Lack of symmetry between $X$ and $Z$ perturbations comes from the open boundary conditions: it might be seen in perturbation theory that $X$ and $Z$ magnetic field excitations at the boundary give contributions to energy in different orders of perturbation theory). To obtain a more accurate estimate of the thermodynamic critical points along these cuts, we perform a simple finite-size extrapolation of the position of extracted $\frac{\textrm{d}O_Z}{\textrm{d}h_z}$ and $\frac{\textrm{d}S_{\textrm{Renyi}}}{\textrm{d}h_z}$ peaks (see Fig. \ref{fig:fss}). In particular, we fit the peak location to a power law in inverse system size, $h_\textrm{peak} = h_\textrm{crit}+b(\frac{1}{L})^x$ and extrapolate $1/L \rightarrow 0$. Uncertainty in the extrapolation determines the uncertainty in critical point extraction (errorbars in Fig. 1(c) in the main text).

\section{Interpretability}
We proceed to discussing interpretability of the approximately-symmetric neural network. We first discuss some caveats on the approximate-to-exact symmetries mapping proposed as the operational principle of the neural network. We second proceed to discuss the network invariance error (a quantity well-known in the machine-learning community \cite{marcfinzithesis}) and demonstrate that its derivative appears to exhibit divergent behavior at the same location as the phase transition from the quantum spin liquid to a trivial phase.  

\subsection{Approximate to exact symmetries mapping}

In the main text we have argued that our approximately symmetric architecture operates mainly by mapping ``fattened" Wilson loops to their known form at the fixed point within the non-symmetric block, and later solves the fully symmetric problem in the symmetric block. We have shown this strictly for an independent neural network training procedure (see main text). We expect this conclusion to largely hold for training both blocks together. One subtlety of the joint training is that the non-symmetric block is capable of learning some symmetric features of the state as well. One of the consequences of this is that after such training on $h_z \neq 0$ (and $h_y=0$) simply setting weights of the non-invariant layer to an identity does not produce the $h_z=0$ ground state. We leave it for future investigations how to fully decouple features learnt by two blocks of the network during the joint training. 

\subsection{Invariance error} 
In the definition of the approximate group invariance $\mathbb{E}_{s \sim p(s)} \mathbb{E}_{g \in G} | \psi_{gs} - \psi_s | < \epsilon$, we defined a measure of the magnitude of the symmetry breaking, $\epsilon$. Here we discuss the dynamics and asymptotic values of the closely related quantity $\tilde \epsilon = \mathbb{E}_{s \sim p(s)} \mathbb{E}_{g \in G} \|\log \psi_{gs} - \log{\psi_s}\|$ as learnt by the neural network. We note that: (i) value of $\tilde \epsilon$ approaches the same value irrespective of the network hyperparameters over the course of optimization, (ii) $\tilde \epsilon$ appears to have a divergent derivative as one tunes the \textit{symmetry-violating} field in a point closely matching the more conventional observables of the toric code phase transitions such as BFFM order parameters (see main text). 

\begin{figure}
\includegraphics[width=1\columnwidth]{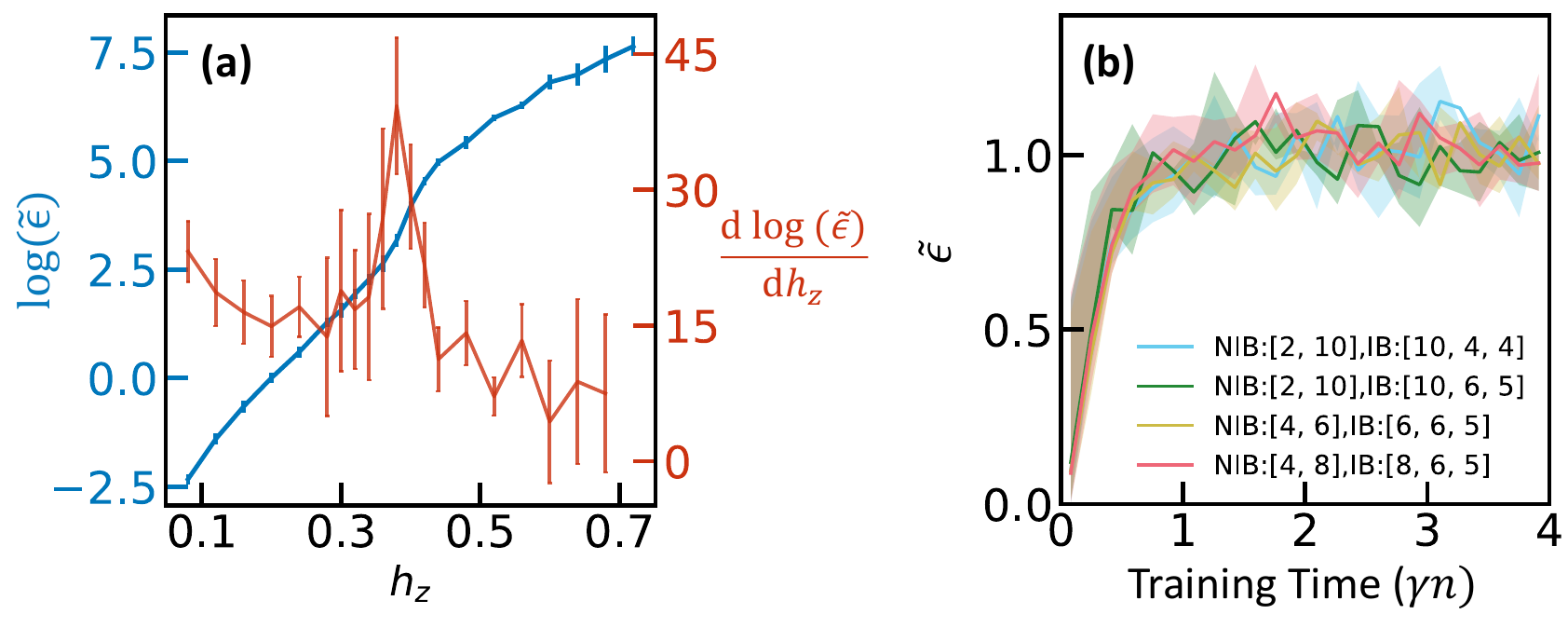}
\caption{(a) Logarithm of an invariance error $\tilde \epsilon$ (blue line) and its derivative (red line) for $L=8$ for $h_x=0.14,h_y=0$ as a function of the group symmetry breaking field $(h_z)$, displaying a peak at a point consistent at the phase transition point identified by more conventional observables at the same system size (see Fig. 3(a) and 3(c) in the main text). (b) Invariance error $\tilde \epsilon$ as a function of time for $(h_x=0.14,h_y=0.0, h_z=0.2)$ for $L=8$. Up to sampling errors, for any (successful) optimization, it converges to the fixed value, dependent only on the value of the $h_z$ field. NIB denotes non-invariant block, IB denotes invariant-block and numbers specify the channel configuration in every layer.}
\label{fig:interpretability_suppl}
\end{figure}

\paragraph{Invariance error universality} In the ``Combo" approximately-symmetric architecture discussed in the main text, one increases the (maximum) degree of non-invariance implicitly by increasing the number of channels in the non-invariant layer. We note, however, that $\tilde \epsilon$ (and correspondingly $\epsilon$) converges to the same value towards the end of optimization (see Fig. \ref{fig:interpretability_suppl}b), regardless of the number of channel configuration (assuming that optimization has succeeded; this might require a certain minimum number of non-invariant channels). 
\paragraph{Invariance error divergence} Furthermore, we found that the invariance error $\tilde \epsilon$ significantly increases with the increasing values of the $h_z$ field. To investigate this behaviour further we plot the $\log \tilde \epsilon$ as a function of $h_z$ field in the Fig. \ref{fig:interpretability_suppl}a. First, as expected, at $h_z=0$ the approximately-symmetric architecture learns to stay fully invariant (despite of extra flexibility). Second, we observe that $\mathrm{d} \log \tilde \epsilon/\mathrm{d} h_z$ reveals a peak matching the critical point found by more conventional observables (as in Fig. 3 main text). Such behavior is most likely comes from the connection of $\epsilon$ invariance error to the extent by which $X$-Wilson loops act non-trivially on a system's ground state: $\epsilon \sim || A_v |\psi \rangle - |\psi \rangle ||$. It should be noted that, as expected, $\tilde \epsilon$ exhibits no such divergent behaviour when tuning \textit{symmetry-preserving} terms (e.g., $h_x$ magnetic field) -- under which the network stays fully invariant.   

\section{Approximately-symmetric NQS for Rydberg spin liquids}

\begin{figure}[h]
    \includegraphics[width=1.0\columnwidth]{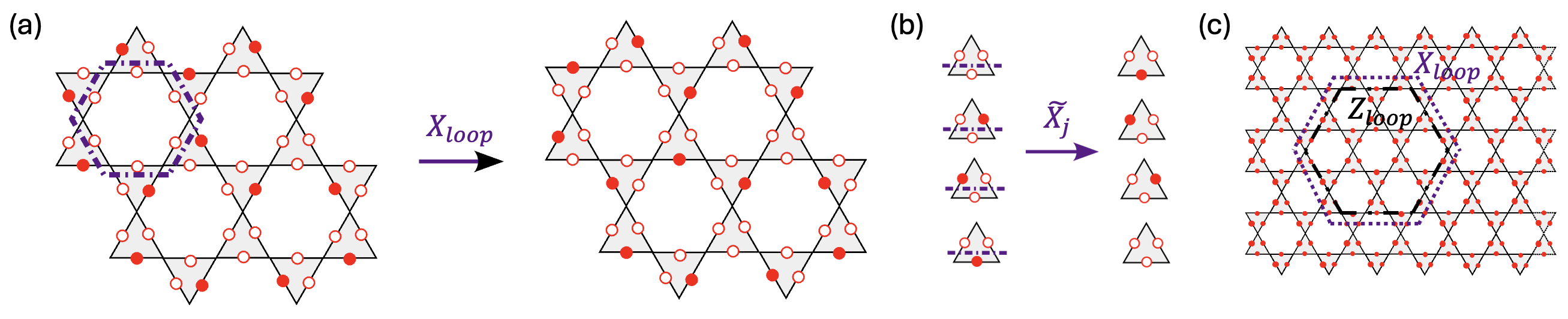} 
    \caption{(a) $X_{loop} = \prod_{j \in \textrm{hexagon}} \tilde{X}_j$ operators used to construct gauge invariant non-linearites for the ruby model. Red filled/empty circles denote Rydberg atoms in an excited/ground state (b) Action of  $\tilde{X}_j$ within each triangle (c) Loop operators used to calculate BFFM order parameters on $L=6$.}
    \label{fig:pxp_supp}
\end{figure}

Here we provide more information about approximately-symmetric neural network applied to a PXP limit of the Rydberg Hamiltonian on a ruby lattice. First, we efficiently incorporate the Rydberg blockade within each triangle by reducing the local Hilbert space to that of a spin-\(3/2\), since double-occupancy configurations cost infinite energy in the PXP limit. The resulting local basis over a triangle is \( \{ |ggg\rangle, |egg\rangle, |geg\rangle, |gge\rangle \} \), effectively defining a spin-\(3/2\) system on a honeycomb lattice. Second, we use an approximately-symmetric neural network consisting of a gauge invariant block, non-gauge-symmetric CNN added together with the non-gauge-symmetric mean-field ansatz (thus forming the ``RPP" architecture discussed before). Relevant gauge invariant Wilson operators -- with respect to which gauge-invariant product non-linearities are constructed -- are shown in Fig. \ref{fig:pxp_supp}(a) and Fig. \ref{fig:pxp_supp}(b) - see also Ref. \cite{verresen2021prediction}. The mean-field ansatz $\textrm{MF}= \sum_i c_i n_i$ is added for numerical stability. We symmetrize the input to the neural network with lattice symmetries (horizontal and vertical parity). We use, single layer, $4-20$ features in the non-symmetric part (in addition to the mean-field features), GELU activation functions and stochastic reconfiguration with the SVD-based solution to the set of linear equations. Finally, on small system sizes we benchmark our architecture against RBMs enhanced with spin-\(3/2\) blockade constraints and a mean-field ansatz respecting all lattice symmetries (including inversion around the \( x \) and \( y \) axes). 

We find evidence of the quantum spin liquid phase by inspecting Wilson loop operators for large system sizes ($L\geq 6$) and finite size versions of BFFM order parameters (on $L=6$ they evaluate to $\langle X_{BFFM} \rangle \sim 0.01$ for equal-superposition probing loop passing through $12$ spins-$1/2$ and $\langle Z_{BFFM} \rangle \sim 0.07$ for Gauss' law probing loop passing through $18$ spins-$1/2$), in agreement with the predictions from iDMRG \cite{verresen2021prediction}. Loops used to calculate BFFM order parameters are defined in Fig. \ref{fig:pxp_supp}(c).

%


\begin{thebibliography}{111}%
\makeatletter
\providecommand \@ifxundefined [1]{%
 \@ifx{#1\undefined}
}%
\providecommand \@ifnum [1]{%
 \ifnum #1\expandafter \@firstoftwo
 \else \expandafter \@secondoftwo
 \fi
}%
\providecommand \@ifx [1]{%
 \ifx #1\expandafter \@firstoftwo
 \else \expandafter \@secondoftwo
 \fi
}%
\providecommand \natexlab [1]{#1}%
\providecommand \enquote  [1]{``#1''}%
\providecommand \bibnamefont  [1]{#1}%
\providecommand \bibfnamefont [1]{#1}%
\providecommand \citenamefont [1]{#1}%
\providecommand \href@noop [0]{\@secondoftwo}%
\providecommand \href [0]{\begingroup \@sanitize@url \@href}%
\providecommand \@href[1]{\@@startlink{#1}\@@href}%
\providecommand \@@href[1]{\endgroup#1\@@endlink}%
\providecommand \@sanitize@url [0]{\catcode `\\12\catcode `\$12\catcode `\&12\catcode `\#12\catcode `\^12\catcode `\_12\catcode `\%12\relax}%
\providecommand \@@startlink[1]{}%
\providecommand \@@endlink[0]{}%
\providecommand \url  [0]{\begingroup\@sanitize@url \@url }%
\providecommand \@url [1]{\endgroup\@href {#1}{\urlprefix }}%
\providecommand \urlprefix  [0]{URL }%
\providecommand \Eprint [0]{\href }%
\providecommand \doibase [0]{https://doi.org/}%
\providecommand \selectlanguage [0]{\@gobble}%
\providecommand \bibinfo  [0]{\@secondoftwo}%
\providecommand \bibfield  [0]{\@secondoftwo}%
\providecommand \translation [1]{[#1]}%
\providecommand \BibitemOpen [0]{}%
\providecommand \bibitemStop [0]{}%
\providecommand \bibitemNoStop [0]{.\EOS\space}%
\providecommand \EOS [0]{\spacefactor3000\relax}%
\providecommand \BibitemShut  [1]{\csname bibitem#1\endcsname}%
\let\auto@bib@innerbib\@empty
\bibitem [{\citenamefont {Anderson}(1973)}]{anderson1973resonating}%
  \BibitemOpen
  \bibfield  {author} {\bibinfo {author} {\bibfnamefont {P.~W.}\ \bibnamefont {Anderson}},\ }\bibfield  {title} {\bibinfo {title} {Resonating valence bonds: A new kind of insulator?},\ }\href@noop {} {\bibfield  {journal} {\bibinfo  {journal} {Materials Research Bulletin}\ }\textbf {\bibinfo {volume} {8}},\ \bibinfo {pages} {153} (\bibinfo {year} {1973})}\BibitemShut {NoStop}%
\bibitem [{\citenamefont {Kitaev}(2006)}]{kitaev2006anyons}%
  \BibitemOpen
  \bibfield  {author} {\bibinfo {author} {\bibfnamefont {A.}~\bibnamefont {Kitaev}},\ }\bibfield  {title} {\bibinfo {title} {Anyons in an exactly solved model and beyond},\ }\href@noop {} {\bibfield  {journal} {\bibinfo  {journal} {Annals of Physics}\ }\textbf {\bibinfo {volume} {321}},\ \bibinfo {pages} {2} (\bibinfo {year} {2006})}\BibitemShut {NoStop}%
\bibitem [{\citenamefont {Savary}\ and\ \citenamefont {Balents}(2016)}]{savary2016quantum}%
  \BibitemOpen
  \bibfield  {author} {\bibinfo {author} {\bibfnamefont {L.}~\bibnamefont {Savary}}\ and\ \bibinfo {author} {\bibfnamefont {L.}~\bibnamefont {Balents}},\ }\bibfield  {title} {\bibinfo {title} {Quantum spin liquids: a review},\ }\href@noop {} {\bibfield  {journal} {\bibinfo  {journal} {Reports on Progress in Physics}\ }\textbf {\bibinfo {volume} {80}},\ \bibinfo {pages} {016502} (\bibinfo {year} {2016})}\BibitemShut {NoStop}%
\bibitem [{\citenamefont {Verresen}\ \emph {et~al.}(2021)\citenamefont {Verresen}, \citenamefont {Lukin},\ and\ \citenamefont {Vishwanath}}]{verresen2021prediction}%
  \BibitemOpen
  \bibfield  {author} {\bibinfo {author} {\bibfnamefont {R.}~\bibnamefont {Verresen}}, \bibinfo {author} {\bibfnamefont {M.~D.}\ \bibnamefont {Lukin}},\ and\ \bibinfo {author} {\bibfnamefont {A.}~\bibnamefont {Vishwanath}},\ }\bibfield  {title} {\bibinfo {title} {Prediction of toric code topological order from rydberg blockade},\ }\href@noop {} {\bibfield  {journal} {\bibinfo  {journal} {Physical Review X}\ }\textbf {\bibinfo {volume} {11}},\ \bibinfo {pages} {031005} (\bibinfo {year} {2021})}\BibitemShut {NoStop}%
\bibitem [{\citenamefont {Semeghini}\ \emph {et~al.}(2021)\citenamefont {Semeghini}, \citenamefont {Levine}, \citenamefont {Keesling}, \citenamefont {Ebadi}, \citenamefont {Wang}, \citenamefont {Bluvstein}, \citenamefont {Verresen}, \citenamefont {Pichler}, \citenamefont {Kalinowski}, \citenamefont {Samajdar} \emph {et~al.}}]{semeghini2021probing}%
  \BibitemOpen
  \bibfield  {author} {\bibinfo {author} {\bibfnamefont {G.}~\bibnamefont {Semeghini}}, \bibinfo {author} {\bibfnamefont {H.}~\bibnamefont {Levine}}, \bibinfo {author} {\bibfnamefont {A.}~\bibnamefont {Keesling}}, \bibinfo {author} {\bibfnamefont {S.}~\bibnamefont {Ebadi}}, \bibinfo {author} {\bibfnamefont {T.~T.}\ \bibnamefont {Wang}}, \bibinfo {author} {\bibfnamefont {D.}~\bibnamefont {Bluvstein}}, \bibinfo {author} {\bibfnamefont {R.}~\bibnamefont {Verresen}}, \bibinfo {author} {\bibfnamefont {H.}~\bibnamefont {Pichler}}, \bibinfo {author} {\bibfnamefont {M.}~\bibnamefont {Kalinowski}}, \bibinfo {author} {\bibfnamefont {R.}~\bibnamefont {Samajdar}}, \emph {et~al.},\ }\bibfield  {title} {\bibinfo {title} {Probing topological spin liquids on a programmable quantum simulator},\ }\href@noop {} {\bibfield  {journal} {\bibinfo  {journal} {Science}\ }\textbf {\bibinfo {volume} {374}},\ \bibinfo {pages} {1242} (\bibinfo {year} {2021})}\BibitemShut {NoStop}%
\bibitem [{\citenamefont {Satzinger}\ \emph {et~al.}(2021)\citenamefont {Satzinger}, \citenamefont {Liu}, \citenamefont {Smith}, \citenamefont {Knapp}, \citenamefont {Newman}, \citenamefont {Jones}, \citenamefont {Chen}, \citenamefont {Quintana}, \citenamefont {Mi}, \citenamefont {Dunsworth} \emph {et~al.}}]{googleTC}%
  \BibitemOpen
  \bibfield  {author} {\bibinfo {author} {\bibfnamefont {K.}~\bibnamefont {Satzinger}}, \bibinfo {author} {\bibfnamefont {Y.-J.}\ \bibnamefont {Liu}}, \bibinfo {author} {\bibfnamefont {A.}~\bibnamefont {Smith}}, \bibinfo {author} {\bibfnamefont {C.}~\bibnamefont {Knapp}}, \bibinfo {author} {\bibfnamefont {M.}~\bibnamefont {Newman}}, \bibinfo {author} {\bibfnamefont {C.}~\bibnamefont {Jones}}, \bibinfo {author} {\bibfnamefont {Z.}~\bibnamefont {Chen}}, \bibinfo {author} {\bibfnamefont {C.}~\bibnamefont {Quintana}}, \bibinfo {author} {\bibfnamefont {X.}~\bibnamefont {Mi}}, \bibinfo {author} {\bibfnamefont {A.}~\bibnamefont {Dunsworth}}, \emph {et~al.},\ }\bibfield  {title} {\bibinfo {title} {Realizing topologically ordered states on a quantum processor},\ }\href@noop {} {\bibfield  {journal} {\bibinfo  {journal} {Science}\ }\textbf {\bibinfo {volume} {374}},\ \bibinfo {pages} {1237} (\bibinfo {year} {2021})}\BibitemShut {NoStop}%
\bibitem [{\citenamefont {Iqbal}\ \emph {et~al.}(2024)\citenamefont {Iqbal}, \citenamefont {Tantivasadakarn}, \citenamefont {Verresen}, \citenamefont {Campbell}, \citenamefont {Dreiling}, \citenamefont {Figgatt}, \citenamefont {Gaebler}, \citenamefont {Johansen}, \citenamefont {Mills}, \citenamefont {Moses} \emph {et~al.}}]{iqbal2024non}%
  \BibitemOpen
  \bibfield  {author} {\bibinfo {author} {\bibfnamefont {M.}~\bibnamefont {Iqbal}}, \bibinfo {author} {\bibfnamefont {N.}~\bibnamefont {Tantivasadakarn}}, \bibinfo {author} {\bibfnamefont {R.}~\bibnamefont {Verresen}}, \bibinfo {author} {\bibfnamefont {S.~L.}\ \bibnamefont {Campbell}}, \bibinfo {author} {\bibfnamefont {J.~M.}\ \bibnamefont {Dreiling}}, \bibinfo {author} {\bibfnamefont {C.}~\bibnamefont {Figgatt}}, \bibinfo {author} {\bibfnamefont {J.~P.}\ \bibnamefont {Gaebler}}, \bibinfo {author} {\bibfnamefont {J.}~\bibnamefont {Johansen}}, \bibinfo {author} {\bibfnamefont {M.}~\bibnamefont {Mills}}, \bibinfo {author} {\bibfnamefont {S.~A.}\ \bibnamefont {Moses}}, \emph {et~al.},\ }\bibfield  {title} {\bibinfo {title} {Non-abelian topological order and anyons on a trapped-ion processor},\ }\href@noop {} {\bibfield  {journal} {\bibinfo  {journal} {Nature}\ }\textbf {\bibinfo {volume} {626}},\ \bibinfo {pages} {505} (\bibinfo {year} {2024})}\BibitemShut {NoStop}%
\bibitem [{goo(2023)}]{google2023non}%
  \BibitemOpen
  \bibfield  {title} {\bibinfo {title} {Non-abelian braiding of graph vertices in a superconducting processor},\ }\href@noop {} {\bibfield  {journal} {\bibinfo  {journal} {Nature}\ }\textbf {\bibinfo {volume} {618}},\ \bibinfo {pages} {264} (\bibinfo {year} {2023})}\BibitemShut {NoStop}%
\bibitem [{\citenamefont {Broholm}\ \emph {et~al.}(2020)\citenamefont {Broholm}, \citenamefont {Cava}, \citenamefont {Kivelson}, \citenamefont {Nocera}, \citenamefont {Norman},\ and\ \citenamefont {Senthil}}]{broholm2020}%
  \BibitemOpen
  \bibfield  {author} {\bibinfo {author} {\bibfnamefont {C.}~\bibnamefont {Broholm}}, \bibinfo {author} {\bibfnamefont {R.~J.}\ \bibnamefont {Cava}}, \bibinfo {author} {\bibfnamefont {S.~A.}\ \bibnamefont {Kivelson}}, \bibinfo {author} {\bibfnamefont {D.~G.}\ \bibnamefont {Nocera}}, \bibinfo {author} {\bibfnamefont {M.~R.}\ \bibnamefont {Norman}},\ and\ \bibinfo {author} {\bibfnamefont {T.}~\bibnamefont {Senthil}},\ }\bibfield  {title} {\bibinfo {title} {Quantum spin liquids},\ }\href {https://doi.org/10.1126/science.aay0668} {\bibfield  {journal} {\bibinfo  {journal} {Science}\ }\textbf {\bibinfo {volume} {367}},\ \bibinfo {pages} {eaay0668} (\bibinfo {year} {2020})},\ \Eprint {https://arxiv.org/abs/https://www.science.org/doi/pdf/10.1126/science.aay0668} {https://www.science.org/doi/pdf/10.1126/science.aay0668} \BibitemShut {NoStop}%
\bibitem [{\citenamefont {Xu}\ \emph {et~al.}(2023)\citenamefont {Xu}, \citenamefont {Bag}, \citenamefont {Sherman}, \citenamefont {Yadav}, \citenamefont {Kolesnikov}, \citenamefont {Podlesnyak}, \citenamefont {Moore},\ and\ \citenamefont {Haravifard}}]{xu2023realization}%
  \BibitemOpen
  \bibfield  {author} {\bibinfo {author} {\bibfnamefont {S.}~\bibnamefont {Xu}}, \bibinfo {author} {\bibfnamefont {R.}~\bibnamefont {Bag}}, \bibinfo {author} {\bibfnamefont {N.~E.}\ \bibnamefont {Sherman}}, \bibinfo {author} {\bibfnamefont {L.}~\bibnamefont {Yadav}}, \bibinfo {author} {\bibfnamefont {A.~I.}\ \bibnamefont {Kolesnikov}}, \bibinfo {author} {\bibfnamefont {A.~A.}\ \bibnamefont {Podlesnyak}}, \bibinfo {author} {\bibfnamefont {J.~E.}\ \bibnamefont {Moore}},\ and\ \bibinfo {author} {\bibfnamefont {S.}~\bibnamefont {Haravifard}},\ }\bibfield  {title} {\bibinfo {title} {Realization of u (1) dirac quantum spin liquid in ybzn2gao5},\ }\href@noop {} {\bibfield  {journal} {\bibinfo  {journal} {arXiv preprint arXiv:2305.20040}\ } (\bibinfo {year} {2023})}\BibitemShut {NoStop}%
\bibitem [{\citenamefont {Scheie}\ \emph {et~al.}(2024)\citenamefont {Scheie}, \citenamefont {Ghioldi}, \citenamefont {Xing}, \citenamefont {Paddison}, \citenamefont {Sherman}, \citenamefont {Dupont}, \citenamefont {Sanjeewa}, \citenamefont {Lee}, \citenamefont {Woods}, \citenamefont {Abernathy} \emph {et~al.}}]{scheie2024proximate}%
  \BibitemOpen
  \bibfield  {author} {\bibinfo {author} {\bibfnamefont {A.}~\bibnamefont {Scheie}}, \bibinfo {author} {\bibfnamefont {E.}~\bibnamefont {Ghioldi}}, \bibinfo {author} {\bibfnamefont {J.}~\bibnamefont {Xing}}, \bibinfo {author} {\bibfnamefont {J.}~\bibnamefont {Paddison}}, \bibinfo {author} {\bibfnamefont {N.}~\bibnamefont {Sherman}}, \bibinfo {author} {\bibfnamefont {M.}~\bibnamefont {Dupont}}, \bibinfo {author} {\bibfnamefont {L.}~\bibnamefont {Sanjeewa}}, \bibinfo {author} {\bibfnamefont {S.}~\bibnamefont {Lee}}, \bibinfo {author} {\bibfnamefont {A.}~\bibnamefont {Woods}}, \bibinfo {author} {\bibfnamefont {D.}~\bibnamefont {Abernathy}}, \emph {et~al.},\ }\bibfield  {title} {\bibinfo {title} {Proximate spin liquid and fractionalization in the triangular antiferromagnet kybse2},\ }\href@noop {} {\bibfield  {journal} {\bibinfo  {journal} {Nature Physics}\ }\textbf {\bibinfo {volume} {20}},\ \bibinfo {pages} {74} (\bibinfo {year} {2024})}\BibitemShut {NoStop}%
\bibitem [{\citenamefont {Zhang}\ \emph {et~al.}(2024)\citenamefont {Zhang}, \citenamefont {He}, \citenamefont {Zhang}, \citenamefont {Chen}, \citenamefont {Jia}, \citenamefont {Hou}, \citenamefont {Ji}, \citenamefont {Yang}, \citenamefont {Zhang}, \citenamefont {Liu}, \citenamefont {Gao}, \citenamefont {Jung},\ and\ \citenamefont {Wang}}]{zhang:2024}%
  \BibitemOpen
  \bibfield  {author} {\bibinfo {author} {\bibfnamefont {Q.}~\bibnamefont {Zhang}}, \bibinfo {author} {\bibfnamefont {W.-Y.}\ \bibnamefont {He}}, \bibinfo {author} {\bibfnamefont {Y.}~\bibnamefont {Zhang}}, \bibinfo {author} {\bibfnamefont {Y.}~\bibnamefont {Chen}}, \bibinfo {author} {\bibfnamefont {L.}~\bibnamefont {Jia}}, \bibinfo {author} {\bibfnamefont {Y.}~\bibnamefont {Hou}}, \bibinfo {author} {\bibfnamefont {H.}~\bibnamefont {Ji}}, \bibinfo {author} {\bibfnamefont {H.}~\bibnamefont {Yang}}, \bibinfo {author} {\bibfnamefont {T.}~\bibnamefont {Zhang}}, \bibinfo {author} {\bibfnamefont {L.}~\bibnamefont {Liu}}, \bibinfo {author} {\bibfnamefont {H.-J.}\ \bibnamefont {Gao}}, \bibinfo {author} {\bibfnamefont {T.~A.}\ \bibnamefont {Jung}},\ and\ \bibinfo {author} {\bibfnamefont {Y.}~\bibnamefont {Wang}},\ }\bibfield  {title} {\bibinfo {title} {Quantum spin liquid signatures in monolayer {{1T-NbSe2}}},\ }\href {https://doi.org/10.1038/s41467-024-46612-1} {\bibfield  {journal} {\bibinfo  {journal} {Nature
  Communications}\ }\textbf {\bibinfo {volume} {15}},\ \bibinfo {pages} {2336} (\bibinfo {year} {2024})}\BibitemShut {NoStop}%
\bibitem [{\citenamefont {Hastings}\ and\ \citenamefont {Wen}(2005)}]{hastingswen}%
  \BibitemOpen
  \bibfield  {author} {\bibinfo {author} {\bibfnamefont {M.~B.}\ \bibnamefont {Hastings}}\ and\ \bibinfo {author} {\bibfnamefont {X.-G.}\ \bibnamefont {Wen}},\ }\bibfield  {title} {\bibinfo {title} {Quasiadiabatic continuation of quantum states: The stability of topological ground-state degeneracy and emergent gauge invariance},\ }\href@noop {} {\bibfield  {journal} {\bibinfo  {journal} {Physical review b}\ }\textbf {\bibinfo {volume} {72}},\ \bibinfo {pages} {045141} (\bibinfo {year} {2005})}\BibitemShut {NoStop}%
\bibitem [{\citenamefont {Schollwöck}(2011)}]{SCHOLLWOCK201196}%
  \BibitemOpen
  \bibfield  {author} {\bibinfo {author} {\bibfnamefont {U.}~\bibnamefont {Schollwöck}},\ }\bibfield  {title} {\bibinfo {title} {The density-matrix renormalization group in the age of matrix product states},\ }\href {https://doi.org/https://doi.org/10.1016/j.aop.2010.09.012} {\bibfield  {journal} {\bibinfo  {journal} {Annals of Physics}\ }\textbf {\bibinfo {volume} {326}},\ \bibinfo {pages} {96} (\bibinfo {year} {2011})},\ \bibinfo {note} {january 2011 Special Issue}\BibitemShut {NoStop}%
\bibitem [{\citenamefont {Haegeman}\ \emph {et~al.}(2011)\citenamefont {Haegeman}, \citenamefont {Cirac}, \citenamefont {Osborne}, \citenamefont {Pi{\v{z}}orn}, \citenamefont {Verschelde},\ and\ \citenamefont {Verstraete}}]{haegeman2011time}%
  \BibitemOpen
  \bibfield  {author} {\bibinfo {author} {\bibfnamefont {J.}~\bibnamefont {Haegeman}}, \bibinfo {author} {\bibfnamefont {J.~I.}\ \bibnamefont {Cirac}}, \bibinfo {author} {\bibfnamefont {T.~J.}\ \bibnamefont {Osborne}}, \bibinfo {author} {\bibfnamefont {I.}~\bibnamefont {Pi{\v{z}}orn}}, \bibinfo {author} {\bibfnamefont {H.}~\bibnamefont {Verschelde}},\ and\ \bibinfo {author} {\bibfnamefont {F.}~\bibnamefont {Verstraete}},\ }\bibfield  {title} {\bibinfo {title} {Time-dependent variational principle for quantum lattices},\ }\href@noop {} {\bibfield  {journal} {\bibinfo  {journal} {Physical review letters}\ }\textbf {\bibinfo {volume} {107}},\ \bibinfo {pages} {070601} (\bibinfo {year} {2011})}\BibitemShut {NoStop}%
\bibitem [{\citenamefont {Ba{\~n}uls}(2023)}]{banuls2023tensor}%
  \BibitemOpen
  \bibfield  {author} {\bibinfo {author} {\bibfnamefont {M.~C.}\ \bibnamefont {Ba{\~n}uls}},\ }\bibfield  {title} {\bibinfo {title} {Tensor network algorithms: A route map},\ }\href@noop {} {\bibfield  {journal} {\bibinfo  {journal} {Annual Review of Condensed Matter Physics}\ }\textbf {\bibinfo {volume} {14}},\ \bibinfo {pages} {173} (\bibinfo {year} {2023})}\BibitemShut {NoStop}%
\bibitem [{\citenamefont {McMillan}(1965)}]{McMillan1965}%
  \BibitemOpen
  \bibfield  {author} {\bibinfo {author} {\bibfnamefont {W.~L.}\ \bibnamefont {McMillan}},\ }\bibfield  {title} {\bibinfo {title} {Ground state of liquid ${\mathrm{he}}^{4}$},\ }\href {https://doi.org/10.1103/PhysRev.138.A442} {\bibfield  {journal} {\bibinfo  {journal} {Phys. Rev.}\ }\textbf {\bibinfo {volume} {138}},\ \bibinfo {pages} {A442} (\bibinfo {year} {1965})}\BibitemShut {NoStop}%
\bibitem [{\citenamefont {Ceperley}\ \emph {et~al.}(1977)\citenamefont {Ceperley}, \citenamefont {Chester},\ and\ \citenamefont {Kalos}}]{Ceperley1977}%
  \BibitemOpen
  \bibfield  {author} {\bibinfo {author} {\bibfnamefont {D.}~\bibnamefont {Ceperley}}, \bibinfo {author} {\bibfnamefont {G.~V.}\ \bibnamefont {Chester}},\ and\ \bibinfo {author} {\bibfnamefont {M.~H.}\ \bibnamefont {Kalos}},\ }\bibfield  {title} {\bibinfo {title} {Monte carlo simulation of a many-fermion study},\ }\href {https://doi.org/10.1103/PhysRevB.16.3081} {\bibfield  {journal} {\bibinfo  {journal} {Phys. Rev. B}\ }\textbf {\bibinfo {volume} {16}},\ \bibinfo {pages} {3081} (\bibinfo {year} {1977})}\BibitemShut {NoStop}%
\bibitem [{\citenamefont {Kent}\ \emph {et~al.}(1999)\citenamefont {Kent}, \citenamefont {Needs},\ and\ \citenamefont {Rajagopal}}]{Kent1999}%
  \BibitemOpen
  \bibfield  {author} {\bibinfo {author} {\bibfnamefont {P.~R.~C.}\ \bibnamefont {Kent}}, \bibinfo {author} {\bibfnamefont {R.~J.}\ \bibnamefont {Needs}},\ and\ \bibinfo {author} {\bibfnamefont {G.}~\bibnamefont {Rajagopal}},\ }\bibfield  {title} {\bibinfo {title} {Monte carlo energy and variance-minimization techniques for optimizing many-body wave functions},\ }\href {https://doi.org/10.1103/PhysRevB.59.12344} {\bibfield  {journal} {\bibinfo  {journal} {Phys. Rev. B}\ }\textbf {\bibinfo {volume} {59}},\ \bibinfo {pages} {12344} (\bibinfo {year} {1999})}\BibitemShut {NoStop}%
\bibitem [{\citenamefont {Foulkes}\ \emph {et~al.}(2001)\citenamefont {Foulkes}, \citenamefont {Mitas}, \citenamefont {Needs},\ and\ \citenamefont {Rajagopal}}]{Foulkes2001}%
  \BibitemOpen
  \bibfield  {author} {\bibinfo {author} {\bibfnamefont {W.~M.~C.}\ \bibnamefont {Foulkes}}, \bibinfo {author} {\bibfnamefont {L.}~\bibnamefont {Mitas}}, \bibinfo {author} {\bibfnamefont {R.~J.}\ \bibnamefont {Needs}},\ and\ \bibinfo {author} {\bibfnamefont {G.}~\bibnamefont {Rajagopal}},\ }\bibfield  {title} {\bibinfo {title} {Quantum monte carlo simulations of solids},\ }\href {https://doi.org/10.1103/RevModPhys.73.33} {\bibfield  {journal} {\bibinfo  {journal} {Rev. Mod. Phys.}\ }\textbf {\bibinfo {volume} {73}},\ \bibinfo {pages} {33} (\bibinfo {year} {2001})}\BibitemShut {NoStop}%
\bibitem [{\citenamefont {Carleo}\ and\ \citenamefont {Troyer}(2017)}]{carleo2017solving}%
  \BibitemOpen
  \bibfield  {author} {\bibinfo {author} {\bibfnamefont {G.}~\bibnamefont {Carleo}}\ and\ \bibinfo {author} {\bibfnamefont {M.}~\bibnamefont {Troyer}},\ }\bibfield  {title} {\bibinfo {title} {Solving the quantum many-body problem with artificial neural networks},\ }\href@noop {} {\bibfield  {journal} {\bibinfo  {journal} {Science}\ }\textbf {\bibinfo {volume} {355}},\ \bibinfo {pages} {602} (\bibinfo {year} {2017})}\BibitemShut {NoStop}%
\bibitem [{\citenamefont {Hornik}\ \emph {et~al.}(1989)\citenamefont {Hornik}, \citenamefont {Stinchcombe},\ and\ \citenamefont {White}}]{hornik1989multilayer}%
  \BibitemOpen
  \bibfield  {author} {\bibinfo {author} {\bibfnamefont {K.}~\bibnamefont {Hornik}}, \bibinfo {author} {\bibfnamefont {M.}~\bibnamefont {Stinchcombe}},\ and\ \bibinfo {author} {\bibfnamefont {H.}~\bibnamefont {White}},\ }\bibfield  {title} {\bibinfo {title} {Multilayer feedforward networks are universal approximators},\ }\href@noop {} {\bibfield  {journal} {\bibinfo  {journal} {Neural networks}\ }\textbf {\bibinfo {volume} {2}},\ \bibinfo {pages} {359} (\bibinfo {year} {1989})}\BibitemShut {NoStop}%
\bibitem [{\citenamefont {Sharir}\ \emph {et~al.}(2022)\citenamefont {Sharir}, \citenamefont {Shashua},\ and\ \citenamefont {Carleo}}]{sharir2022neural}%
  \BibitemOpen
  \bibfield  {author} {\bibinfo {author} {\bibfnamefont {O.}~\bibnamefont {Sharir}}, \bibinfo {author} {\bibfnamefont {A.}~\bibnamefont {Shashua}},\ and\ \bibinfo {author} {\bibfnamefont {G.}~\bibnamefont {Carleo}},\ }\bibfield  {title} {\bibinfo {title} {Neural tensor contractions and the expressive power of deep neural quantum states},\ }\href@noop {} {\bibfield  {journal} {\bibinfo  {journal} {Physical Review B}\ }\textbf {\bibinfo {volume} {106}},\ \bibinfo {pages} {205136} (\bibinfo {year} {2022})}\BibitemShut {NoStop}%
\bibitem [{\citenamefont {Sharir}\ \emph {et~al.}(2020)\citenamefont {Sharir}, \citenamefont {Levine}, \citenamefont {Wies}, \citenamefont {Carleo},\ and\ \citenamefont {Shashua}}]{sharir2020deep}%
  \BibitemOpen
  \bibfield  {author} {\bibinfo {author} {\bibfnamefont {O.}~\bibnamefont {Sharir}}, \bibinfo {author} {\bibfnamefont {Y.}~\bibnamefont {Levine}}, \bibinfo {author} {\bibfnamefont {N.}~\bibnamefont {Wies}}, \bibinfo {author} {\bibfnamefont {G.}~\bibnamefont {Carleo}},\ and\ \bibinfo {author} {\bibfnamefont {A.}~\bibnamefont {Shashua}},\ }\bibfield  {title} {\bibinfo {title} {Deep autoregressive models for the efficient variational simulation of many-body quantum systems},\ }\href@noop {} {\bibfield  {journal} {\bibinfo  {journal} {Physical review letters}\ }\textbf {\bibinfo {volume} {124}},\ \bibinfo {pages} {020503} (\bibinfo {year} {2020})}\BibitemShut {NoStop}%
\bibitem [{\citenamefont {Sprague}\ and\ \citenamefont {Czischek}(2023)}]{sprague2023variational}%
  \BibitemOpen
  \bibfield  {author} {\bibinfo {author} {\bibfnamefont {K.}~\bibnamefont {Sprague}}\ and\ \bibinfo {author} {\bibfnamefont {S.}~\bibnamefont {Czischek}},\ }\bibfield  {title} {\bibinfo {title} {Variational monte carlo with large patched transformers},\ }\href@noop {} {\bibfield  {journal} {\bibinfo  {journal} {arXiv preprint arXiv:2306.03921}\ } (\bibinfo {year} {2023})}\BibitemShut {NoStop}%
\bibitem [{\citenamefont {Patil}(2023)}]{patil2023quantum}%
  \BibitemOpen
  \bibfield  {author} {\bibinfo {author} {\bibfnamefont {P.}~\bibnamefont {Patil}},\ }\bibfield  {title} {\bibinfo {title} {Quantum monte carlo simulations in the restricted hilbert space of rydberg atom arrays},\ }\href@noop {} {\bibfield  {journal} {\bibinfo  {journal} {arXiv preprint arXiv:2309.00482}\ } (\bibinfo {year} {2023})}\BibitemShut {NoStop}%
\bibitem [{\citenamefont {Viteritti}\ \emph {et~al.}(2022)\citenamefont {Viteritti}, \citenamefont {Ferrari},\ and\ \citenamefont {Becca}}]{viteritti2022accuracy}%
  \BibitemOpen
  \bibfield  {author} {\bibinfo {author} {\bibfnamefont {L.~L.}\ \bibnamefont {Viteritti}}, \bibinfo {author} {\bibfnamefont {F.}~\bibnamefont {Ferrari}},\ and\ \bibinfo {author} {\bibfnamefont {F.}~\bibnamefont {Becca}},\ }\bibfield  {title} {\bibinfo {title} {Accuracy of restricted boltzmann machines for the one-dimensional $ j\_1-j\_2 $ heisenberg model},\ }\href@noop {} {\bibfield  {journal} {\bibinfo  {journal} {SciPost Physics}\ }\textbf {\bibinfo {volume} {12}},\ \bibinfo {pages} {166} (\bibinfo {year} {2022})}\BibitemShut {NoStop}%
\bibitem [{\citenamefont {Valenti}\ \emph {et~al.}(2022)\citenamefont {Valenti}, \citenamefont {Greplova}, \citenamefont {Lindner},\ and\ \citenamefont {Huber}}]{valenti2022correlation}%
  \BibitemOpen
  \bibfield  {author} {\bibinfo {author} {\bibfnamefont {A.}~\bibnamefont {Valenti}}, \bibinfo {author} {\bibfnamefont {E.}~\bibnamefont {Greplova}}, \bibinfo {author} {\bibfnamefont {N.~H.}\ \bibnamefont {Lindner}},\ and\ \bibinfo {author} {\bibfnamefont {S.~D.}\ \bibnamefont {Huber}},\ }\bibfield  {title} {\bibinfo {title} {Correlation-enhanced neural networks as interpretable variational quantum states},\ }\href@noop {} {\bibfield  {journal} {\bibinfo  {journal} {Physical Review Research}\ }\textbf {\bibinfo {volume} {4}},\ \bibinfo {pages} {L012010} (\bibinfo {year} {2022})}\BibitemShut {NoStop}%
\bibitem [{\citenamefont {Choo}\ \emph {et~al.}(2019)\citenamefont {Choo}, \citenamefont {Neupert},\ and\ \citenamefont {Carleo}}]{choo2019two}%
  \BibitemOpen
  \bibfield  {author} {\bibinfo {author} {\bibfnamefont {K.}~\bibnamefont {Choo}}, \bibinfo {author} {\bibfnamefont {T.}~\bibnamefont {Neupert}},\ and\ \bibinfo {author} {\bibfnamefont {G.}~\bibnamefont {Carleo}},\ }\bibfield  {title} {\bibinfo {title} {Two-dimensional frustrated j 1- j 2 model studied with neural network quantum states},\ }\href@noop {} {\bibfield  {journal} {\bibinfo  {journal} {Physical Review B}\ }\textbf {\bibinfo {volume} {100}},\ \bibinfo {pages} {125124} (\bibinfo {year} {2019})}\BibitemShut {NoStop}%
\bibitem [{\citenamefont {Roth}\ \emph {et~al.}(2023)\citenamefont {Roth}, \citenamefont {Szab{\'o}},\ and\ \citenamefont {MacDonald}}]{roth2023high}%
  \BibitemOpen
  \bibfield  {author} {\bibinfo {author} {\bibfnamefont {C.}~\bibnamefont {Roth}}, \bibinfo {author} {\bibfnamefont {A.}~\bibnamefont {Szab{\'o}}},\ and\ \bibinfo {author} {\bibfnamefont {A.~H.}\ \bibnamefont {MacDonald}},\ }\bibfield  {title} {\bibinfo {title} {High-accuracy variational monte carlo for frustrated magnets with deep neural networks},\ }\href@noop {} {\bibfield  {journal} {\bibinfo  {journal} {Physical Review B}\ }\textbf {\bibinfo {volume} {108}},\ \bibinfo {pages} {054410} (\bibinfo {year} {2023})}\BibitemShut {NoStop}%
\bibitem [{\citenamefont {Chen}\ \emph {et~al.}(2023{\natexlab{a}})\citenamefont {Chen}, \citenamefont {Newhouse}, \citenamefont {Chen}, \citenamefont {Luo},\ and\ \citenamefont {Solja{\v{c}}i{\'c}}}]{chen2023autoregressive}%
  \BibitemOpen
  \bibfield  {author} {\bibinfo {author} {\bibfnamefont {Z.}~\bibnamefont {Chen}}, \bibinfo {author} {\bibfnamefont {L.}~\bibnamefont {Newhouse}}, \bibinfo {author} {\bibfnamefont {E.}~\bibnamefont {Chen}}, \bibinfo {author} {\bibfnamefont {D.}~\bibnamefont {Luo}},\ and\ \bibinfo {author} {\bibfnamefont {M.}~\bibnamefont {Solja{\v{c}}i{\'c}}},\ }\bibfield  {title} {\bibinfo {title} {Autoregressive neural tensornet: Bridging neural networks and tensor networks for quantum many-body simulation},\ }\href@noop {} {\bibfield  {journal} {\bibinfo  {journal} {arXiv preprint arXiv:2304.01996}\ } (\bibinfo {year} {2023}{\natexlab{a}})}\BibitemShut {NoStop}%
\bibitem [{\citenamefont {Beck}\ \emph {et~al.}(2024)\citenamefont {Beck}, \citenamefont {Bodky}, \citenamefont {Motruk}, \citenamefont {M{\"u}ller}, \citenamefont {Thomale},\ and\ \citenamefont {Ghosh}}]{beck2024phase}%
  \BibitemOpen
  \bibfield  {author} {\bibinfo {author} {\bibfnamefont {J.}~\bibnamefont {Beck}}, \bibinfo {author} {\bibfnamefont {J.}~\bibnamefont {Bodky}}, \bibinfo {author} {\bibfnamefont {J.}~\bibnamefont {Motruk}}, \bibinfo {author} {\bibfnamefont {T.}~\bibnamefont {M{\"u}ller}}, \bibinfo {author} {\bibfnamefont {R.}~\bibnamefont {Thomale}},\ and\ \bibinfo {author} {\bibfnamefont {P.}~\bibnamefont {Ghosh}},\ }\bibfield  {title} {\bibinfo {title} {Phase diagram of the j-j d heisenberg model on the maple leaf lattice: Neural networks and density matrix renormalization group},\ }\href@noop {} {\bibfield  {journal} {\bibinfo  {journal} {Physical Review B}\ }\textbf {\bibinfo {volume} {109}},\ \bibinfo {pages} {184422} (\bibinfo {year} {2024})}\BibitemShut {NoStop}%
\bibitem [{\citenamefont {Roth}\ and\ \citenamefont {MacDonald}()}]{roth2104group}%
  \BibitemOpen
  \bibfield  {author} {\bibinfo {author} {\bibfnamefont {C.}~\bibnamefont {Roth}}\ and\ \bibinfo {author} {\bibfnamefont {A.~H.}\ \bibnamefont {MacDonald}},\ }\bibfield  {title} {\bibinfo {title} {Group convolutional neural networks improve quantum state accuracy. 2021. doi: 10.48550},\ }\href@noop {} {\bibinfo  {journal} {arXiv preprint arXiv.2104.05085}\ }\BibitemShut {NoStop}%
\bibitem [{\citenamefont {Zhang}\ \emph {et~al.}(2022)\citenamefont {Zhang}, \citenamefont {Lederer}, \citenamefont {Choo}, \citenamefont {Neupert}, \citenamefont {Carleo},\ and\ \citenamefont {Kim}}]{zhang2022hamiltonian}%
  \BibitemOpen
\bibfield  {journal} {  }\bibfield  {author} {\bibinfo {author} {\bibfnamefont {K.}~\bibnamefont {Zhang}}, \bibinfo {author} {\bibfnamefont {S.}~\bibnamefont {Lederer}}, \bibinfo {author} {\bibfnamefont {K.}~\bibnamefont {Choo}}, \bibinfo {author} {\bibfnamefont {T.}~\bibnamefont {Neupert}}, \bibinfo {author} {\bibfnamefont {G.}~\bibnamefont {Carleo}},\ and\ \bibinfo {author} {\bibfnamefont {E.-A.}\ \bibnamefont {Kim}},\ }\bibfield  {title} {\bibinfo {title} {Hamiltonian reconstruction as metric for variational studies},\ }\href@noop {} {\bibfield  {journal} {\bibinfo  {journal} {SciPost Physics}\ }\textbf {\bibinfo {volume} {13}},\ \bibinfo {pages} {063} (\bibinfo {year} {2022})}\BibitemShut {NoStop}%
\bibitem [{\citenamefont {Duric}\ \emph {et~al.}(2024)\citenamefont {Duric}, \citenamefont {Chung}, \citenamefont {Yang},\ and\ \citenamefont {Sengupta}}]{duric2024spin}%
  \BibitemOpen
  \bibfield  {author} {\bibinfo {author} {\bibfnamefont {T.}~\bibnamefont {Duric}}, \bibinfo {author} {\bibfnamefont {J.~H.}\ \bibnamefont {Chung}}, \bibinfo {author} {\bibfnamefont {B.}~\bibnamefont {Yang}},\ and\ \bibinfo {author} {\bibfnamefont {P.}~\bibnamefont {Sengupta}},\ }\bibfield  {title} {\bibinfo {title} {Spin-1/2 kagome heisenberg antiferromagnet: Machine learning discovery of the spinon pair density wave ground state},\ }\href@noop {} {\bibfield  {journal} {\bibinfo  {journal} {arXiv preprint arXiv:2401.02866}\ } (\bibinfo {year} {2024})}\BibitemShut {NoStop}%
\bibitem [{\citenamefont {Cohen}\ and\ \citenamefont {Welling}(2016{\natexlab{a}})}]{cohen2016group}%
  \BibitemOpen
  \bibfield  {author} {\bibinfo {author} {\bibfnamefont {T.}~\bibnamefont {Cohen}}\ and\ \bibinfo {author} {\bibfnamefont {M.}~\bibnamefont {Welling}},\ }\bibfield  {title} {\bibinfo {title} {Group equivariant convolutional networks},\ }in\ \href@noop {} {\emph {\bibinfo {booktitle} {International conference on machine learning}}}\ (\bibinfo {organization} {PMLR},\ \bibinfo {year} {2016})\ pp.\ \bibinfo {pages} {2990--2999}\BibitemShut {NoStop}%
\bibitem [{\citenamefont {Reh}\ \emph {et~al.}(2023)\citenamefont {Reh}, \citenamefont {Schmitt},\ and\ \citenamefont {G{\"a}rttner}}]{reh2023optimizing}%
  \BibitemOpen
  \bibfield  {author} {\bibinfo {author} {\bibfnamefont {M.}~\bibnamefont {Reh}}, \bibinfo {author} {\bibfnamefont {M.}~\bibnamefont {Schmitt}},\ and\ \bibinfo {author} {\bibfnamefont {M.}~\bibnamefont {G{\"a}rttner}},\ }\bibfield  {title} {\bibinfo {title} {Optimizing design choices for neural quantum states},\ }\href@noop {} {\bibfield  {journal} {\bibinfo  {journal} {Physical Review B}\ }\textbf {\bibinfo {volume} {107}},\ \bibinfo {pages} {195115} (\bibinfo {year} {2023})}\BibitemShut {NoStop}%
\bibitem [{\citenamefont {Luo}\ \emph {et~al.}(2021)\citenamefont {Luo}, \citenamefont {Carleo}, \citenamefont {Clark},\ and\ \citenamefont {Stokes}}]{luo2021gauge}%
  \BibitemOpen
  \bibfield  {author} {\bibinfo {author} {\bibfnamefont {D.}~\bibnamefont {Luo}}, \bibinfo {author} {\bibfnamefont {G.}~\bibnamefont {Carleo}}, \bibinfo {author} {\bibfnamefont {B.~K.}\ \bibnamefont {Clark}},\ and\ \bibinfo {author} {\bibfnamefont {J.}~\bibnamefont {Stokes}},\ }\bibfield  {title} {\bibinfo {title} {Gauge equivariant neural networks for quantum lattice gauge theories},\ }\href@noop {} {\bibfield  {journal} {\bibinfo  {journal} {Physical review letters}\ }\textbf {\bibinfo {volume} {127}},\ \bibinfo {pages} {276402} (\bibinfo {year} {2021})}\BibitemShut {NoStop}%
\bibitem [{\citenamefont {Luo}\ \emph {et~al.}(2022)\citenamefont {Luo}, \citenamefont {Yuan}, \citenamefont {Stokes},\ and\ \citenamefont {Clark}}]{luo2022gauge}%
  \BibitemOpen
  \bibfield  {author} {\bibinfo {author} {\bibfnamefont {D.}~\bibnamefont {Luo}}, \bibinfo {author} {\bibfnamefont {S.}~\bibnamefont {Yuan}}, \bibinfo {author} {\bibfnamefont {J.}~\bibnamefont {Stokes}},\ and\ \bibinfo {author} {\bibfnamefont {B.}~\bibnamefont {Clark}},\ }\bibfield  {title} {\bibinfo {title} {Gauge equivariant neural networks for 2+ 1d u (1) gauge theory simulations in hamiltonian formulation},\ }in\ \href@noop {} {\emph {\bibinfo {booktitle} {NeurIPS 2022 AI for Science: Progress and Promises}}}\ (\bibinfo {year} {2022})\BibitemShut {NoStop}%
\bibitem [{\citenamefont {Luo}\ \emph {et~al.}(2023)\citenamefont {Luo}, \citenamefont {Chen}, \citenamefont {Hu}, \citenamefont {Zhao}, \citenamefont {Hur},\ and\ \citenamefont {Clark}}]{luo2023gauge}%
  \BibitemOpen
  \bibfield  {author} {\bibinfo {author} {\bibfnamefont {D.}~\bibnamefont {Luo}}, \bibinfo {author} {\bibfnamefont {Z.}~\bibnamefont {Chen}}, \bibinfo {author} {\bibfnamefont {K.}~\bibnamefont {Hu}}, \bibinfo {author} {\bibfnamefont {Z.}~\bibnamefont {Zhao}}, \bibinfo {author} {\bibfnamefont {V.~M.}\ \bibnamefont {Hur}},\ and\ \bibinfo {author} {\bibfnamefont {B.~K.}\ \bibnamefont {Clark}},\ }\bibfield  {title} {\bibinfo {title} {Gauge-invariant and anyonic-symmetric autoregressive neural network for quantum lattice models},\ }\href@noop {} {\bibfield  {journal} {\bibinfo  {journal} {Physical Review Research}\ }\textbf {\bibinfo {volume} {5}},\ \bibinfo {pages} {013216} (\bibinfo {year} {2023})}\BibitemShut {NoStop}%
\bibitem [{exa()}]{exactsoluble}%
  \BibitemOpen
  \href@noop {} {}\bibinfo {note} {Of course, for some spin liquids, there may not even be exactly solvable lattice models for which the exact symmetry operators are known.}\BibitemShut {Stop}%
\bibitem [{\citenamefont {Finzi}\ \emph {et~al.}(2021{\natexlab{a}})\citenamefont {Finzi}, \citenamefont {Benton},\ and\ \citenamefont {Wilson}}]{finzi2021residual}%
  \BibitemOpen
  \bibfield  {author} {\bibinfo {author} {\bibfnamefont {M.}~\bibnamefont {Finzi}}, \bibinfo {author} {\bibfnamefont {G.}~\bibnamefont {Benton}},\ and\ \bibinfo {author} {\bibfnamefont {A.~G.}\ \bibnamefont {Wilson}},\ }\bibfield  {title} {\bibinfo {title} {Residual pathway priors for soft equivariance constraints},\ }\href@noop {} {\bibfield  {journal} {\bibinfo  {journal} {Advances in Neural Information Processing Systems}\ }\textbf {\bibinfo {volume} {34}},\ \bibinfo {pages} {30037} (\bibinfo {year} {2021}{\natexlab{a}})}\BibitemShut {NoStop}%
\bibitem [{\citenamefont {Wang}\ \emph {et~al.}(2022)\citenamefont {Wang}, \citenamefont {Walters},\ and\ \citenamefont {Yu}}]{wang2022approximately}%
  \BibitemOpen
  \bibfield  {author} {\bibinfo {author} {\bibfnamefont {R.}~\bibnamefont {Wang}}, \bibinfo {author} {\bibfnamefont {R.}~\bibnamefont {Walters}},\ and\ \bibinfo {author} {\bibfnamefont {R.}~\bibnamefont {Yu}},\ }\bibfield  {title} {\bibinfo {title} {Approximately equivariant networks for imperfectly symmetric dynamics},\ }in\ \href@noop {} {\emph {\bibinfo {booktitle} {International Conference on Machine Learning}}}\ (\bibinfo {organization} {PMLR},\ \bibinfo {year} {2022})\ pp.\ \bibinfo {pages} {23078--23091}\BibitemShut {NoStop}%
\bibitem [{\citenamefont {Wu}\ \emph {et~al.}(2012)\citenamefont {Wu}, \citenamefont {Deng},\ and\ \citenamefont {Prokof'ev}}]{wu2012phase}%
  \BibitemOpen
  \bibfield  {author} {\bibinfo {author} {\bibfnamefont {F.}~\bibnamefont {Wu}}, \bibinfo {author} {\bibfnamefont {Y.}~\bibnamefont {Deng}},\ and\ \bibinfo {author} {\bibfnamefont {N.}~\bibnamefont {Prokof'ev}},\ }\bibfield  {title} {\bibinfo {title} {Phase diagram of the toric code model in a parallel magnetic field},\ }\href@noop {} {\bibfield  {journal} {\bibinfo  {journal} {Physical Review B}\ }\textbf {\bibinfo {volume} {85}},\ \bibinfo {pages} {195104} (\bibinfo {year} {2012})}\BibitemShut {NoStop}%
\bibitem [{\citenamefont {Trebst}\ \emph {et~al.}(2007)\citenamefont {Trebst}, \citenamefont {Werner}, \citenamefont {Troyer}, \citenamefont {Shtengel},\ and\ \citenamefont {Nayak}}]{trebst2007breakdown}%
  \BibitemOpen
  \bibfield  {author} {\bibinfo {author} {\bibfnamefont {S.}~\bibnamefont {Trebst}}, \bibinfo {author} {\bibfnamefont {P.}~\bibnamefont {Werner}}, \bibinfo {author} {\bibfnamefont {M.}~\bibnamefont {Troyer}}, \bibinfo {author} {\bibfnamefont {K.}~\bibnamefont {Shtengel}},\ and\ \bibinfo {author} {\bibfnamefont {C.}~\bibnamefont {Nayak}},\ }\bibfield  {title} {\bibinfo {title} {Breakdown of a topological phase: Quantum phase transition in a loop gas model with tension},\ }\href@noop {} {\bibfield  {journal} {\bibinfo  {journal} {Physical review letters}\ }\textbf {\bibinfo {volume} {98}},\ \bibinfo {pages} {070602} (\bibinfo {year} {2007})}\BibitemShut {NoStop}%
\bibitem [{\citenamefont {Fradkin}\ and\ \citenamefont {Shenker}(1979)}]{FradkinShenker1979}%
  \BibitemOpen
  \bibfield  {author} {\bibinfo {author} {\bibfnamefont {E.}~\bibnamefont {Fradkin}}\ and\ \bibinfo {author} {\bibfnamefont {S.~H.}\ \bibnamefont {Shenker}},\ }\bibfield  {title} {\bibinfo {title} {Phase diagrams of lattice gauge theories with higgs fields},\ }\href {https://doi.org/10.1103/PhysRevD.19.3682} {\bibfield  {journal} {\bibinfo  {journal} {Phys. Rev. D}\ }\textbf {\bibinfo {volume} {19}},\ \bibinfo {pages} {3682} (\bibinfo {year} {1979})}\BibitemShut {NoStop}%
\bibitem [{\citenamefont {Sorella}(1998)}]{sorella1998green}%
  \BibitemOpen
  \bibfield  {author} {\bibinfo {author} {\bibfnamefont {S.}~\bibnamefont {Sorella}},\ }\bibfield  {title} {\bibinfo {title} {Green function monte carlo with stochastic reconfiguration},\ }\href@noop {} {\bibfield  {journal} {\bibinfo  {journal} {Physical review letters}\ }\textbf {\bibinfo {volume} {80}},\ \bibinfo {pages} {4558} (\bibinfo {year} {1998})}\BibitemShut {NoStop}%
\bibitem [{\citenamefont {Amari}(1998)}]{amari1998natural}%
  \BibitemOpen
  \bibfield  {author} {\bibinfo {author} {\bibfnamefont {S.-I.}\ \bibnamefont {Amari}},\ }\bibfield  {title} {\bibinfo {title} {Natural gradient works efficiently in learning},\ }\href@noop {} {\bibfield  {journal} {\bibinfo  {journal} {Neural computation}\ }\textbf {\bibinfo {volume} {10}},\ \bibinfo {pages} {251} (\bibinfo {year} {1998})}\BibitemShut {NoStop}%
\bibitem [{\citenamefont {Stokes}\ \emph {et~al.}(2020)\citenamefont {Stokes}, \citenamefont {Izaac}, \citenamefont {Killoran},\ and\ \citenamefont {Carleo}}]{stokes2020quantum}%
  \BibitemOpen
  \bibfield  {author} {\bibinfo {author} {\bibfnamefont {J.}~\bibnamefont {Stokes}}, \bibinfo {author} {\bibfnamefont {J.}~\bibnamefont {Izaac}}, \bibinfo {author} {\bibfnamefont {N.}~\bibnamefont {Killoran}},\ and\ \bibinfo {author} {\bibfnamefont {G.}~\bibnamefont {Carleo}},\ }\bibfield  {title} {\bibinfo {title} {Quantum natural gradient},\ }\href@noop {} {\bibfield  {journal} {\bibinfo  {journal} {Quantum}\ }\textbf {\bibinfo {volume} {4}},\ \bibinfo {pages} {269} (\bibinfo {year} {2020})}\BibitemShut {NoStop}%
\bibitem [{exp()}]{expectation}%
  \BibitemOpen
  \href@noop {} {}\bibinfo {note} {Here the expectation $\mathbb{E}_s$ is calculated with respect to the probability $p(s)=|\psi(s)|^2$}\BibitemShut {NoStop}%
\bibitem [{\citenamefont {Favoni}\ \emph {et~al.}(2022)\citenamefont {Favoni}, \citenamefont {Ipp}, \citenamefont {M{\"u}ller},\ and\ \citenamefont {Schuh}}]{favoni2022lattice}%
  \BibitemOpen
  \bibfield  {author} {\bibinfo {author} {\bibfnamefont {M.}~\bibnamefont {Favoni}}, \bibinfo {author} {\bibfnamefont {A.}~\bibnamefont {Ipp}}, \bibinfo {author} {\bibfnamefont {D.~I.}\ \bibnamefont {M{\"u}ller}},\ and\ \bibinfo {author} {\bibfnamefont {D.}~\bibnamefont {Schuh}},\ }\bibfield  {title} {\bibinfo {title} {Lattice gauge equivariant convolutional neural networks},\ }\href@noop {} {\bibfield  {journal} {\bibinfo  {journal} {Physical Review Letters}\ }\textbf {\bibinfo {volume} {128}},\ \bibinfo {pages} {032003} (\bibinfo {year} {2022})}\BibitemShut {NoStop}%
\bibitem [{\citenamefont {Vicentini}\ \emph {et~al.}(2022)\citenamefont {Vicentini}, \citenamefont {Hofmann}, \citenamefont {Szab{\'o}}, \citenamefont {Wu}, \citenamefont {Roth}, \citenamefont {Giuliani}, \citenamefont {Pescia}, \citenamefont {Nys}, \citenamefont {Vargas-Calder{\'o}n}, \citenamefont {Astrakhantsev} \emph {et~al.}}]{vicentini2022netket}%
  \BibitemOpen
  \bibfield  {author} {\bibinfo {author} {\bibfnamefont {F.}~\bibnamefont {Vicentini}}, \bibinfo {author} {\bibfnamefont {D.}~\bibnamefont {Hofmann}}, \bibinfo {author} {\bibfnamefont {A.}~\bibnamefont {Szab{\'o}}}, \bibinfo {author} {\bibfnamefont {D.}~\bibnamefont {Wu}}, \bibinfo {author} {\bibfnamefont {C.}~\bibnamefont {Roth}}, \bibinfo {author} {\bibfnamefont {C.}~\bibnamefont {Giuliani}}, \bibinfo {author} {\bibfnamefont {G.}~\bibnamefont {Pescia}}, \bibinfo {author} {\bibfnamefont {J.}~\bibnamefont {Nys}}, \bibinfo {author} {\bibfnamefont {V.}~\bibnamefont {Vargas-Calder{\'o}n}}, \bibinfo {author} {\bibfnamefont {N.}~\bibnamefont {Astrakhantsev}}, \emph {et~al.},\ }\bibfield  {title} {\bibinfo {title} {Netket 3: machine learning toolbox for many-body quantum systems},\ }\href@noop {} {\bibfield  {journal} {\bibinfo  {journal} {SciPost Physics Codebases}\ ,\ \bibinfo {pages} {007}} (\bibinfo {year} {2022})}\BibitemShut {NoStop}%
\bibitem [{\citenamefont {White}(1992)}]{White1992}%
  \BibitemOpen
  \bibfield  {author} {\bibinfo {author} {\bibfnamefont {S.~R.}\ \bibnamefont {White}},\ }\bibfield  {title} {\bibinfo {title} {Density matrix formulation for quantum renormalization groups},\ }\href {https://doi.org/10.1103/PhysRevLett.69.2863} {\bibfield  {journal} {\bibinfo  {journal} {Phys. Rev. Lett.}\ }\textbf {\bibinfo {volume} {69}},\ \bibinfo {pages} {2863} (\bibinfo {year} {1992})}\BibitemShut {NoStop}%
\bibitem [{\citenamefont {White}(1993)}]{White1993}%
  \BibitemOpen
  \bibfield  {author} {\bibinfo {author} {\bibfnamefont {S.~R.}\ \bibnamefont {White}},\ }\bibfield  {title} {\bibinfo {title} {Density-matrix algorithms for quantum renormalization groups},\ }\href {https://doi.org/10.1103/PhysRevB.48.10345} {\bibfield  {journal} {\bibinfo  {journal} {Phys. Rev. B}\ }\textbf {\bibinfo {volume} {48}},\ \bibinfo {pages} {10345} (\bibinfo {year} {1993})}\BibitemShut {NoStop}%
\bibitem [{\citenamefont {Schollw\"ock}(2005)}]{Schollwoeck2005}%
  \BibitemOpen
  \bibfield  {author} {\bibinfo {author} {\bibfnamefont {U.}~\bibnamefont {Schollw\"ock}},\ }\bibfield  {title} {\bibinfo {title} {The density-matrix renormalization group},\ }\href {https://doi.org/10.1103/RevModPhys.77.259} {\bibfield  {journal} {\bibinfo  {journal} {Rev. Mod. Phys.}\ }\textbf {\bibinfo {volume} {77}},\ \bibinfo {pages} {259} (\bibinfo {year} {2005})}\BibitemShut {NoStop}%
\bibitem [{\citenamefont {Linsel}\ \emph {et~al.}(2024)\citenamefont {Linsel}, \citenamefont {Bohrdt}, \citenamefont {Homeier}, \citenamefont {Pollet},\ and\ \citenamefont {Grusdt}}]{linsel2024percolation}%
  \BibitemOpen
  \bibfield  {author} {\bibinfo {author} {\bibfnamefont {S.~M.}\ \bibnamefont {Linsel}}, \bibinfo {author} {\bibfnamefont {A.}~\bibnamefont {Bohrdt}}, \bibinfo {author} {\bibfnamefont {L.}~\bibnamefont {Homeier}}, \bibinfo {author} {\bibfnamefont {L.}~\bibnamefont {Pollet}},\ and\ \bibinfo {author} {\bibfnamefont {F.}~\bibnamefont {Grusdt}},\ }\href@noop {} {\bibinfo {title} {Percolation as a confinement order parameter in $\mathbb{Z}_2$ lattice gauge theories}} (\bibinfo {year} {2024}),\ \Eprint {https://arxiv.org/abs/2401.08770} {arXiv:2401.08770 [quant-ph]} \BibitemShut {NoStop}%
\bibitem [{\citenamefont {Greitemann}\ and\ \citenamefont {Pollet}(2018)}]{greitemann2018lecture}%
  \BibitemOpen
  \bibfield  {author} {\bibinfo {author} {\bibfnamefont {J.}~\bibnamefont {Greitemann}}\ and\ \bibinfo {author} {\bibfnamefont {L.}~\bibnamefont {Pollet}},\ }\bibfield  {title} {\bibinfo {title} {Lecture notes on diagrammatic monte carlo for the fr{\"o}hlich polaron},\ }\href@noop {} {\bibfield  {journal} {\bibinfo  {journal} {SciPost Physics Lecture Notes}\ ,\ \bibinfo {pages} {002}} (\bibinfo {year} {2018})}\BibitemShut {NoStop}%
\bibitem [{\citenamefont {Sahay}\ \emph {et~al.}(2022)\citenamefont {Sahay}, \citenamefont {Vishwanath},\ and\ \citenamefont {Verresen}}]{sahay2022quantum}%
  \BibitemOpen
  \bibfield  {author} {\bibinfo {author} {\bibfnamefont {R.}~\bibnamefont {Sahay}}, \bibinfo {author} {\bibfnamefont {A.}~\bibnamefont {Vishwanath}},\ and\ \bibinfo {author} {\bibfnamefont {R.}~\bibnamefont {Verresen}},\ }\bibfield  {title} {\bibinfo {title} {Quantum spin puddles and lakes: Nisq-era spin liquids from non-equilibrium dynamics},\ }\href@noop {} {\bibfield  {journal} {\bibinfo  {journal} {arXiv preprint arXiv:2211.01381}\ } (\bibinfo {year} {2022})}\BibitemShut {NoStop}%
\bibitem [{\citenamefont {Giudici}\ \emph {et~al.}(2022)\citenamefont {Giudici}, \citenamefont {Lukin},\ and\ \citenamefont {Pichler}}]{giudici2022dynamical}%
  \BibitemOpen
  \bibfield  {author} {\bibinfo {author} {\bibfnamefont {G.}~\bibnamefont {Giudici}}, \bibinfo {author} {\bibfnamefont {M.~D.}\ \bibnamefont {Lukin}},\ and\ \bibinfo {author} {\bibfnamefont {H.}~\bibnamefont {Pichler}},\ }\bibfield  {title} {\bibinfo {title} {Dynamical preparation of quantum spin liquids in rydberg atom arrays},\ }\href@noop {} {\bibfield  {journal} {\bibinfo  {journal} {Physical Review Letters}\ }\textbf {\bibinfo {volume} {129}},\ \bibinfo {pages} {090401} (\bibinfo {year} {2022})}\BibitemShut {NoStop}%
\bibitem [{\citenamefont {Wang}\ and\ \citenamefont {Pollet}(2025)}]{wang2025renormalized}%
  \BibitemOpen
  \bibfield  {author} {\bibinfo {author} {\bibfnamefont {Z.}~\bibnamefont {Wang}}\ and\ \bibinfo {author} {\bibfnamefont {L.}~\bibnamefont {Pollet}},\ }\bibfield  {title} {\bibinfo {title} {Renormalized classical spin liquid on the ruby lattice},\ }\href@noop {} {\bibfield  {journal} {\bibinfo  {journal} {Physical Review Letters}\ }\textbf {\bibinfo {volume} {134}},\ \bibinfo {pages} {086601} (\bibinfo {year} {2025})}\BibitemShut {NoStop}%
\bibitem [{\citenamefont {Manetsch}\ \emph {et~al.}(2024)\citenamefont {Manetsch}, \citenamefont {Nomura}, \citenamefont {Bataille}, \citenamefont {Leung}, \citenamefont {Lv},\ and\ \citenamefont {Endres}}]{manetsch2024tweezer}%
  \BibitemOpen
  \bibfield  {author} {\bibinfo {author} {\bibfnamefont {H.~J.}\ \bibnamefont {Manetsch}}, \bibinfo {author} {\bibfnamefont {G.}~\bibnamefont {Nomura}}, \bibinfo {author} {\bibfnamefont {E.}~\bibnamefont {Bataille}}, \bibinfo {author} {\bibfnamefont {K.~H.}\ \bibnamefont {Leung}}, \bibinfo {author} {\bibfnamefont {X.}~\bibnamefont {Lv}},\ and\ \bibinfo {author} {\bibfnamefont {M.}~\bibnamefont {Endres}},\ }\bibfield  {title} {\bibinfo {title} {A tweezer array with 6100 highly coherent atomic qubits},\ }\href@noop {} {\bibfield  {journal} {\bibinfo  {journal} {arXiv preprint arXiv:2403.12021}\ } (\bibinfo {year} {2024})}\BibitemShut {NoStop}%
\bibitem [{\citenamefont {Chaloupka}\ \emph {et~al.}(2010)\citenamefont {Chaloupka}, \citenamefont {Jackeli},\ and\ \citenamefont {Khaliullin}}]{Chaloupka2010}%
  \BibitemOpen
  \bibfield  {author} {\bibinfo {author} {\bibfnamefont {J.~c.~v.}\ \bibnamefont {Chaloupka}}, \bibinfo {author} {\bibfnamefont {G.}~\bibnamefont {Jackeli}},\ and\ \bibinfo {author} {\bibfnamefont {G.}~\bibnamefont {Khaliullin}},\ }\bibfield  {title} {\bibinfo {title} {Kitaev-heisenberg model on a honeycomb lattice: Possible exotic phases in iridium oxides ${A}_{2}{\mathrm{iro}}_{3}$},\ }\href {https://doi.org/10.1103/PhysRevLett.105.027204} {\bibfield  {journal} {\bibinfo  {journal} {Phys. Rev. Lett.}\ }\textbf {\bibinfo {volume} {105}},\ \bibinfo {pages} {027204} (\bibinfo {year} {2010})}\BibitemShut {NoStop}%
\bibitem [{\citenamefont {Jiang}\ \emph {et~al.}(2012)\citenamefont {Jiang}, \citenamefont {Yao},\ and\ \citenamefont {Balents}}]{Jiang2012}%
  \BibitemOpen
  \bibfield  {author} {\bibinfo {author} {\bibfnamefont {H.-C.}\ \bibnamefont {Jiang}}, \bibinfo {author} {\bibfnamefont {H.}~\bibnamefont {Yao}},\ and\ \bibinfo {author} {\bibfnamefont {L.}~\bibnamefont {Balents}},\ }\bibfield  {title} {\bibinfo {title} {Spin liquid ground state of the spin-$\frac{1}{2}$ square ${J}_{1}$-${J}_{2}$ heisenberg model},\ }\href {https://doi.org/10.1103/PhysRevB.86.024424} {\bibfield  {journal} {\bibinfo  {journal} {Phys. Rev. B}\ }\textbf {\bibinfo {volume} {86}},\ \bibinfo {pages} {024424} (\bibinfo {year} {2012})}\BibitemShut {NoStop}%
\bibitem [{\citenamefont {Bricmont}\ and\ \citenamefont {Fr{\"o}lich}(1983)}]{bricmont1983order}%
  \BibitemOpen
  \bibfield  {author} {\bibinfo {author} {\bibfnamefont {J.}~\bibnamefont {Bricmont}}\ and\ \bibinfo {author} {\bibfnamefont {J.}~\bibnamefont {Fr{\"o}lich}},\ }\bibfield  {title} {\bibinfo {title} {An order parameter distinguishing between different phases of lattice gauge theories with matter fields},\ }\href@noop {} {\bibfield  {journal} {\bibinfo  {journal} {Physics Letters B}\ }\textbf {\bibinfo {volume} {122}},\ \bibinfo {pages} {73} (\bibinfo {year} {1983})}\BibitemShut {NoStop}%
\bibitem [{\citenamefont {Fredenhagen}\ and\ \citenamefont {Marcu}(1983)}]{fredenhagen1983charged}%
  \BibitemOpen
  \bibfield  {author} {\bibinfo {author} {\bibfnamefont {K.}~\bibnamefont {Fredenhagen}}\ and\ \bibinfo {author} {\bibfnamefont {M.}~\bibnamefont {Marcu}},\ }\bibfield  {title} {\bibinfo {title} {Charged states in {{$\mathbb{Z}$2}} gauge theories},\ }\href {https://doi.org/10.1007/BF01206315} {\bibfield  {journal} {\bibinfo  {journal} {Communications in Mathematical Physics}\ }\textbf {\bibinfo {volume} {92}},\ \bibinfo {pages} {81} (\bibinfo {year} {1983})}\BibitemShut {NoStop}%
\bibitem [{\citenamefont {Gregor}\ \emph {et~al.}(2011)\citenamefont {Gregor}, \citenamefont {Huse}, \citenamefont {Moessner},\ and\ \citenamefont {Sondhi}}]{Gregor_2011}%
  \BibitemOpen
  \bibfield  {author} {\bibinfo {author} {\bibfnamefont {K.}~\bibnamefont {Gregor}}, \bibinfo {author} {\bibfnamefont {D.~A.}\ \bibnamefont {Huse}}, \bibinfo {author} {\bibfnamefont {R.}~\bibnamefont {Moessner}},\ and\ \bibinfo {author} {\bibfnamefont {S.~L.}\ \bibnamefont {Sondhi}},\ }\bibfield  {title} {\bibinfo {title} {Diagnosing deconfinement and topological order},\ }\href {https://doi.org/10.1088/1367-2630/13/2/025009} {\bibfield  {journal} {\bibinfo  {journal} {New Journal of Physics}\ }\textbf {\bibinfo {volume} {13}},\ \bibinfo {pages} {025009} (\bibinfo {year} {2011})}\BibitemShut {NoStop}%
\bibitem [{\citenamefont {Dusuel}\ \emph {et~al.}(2011)\citenamefont {Dusuel}, \citenamefont {Kamfor}, \citenamefont {Or{\'u}s}, \citenamefont {Schmidt},\ and\ \citenamefont {Vidal}}]{dusuel2011robustness}%
  \BibitemOpen
  \bibfield  {author} {\bibinfo {author} {\bibfnamefont {S.}~\bibnamefont {Dusuel}}, \bibinfo {author} {\bibfnamefont {M.}~\bibnamefont {Kamfor}}, \bibinfo {author} {\bibfnamefont {R.}~\bibnamefont {Or{\'u}s}}, \bibinfo {author} {\bibfnamefont {K.~P.}\ \bibnamefont {Schmidt}},\ and\ \bibinfo {author} {\bibfnamefont {J.}~\bibnamefont {Vidal}},\ }\bibfield  {title} {\bibinfo {title} {Robustness of a perturbed topological phase},\ }\href@noop {} {\bibfield  {journal} {\bibinfo  {journal} {Physical review letters}\ }\textbf {\bibinfo {volume} {106}},\ \bibinfo {pages} {107203} (\bibinfo {year} {2011})}\BibitemShut {NoStop}%
\bibitem [{res()}]{rescaling}%
  \BibitemOpen
  \href@noop {} {}\bibinfo {note} {Note the factor of $1/2$ rescaling in the definition of magnetic fields between our work and \cite{dusuel2011robustness}}\BibitemShut {NoStop}%
\bibitem [{\citenamefont {Diaconu}\ and\ \citenamefont {Worrall}(2019)}]{diaconu2019learning}%
  \BibitemOpen
  \bibfield  {author} {\bibinfo {author} {\bibfnamefont {N.}~\bibnamefont {Diaconu}}\ and\ \bibinfo {author} {\bibfnamefont {D.}~\bibnamefont {Worrall}},\ }\bibfield  {title} {\bibinfo {title} {Learning to convolve: A generalized weight-tying approach},\ }in\ \href@noop {} {\emph {\bibinfo {booktitle} {International Conference on Machine Learning}}}\ (\bibinfo {organization} {PMLR},\ \bibinfo {year} {2019})\ pp.\ \bibinfo {pages} {1586--1595}\BibitemShut {NoStop}%
\bibitem [{\citenamefont {Vieijra}\ and\ \citenamefont {Nys}(2021)}]{vieijra2021many}%
  \BibitemOpen
  \bibfield  {author} {\bibinfo {author} {\bibfnamefont {T.}~\bibnamefont {Vieijra}}\ and\ \bibinfo {author} {\bibfnamefont {J.}~\bibnamefont {Nys}},\ }\bibfield  {title} {\bibinfo {title} {Many-body quantum states with exact conservation of non-abelian and lattice symmetries through variational monte carlo},\ }\href@noop {} {\bibfield  {journal} {\bibinfo  {journal} {Physical Review B}\ }\textbf {\bibinfo {volume} {104}},\ \bibinfo {pages} {045123} (\bibinfo {year} {2021})}\BibitemShut {NoStop}%
\bibitem [{\citenamefont {Vieijra}\ \emph {et~al.}(2020)\citenamefont {Vieijra}, \citenamefont {Casert}, \citenamefont {Nys}, \citenamefont {De~Neve}, \citenamefont {Haegeman}, \citenamefont {Ryckebusch},\ and\ \citenamefont {Verstraete}}]{vieijra2020restricted}%
  \BibitemOpen
  \bibfield  {author} {\bibinfo {author} {\bibfnamefont {T.}~\bibnamefont {Vieijra}}, \bibinfo {author} {\bibfnamefont {C.}~\bibnamefont {Casert}}, \bibinfo {author} {\bibfnamefont {J.}~\bibnamefont {Nys}}, \bibinfo {author} {\bibfnamefont {W.}~\bibnamefont {De~Neve}}, \bibinfo {author} {\bibfnamefont {J.}~\bibnamefont {Haegeman}}, \bibinfo {author} {\bibfnamefont {J.}~\bibnamefont {Ryckebusch}},\ and\ \bibinfo {author} {\bibfnamefont {F.}~\bibnamefont {Verstraete}},\ }\bibfield  {title} {\bibinfo {title} {Restricted boltzmann machines for quantum states with non-abelian or anyonic symmetries},\ }\href@noop {} {\bibfield  {journal} {\bibinfo  {journal} {Physical review letters}\ }\textbf {\bibinfo {volume} {124}},\ \bibinfo {pages} {097201} (\bibinfo {year} {2020})}\BibitemShut {NoStop}%
\bibitem [{\citenamefont {Chen}\ \emph {et~al.}(2023{\natexlab{b}})\citenamefont {Chen}, \citenamefont {Newhouse}, \citenamefont {Chen}, \citenamefont {Luo},\ and\ \citenamefont {Soljacic}}]{chen2023antn}%
  \BibitemOpen
  \bibfield  {author} {\bibinfo {author} {\bibfnamefont {Z.}~\bibnamefont {Chen}}, \bibinfo {author} {\bibfnamefont {L.}~\bibnamefont {Newhouse}}, \bibinfo {author} {\bibfnamefont {E.}~\bibnamefont {Chen}}, \bibinfo {author} {\bibfnamefont {D.}~\bibnamefont {Luo}},\ and\ \bibinfo {author} {\bibfnamefont {M.}~\bibnamefont {Soljacic}},\ }\bibfield  {title} {\bibinfo {title} {Antn: Bridging autoregressive neural networks and tensor networks for quantum many-body simulation},\ }\href@noop {} {\bibfield  {journal} {\bibinfo  {journal} {Advances in Neural Information Processing Systems}\ }\textbf {\bibinfo {volume} {36}},\ \bibinfo {pages} {450} (\bibinfo {year} {2023}{\natexlab{b}})}\BibitemShut {NoStop}%
\bibitem [{\citenamefont {Hibat-Allah}\ \emph {et~al.}(2020)\citenamefont {Hibat-Allah}, \citenamefont {Ganahl}, \citenamefont {Hayward}, \citenamefont {Melko},\ and\ \citenamefont {Carrasquilla}}]{hibat2020recurrent}%
  \BibitemOpen
  \bibfield  {author} {\bibinfo {author} {\bibfnamefont {M.}~\bibnamefont {Hibat-Allah}}, \bibinfo {author} {\bibfnamefont {M.}~\bibnamefont {Ganahl}}, \bibinfo {author} {\bibfnamefont {L.~E.}\ \bibnamefont {Hayward}}, \bibinfo {author} {\bibfnamefont {R.~G.}\ \bibnamefont {Melko}},\ and\ \bibinfo {author} {\bibfnamefont {J.}~\bibnamefont {Carrasquilla}},\ }\bibfield  {title} {\bibinfo {title} {Recurrent neural network wave functions},\ }\href@noop {} {\bibfield  {journal} {\bibinfo  {journal} {Physical Review Research}\ }\textbf {\bibinfo {volume} {2}},\ \bibinfo {pages} {023358} (\bibinfo {year} {2020})}\BibitemShut {NoStop}%
\bibitem [{\citenamefont {Jumper}\ \emph {et~al.}(2021)\citenamefont {Jumper}, \citenamefont {Evans}, \citenamefont {Pritzel}, \citenamefont {Green}, \citenamefont {Figurnov}, \citenamefont {Ronneberger}, \citenamefont {Tunyasuvunakool}, \citenamefont {Bates}, \citenamefont {{\v{Z}}{\'\i}dek}, \citenamefont {Potapenko} \emph {et~al.}}]{jumper2021highly}%
  \BibitemOpen
  \bibfield  {author} {\bibinfo {author} {\bibfnamefont {J.}~\bibnamefont {Jumper}}, \bibinfo {author} {\bibfnamefont {R.}~\bibnamefont {Evans}}, \bibinfo {author} {\bibfnamefont {A.}~\bibnamefont {Pritzel}}, \bibinfo {author} {\bibfnamefont {T.}~\bibnamefont {Green}}, \bibinfo {author} {\bibfnamefont {M.}~\bibnamefont {Figurnov}}, \bibinfo {author} {\bibfnamefont {O.}~\bibnamefont {Ronneberger}}, \bibinfo {author} {\bibfnamefont {K.}~\bibnamefont {Tunyasuvunakool}}, \bibinfo {author} {\bibfnamefont {R.}~\bibnamefont {Bates}}, \bibinfo {author} {\bibfnamefont {A.}~\bibnamefont {{\v{Z}}{\'\i}dek}}, \bibinfo {author} {\bibfnamefont {A.}~\bibnamefont {Potapenko}}, \emph {et~al.},\ }\bibfield  {title} {\bibinfo {title} {Highly accurate protein structure prediction with alphafold},\ }\href@noop {} {\bibfield  {journal} {\bibinfo  {journal} {Nature}\ }\textbf {\bibinfo {volume} {596}},\ \bibinfo {pages} {583} (\bibinfo {year} {2021})}\BibitemShut {NoStop}%
\bibitem [{\citenamefont {Kondor}\ \emph {et~al.}(2018)\citenamefont {Kondor}, \citenamefont {Lin},\ and\ \citenamefont {Trivedi}}]{kondor2018clebsch}%
  \BibitemOpen
  \bibfield  {author} {\bibinfo {author} {\bibfnamefont {R.}~\bibnamefont {Kondor}}, \bibinfo {author} {\bibfnamefont {Z.}~\bibnamefont {Lin}},\ and\ \bibinfo {author} {\bibfnamefont {S.}~\bibnamefont {Trivedi}},\ }\bibfield  {title} {\bibinfo {title} {Clebsch--gordan nets: a fully fourier space spherical convolutional neural network},\ }\href@noop {} {\bibfield  {journal} {\bibinfo  {journal} {Advances in Neural Information Processing Systems}\ }\textbf {\bibinfo {volume} {31}} (\bibinfo {year} {2018})}\BibitemShut {NoStop}%
\bibitem [{\citenamefont {QUVA~lab}()}]{S_websiteequivariance}%
  \BibitemOpen
  \bibfield  {author} {\bibinfo {author} {\bibfnamefont {U.~o.~A.}\ \bibnamefont {QUVA~lab}},\ }\href@noop {} {\bibinfo {title} {Group-equivariant neural network demonstration}},\ \bibinfo {howpublished} {\url{https://github.com/dom-kufel/g_equiv_networks/blob/main/conventional_cnn.gif}}\BibitemShut {NoStop}%
\bibitem [{\citenamefont {Choo}\ \emph {et~al.}(2018)\citenamefont {Choo}, \citenamefont {Carleo}, \citenamefont {Regnault},\ and\ \citenamefont {Neupert}}]{choo2018symmetries}%
  \BibitemOpen
  \bibfield  {author} {\bibinfo {author} {\bibfnamefont {K.}~\bibnamefont {Choo}}, \bibinfo {author} {\bibfnamefont {G.}~\bibnamefont {Carleo}}, \bibinfo {author} {\bibfnamefont {N.}~\bibnamefont {Regnault}},\ and\ \bibinfo {author} {\bibfnamefont {T.}~\bibnamefont {Neupert}},\ }\bibfield  {title} {\bibinfo {title} {Symmetries and many-body excitations with neural-network quantum states},\ }\href@noop {} {\bibfield  {journal} {\bibinfo  {journal} {Physical review letters}\ }\textbf {\bibinfo {volume} {121}},\ \bibinfo {pages} {167204} (\bibinfo {year} {2018})}\BibitemShut {NoStop}%
\bibitem [{\citenamefont {Kaba}\ \emph {et~al.}(2023)\citenamefont {Kaba}, \citenamefont {Mondal}, \citenamefont {Zhang}, \citenamefont {Bengio},\ and\ \citenamefont {Ravanbakhsh}}]{kaba2023equivariance}%
  \BibitemOpen
  \bibfield  {author} {\bibinfo {author} {\bibfnamefont {S.-O.}\ \bibnamefont {Kaba}}, \bibinfo {author} {\bibfnamefont {A.~K.}\ \bibnamefont {Mondal}}, \bibinfo {author} {\bibfnamefont {Y.}~\bibnamefont {Zhang}}, \bibinfo {author} {\bibfnamefont {Y.}~\bibnamefont {Bengio}},\ and\ \bibinfo {author} {\bibfnamefont {S.}~\bibnamefont {Ravanbakhsh}},\ }\bibfield  {title} {\bibinfo {title} {Equivariance with learned canonicalization functions},\ }in\ \href@noop {} {\emph {\bibinfo {booktitle} {International Conference on Machine Learning}}}\ (\bibinfo {organization} {PMLR},\ \bibinfo {year} {2023})\ pp.\ \bibinfo {pages} {15546--15566}\BibitemShut {NoStop}%
\bibitem [{\citenamefont {Kaba}\ and\ \citenamefont {Ravanbakhsh}(2023)}]{kaba2023symmetry}%
  \BibitemOpen
  \bibfield  {author} {\bibinfo {author} {\bibfnamefont {S.-O.}\ \bibnamefont {Kaba}}\ and\ \bibinfo {author} {\bibfnamefont {S.}~\bibnamefont {Ravanbakhsh}},\ }\bibfield  {title} {\bibinfo {title} {Symmetry breaking and equivariant neural networks},\ }\href@noop {} {\bibfield  {journal} {\bibinfo  {journal} {arXiv preprint arXiv:2312.09016}\ } (\bibinfo {year} {2023})}\BibitemShut {NoStop}%
\bibitem [{\citenamefont {Finzi}\ \emph {et~al.}(2021{\natexlab{b}})\citenamefont {Finzi}, \citenamefont {Welling},\ and\ \citenamefont {Wilson}}]{finzi2021practical}%
  \BibitemOpen
  \bibfield  {author} {\bibinfo {author} {\bibfnamefont {M.}~\bibnamefont {Finzi}}, \bibinfo {author} {\bibfnamefont {M.}~\bibnamefont {Welling}},\ and\ \bibinfo {author} {\bibfnamefont {A.~G.}\ \bibnamefont {Wilson}},\ }\bibfield  {title} {\bibinfo {title} {A practical method for constructing equivariant multilayer perceptrons for arbitrary matrix groups},\ }in\ \href@noop {} {\emph {\bibinfo {booktitle} {International conference on machine learning}}}\ (\bibinfo {organization} {PMLR},\ \bibinfo {year} {2021})\ pp.\ \bibinfo {pages} {3318--3328}\BibitemShut {NoStop}%
\bibitem [{\citenamefont {Cohen}\ and\ \citenamefont {Welling}(2016{\natexlab{b}})}]{cohen2016steerable}%
  \BibitemOpen
  \bibfield  {author} {\bibinfo {author} {\bibfnamefont {T.~S.}\ \bibnamefont {Cohen}}\ and\ \bibinfo {author} {\bibfnamefont {M.}~\bibnamefont {Welling}},\ }\bibfield  {title} {\bibinfo {title} {Steerable cnns},\ }\href@noop {} {\bibfield  {journal} {\bibinfo  {journal} {arXiv preprint arXiv:1612.08498}\ } (\bibinfo {year} {2016}{\natexlab{b}})}\BibitemShut {NoStop}%
\bibitem [{\citenamefont {Zaheer}\ \emph {et~al.}(2017)\citenamefont {Zaheer}, \citenamefont {Kottur}, \citenamefont {Ravanbakhsh}, \citenamefont {Poczos}, \citenamefont {Salakhutdinov},\ and\ \citenamefont {Smola}}]{zaheer2017deep}%
  \BibitemOpen
  \bibfield  {author} {\bibinfo {author} {\bibfnamefont {M.}~\bibnamefont {Zaheer}}, \bibinfo {author} {\bibfnamefont {S.}~\bibnamefont {Kottur}}, \bibinfo {author} {\bibfnamefont {S.}~\bibnamefont {Ravanbakhsh}}, \bibinfo {author} {\bibfnamefont {B.}~\bibnamefont {Poczos}}, \bibinfo {author} {\bibfnamefont {R.~R.}\ \bibnamefont {Salakhutdinov}},\ and\ \bibinfo {author} {\bibfnamefont {A.~J.}\ \bibnamefont {Smola}},\ }\bibfield  {title} {\bibinfo {title} {Deep sets},\ }\href@noop {} {\bibfield  {journal} {\bibinfo  {journal} {Advances in neural information processing systems}\ }\textbf {\bibinfo {volume} {30}} (\bibinfo {year} {2017})}\BibitemShut {NoStop}%
\bibitem [{\citenamefont {Weiler}\ and\ \citenamefont {Cesa}(2019)}]{weiler2019general}%
  \BibitemOpen
  \bibfield  {author} {\bibinfo {author} {\bibfnamefont {M.}~\bibnamefont {Weiler}}\ and\ \bibinfo {author} {\bibfnamefont {G.}~\bibnamefont {Cesa}},\ }\bibfield  {title} {\bibinfo {title} {General e (2)-equivariant steerable cnns},\ }\href@noop {} {\bibfield  {journal} {\bibinfo  {journal} {Advances in neural information processing systems}\ }\textbf {\bibinfo {volume} {32}} (\bibinfo {year} {2019})}\BibitemShut {NoStop}%
\bibitem [{\citenamefont {Weiler}\ \emph {et~al.}(2018)\citenamefont {Weiler}, \citenamefont {Geiger}, \citenamefont {Welling}, \citenamefont {Boomsma},\ and\ \citenamefont {Cohen}}]{weiler20183d}%
  \BibitemOpen
  \bibfield  {author} {\bibinfo {author} {\bibfnamefont {M.}~\bibnamefont {Weiler}}, \bibinfo {author} {\bibfnamefont {M.}~\bibnamefont {Geiger}}, \bibinfo {author} {\bibfnamefont {M.}~\bibnamefont {Welling}}, \bibinfo {author} {\bibfnamefont {W.}~\bibnamefont {Boomsma}},\ and\ \bibinfo {author} {\bibfnamefont {T.~S.}\ \bibnamefont {Cohen}},\ }\bibfield  {title} {\bibinfo {title} {3d steerable cnns: Learning rotationally equivariant features in volumetric data},\ }\href@noop {} {\bibfield  {journal} {\bibinfo  {journal} {Advances in Neural Information Processing Systems}\ }\textbf {\bibinfo {volume} {31}} (\bibinfo {year} {2018})}\BibitemShut {NoStop}%
\bibitem [{\citenamefont {He}\ \emph {et~al.}(2016)\citenamefont {He}, \citenamefont {Zhang}, \citenamefont {Ren},\ and\ \citenamefont {Sun}}]{he2016deep}%
  \BibitemOpen
  \bibfield  {author} {\bibinfo {author} {\bibfnamefont {K.}~\bibnamefont {He}}, \bibinfo {author} {\bibfnamefont {X.}~\bibnamefont {Zhang}}, \bibinfo {author} {\bibfnamefont {S.}~\bibnamefont {Ren}},\ and\ \bibinfo {author} {\bibfnamefont {J.}~\bibnamefont {Sun}},\ }\bibfield  {title} {\bibinfo {title} {Deep residual learning for image recognition},\ }in\ \href@noop {} {\emph {\bibinfo {booktitle} {Proceedings of the IEEE conference on computer vision and pattern recognition}}}\ (\bibinfo {year} {2016})\ pp.\ \bibinfo {pages} {770--778}\BibitemShut {NoStop}%
\bibitem [{\citenamefont {Weerda}\ and\ \citenamefont {Rizzi}(2024)}]{weerda2024fractional}%
  \BibitemOpen
  \bibfield  {author} {\bibinfo {author} {\bibfnamefont {E.~L.}\ \bibnamefont {Weerda}}\ and\ \bibinfo {author} {\bibfnamefont {M.}~\bibnamefont {Rizzi}},\ }\bibfield  {title} {\bibinfo {title} {Fractional quantum hall states with variational projected entangled-pair states: A study of the bosonic harper-hofstadter model},\ }\href@noop {} {\bibfield  {journal} {\bibinfo  {journal} {Physical Review B}\ }\textbf {\bibinfo {volume} {109}},\ \bibinfo {pages} {L241117} (\bibinfo {year} {2024})}\BibitemShut {NoStop}%
\bibitem [{\citenamefont {Feldmeier}\ \emph {et~al.}(2024)\citenamefont {Feldmeier}, \citenamefont {Maskara}, \citenamefont {K{\"o}yl{\"u}o{\u{g}}lu},\ and\ \citenamefont {Lukin}}]{feldmeier2024quantum}%
  \BibitemOpen
  \bibfield  {author} {\bibinfo {author} {\bibfnamefont {J.}~\bibnamefont {Feldmeier}}, \bibinfo {author} {\bibfnamefont {N.}~\bibnamefont {Maskara}}, \bibinfo {author} {\bibfnamefont {N.~U.}\ \bibnamefont {K{\"o}yl{\"u}o{\u{g}}lu}},\ and\ \bibinfo {author} {\bibfnamefont {M.~D.}\ \bibnamefont {Lukin}},\ }\bibfield  {title} {\bibinfo {title} {Quantum simulation of dynamical gauge theories in periodically driven rydberg atom arrays},\ }\href@noop {} {\bibfield  {journal} {\bibinfo  {journal} {arXiv preprint arXiv:2408.02733}\ } (\bibinfo {year} {2024})}\BibitemShut {NoStop}%
\bibitem [{\citenamefont {Hastings}(2016)}]{hastings2016quantum}%
  \BibitemOpen
  \bibfield  {author} {\bibinfo {author} {\bibfnamefont {M.~B.}\ \bibnamefont {Hastings}},\ }\bibfield  {title} {\bibinfo {title} {How quantum are non-negative wavefunctions?},\ }\href@noop {} {\bibfield  {journal} {\bibinfo  {journal} {Journal of Mathematical Physics}\ }\textbf {\bibinfo {volume} {57}} (\bibinfo {year} {2016})}\BibitemShut {NoStop}%
\bibitem [{\citenamefont {Levin}\ and\ \citenamefont {Wen}(2005)}]{levin2005string}%
  \BibitemOpen
  \bibfield  {author} {\bibinfo {author} {\bibfnamefont {M.~A.}\ \bibnamefont {Levin}}\ and\ \bibinfo {author} {\bibfnamefont {X.-G.}\ \bibnamefont {Wen}},\ }\bibfield  {title} {\bibinfo {title} {String-net condensation: A physical mechanism for topological phases},\ }\href@noop {} {\bibfield  {journal} {\bibinfo  {journal} {Physical Review B—Condensed Matter and Materials Physics}\ }\textbf {\bibinfo {volume} {71}},\ \bibinfo {pages} {045110} (\bibinfo {year} {2005})}\BibitemShut {NoStop}%
\bibitem [{\citenamefont {Wiese}(2014)}]{wiese2014towards}%
  \BibitemOpen
  \bibfield  {author} {\bibinfo {author} {\bibfnamefont {U.-J.}\ \bibnamefont {Wiese}},\ }\bibfield  {title} {\bibinfo {title} {Towards quantum simulating qcd},\ }\href@noop {} {\bibfield  {journal} {\bibinfo  {journal} {Nuclear Physics A}\ }\textbf {\bibinfo {volume} {931}},\ \bibinfo {pages} {246} (\bibinfo {year} {2014})}\BibitemShut {NoStop}%
\bibitem [{\citenamefont {Pillai}\ \emph {et~al.}(2005)\citenamefont {Pillai}, \citenamefont {Suel},\ and\ \citenamefont {Cha}}]{pillai2005perron}%
  \BibitemOpen
  \bibfield  {author} {\bibinfo {author} {\bibfnamefont {S.~U.}\ \bibnamefont {Pillai}}, \bibinfo {author} {\bibfnamefont {T.}~\bibnamefont {Suel}},\ and\ \bibinfo {author} {\bibfnamefont {S.}~\bibnamefont {Cha}},\ }\bibfield  {title} {\bibinfo {title} {The perron-frobenius theorem: some of its applications},\ }\href@noop {} {\bibfield  {journal} {\bibinfo  {journal} {IEEE Signal Processing Magazine}\ }\textbf {\bibinfo {volume} {22}},\ \bibinfo {pages} {62} (\bibinfo {year} {2005})}\BibitemShut {NoStop}%
\bibitem [{\citenamefont {Machaczek}(2024)}]{marcmachaczekthesis}%
  \BibitemOpen
  \bibfield  {author} {\bibinfo {author} {\bibfnamefont {M.}~\bibnamefont {Machaczek}},\ }\href@noop {} {\bibinfo {title} {Neural quantum states for fracton models}},\ \bibinfo {howpublished} {\url{https://zenodo.org/records/10728168}} (\bibinfo {year} {2024})\BibitemShut {NoStop}%
\bibitem [{\citenamefont {Clevert}\ \emph {et~al.}(2015)\citenamefont {Clevert}, \citenamefont {Unterthiner},\ and\ \citenamefont {Hochreiter}}]{clevert2015fast}%
  \BibitemOpen
  \bibfield  {author} {\bibinfo {author} {\bibfnamefont {D.-A.}\ \bibnamefont {Clevert}}, \bibinfo {author} {\bibfnamefont {T.}~\bibnamefont {Unterthiner}},\ and\ \bibinfo {author} {\bibfnamefont {S.}~\bibnamefont {Hochreiter}},\ }\bibfield  {title} {\bibinfo {title} {Fast and accurate deep network learning by exponential linear units (elus)},\ }\href@noop {} {\bibfield  {journal} {\bibinfo  {journal} {arXiv preprint arXiv:1511.07289}\ } (\bibinfo {year} {2015})}\BibitemShut {NoStop}%
\bibitem [{\citenamefont {Chen}\ and\ \citenamefont {Heyl}(2023)}]{chen2023efficient}%
  \BibitemOpen
  \bibfield  {author} {\bibinfo {author} {\bibfnamefont {A.}~\bibnamefont {Chen}}\ and\ \bibinfo {author} {\bibfnamefont {M.}~\bibnamefont {Heyl}},\ }\bibfield  {title} {\bibinfo {title} {Efficient optimization of deep neural quantum states toward machine precision},\ }\href@noop {} {\bibfield  {journal} {\bibinfo  {journal} {arXiv preprint arXiv:2302.01941}\ } (\bibinfo {year} {2023})}\BibitemShut {NoStop}%
\bibitem [{\citenamefont {Bradbury}\ \emph {et~al.}(2018)\citenamefont {Bradbury}, \citenamefont {Frostig}, \citenamefont {Hawkins}, \citenamefont {Johnson}, \citenamefont {Leary}, \citenamefont {Maclaurin}, \citenamefont {Necula}, \citenamefont {Paszke}, \citenamefont {Vander{P}las}, \citenamefont {Wanderman-{M}ilne},\ and\ \citenamefont {Zhang}}]{jax2018github}%
  \BibitemOpen
  \bibfield  {author} {\bibinfo {author} {\bibfnamefont {J.}~\bibnamefont {Bradbury}}, \bibinfo {author} {\bibfnamefont {R.}~\bibnamefont {Frostig}}, \bibinfo {author} {\bibfnamefont {P.}~\bibnamefont {Hawkins}}, \bibinfo {author} {\bibfnamefont {M.~J.}\ \bibnamefont {Johnson}}, \bibinfo {author} {\bibfnamefont {C.}~\bibnamefont {Leary}}, \bibinfo {author} {\bibfnamefont {D.}~\bibnamefont {Maclaurin}}, \bibinfo {author} {\bibfnamefont {G.}~\bibnamefont {Necula}}, \bibinfo {author} {\bibfnamefont {A.}~\bibnamefont {Paszke}}, \bibinfo {author} {\bibfnamefont {J.}~\bibnamefont {Vander{P}las}}, \bibinfo {author} {\bibfnamefont {S.}~\bibnamefont {Wanderman-{M}ilne}},\ and\ \bibinfo {author} {\bibfnamefont {Q.}~\bibnamefont {Zhang}},\ }\href {http://github.com/google/jax} {\bibinfo {title} {{JAX}: composable transformations of {P}ython+{N}um{P}y programs}} (\bibinfo {year} {2018})\BibitemShut {NoStop}%
\bibitem [{\citenamefont {Heek}\ \emph {et~al.}(2023)\citenamefont {Heek}, \citenamefont {Levskaya}, \citenamefont {Oliver}, \citenamefont {Ritter}, \citenamefont {Rondepierre}, \citenamefont {Steiner},\ and\ \citenamefont {van {Z}ee}}]{flax2020github}%
  \BibitemOpen
  \bibfield  {author} {\bibinfo {author} {\bibfnamefont {J.}~\bibnamefont {Heek}}, \bibinfo {author} {\bibfnamefont {A.}~\bibnamefont {Levskaya}}, \bibinfo {author} {\bibfnamefont {A.}~\bibnamefont {Oliver}}, \bibinfo {author} {\bibfnamefont {M.}~\bibnamefont {Ritter}}, \bibinfo {author} {\bibfnamefont {B.}~\bibnamefont {Rondepierre}}, \bibinfo {author} {\bibfnamefont {A.}~\bibnamefont {Steiner}},\ and\ \bibinfo {author} {\bibfnamefont {M.}~\bibnamefont {van {Z}ee}},\ }\href {http://github.com/google/flax} {\bibinfo {title} {{F}lax: A neural network library and ecosystem for {JAX}}} (\bibinfo {year} {2023})\BibitemShut {NoStop}%
\bibitem [{\citenamefont {Fishman}\ \emph {et~al.}(2022{\natexlab{a}})\citenamefont {Fishman}, \citenamefont {White},\ and\ \citenamefont {Stoudenmire}}]{itensor}%
  \BibitemOpen
  \bibfield  {author} {\bibinfo {author} {\bibfnamefont {M.}~\bibnamefont {Fishman}}, \bibinfo {author} {\bibfnamefont {S.~R.}\ \bibnamefont {White}},\ and\ \bibinfo {author} {\bibfnamefont {E.~M.}\ \bibnamefont {Stoudenmire}},\ }\bibfield  {title} {\bibinfo {title} {{The ITensor Software Library for Tensor Network Calculations}},\ }\href {https://doi.org/10.21468/SciPostPhysCodeb.4} {\bibfield  {journal} {\bibinfo  {journal} {SciPost Phys. Codebases}\ ,\ \bibinfo {pages} {4}} (\bibinfo {year} {2022}{\natexlab{a}})}\BibitemShut {NoStop}%
\bibitem [{\citenamefont {Fishman}\ \emph {et~al.}(2022{\natexlab{b}})\citenamefont {Fishman}, \citenamefont {White},\ and\ \citenamefont {Stoudenmire}}]{itensor-r0.3}%
  \BibitemOpen
  \bibfield  {author} {\bibinfo {author} {\bibfnamefont {M.}~\bibnamefont {Fishman}}, \bibinfo {author} {\bibfnamefont {S.~R.}\ \bibnamefont {White}},\ and\ \bibinfo {author} {\bibfnamefont {E.~M.}\ \bibnamefont {Stoudenmire}},\ }\bibfield  {title} {\bibinfo {title} {{Codebase release 0.3 for ITensor}},\ }\href {https://doi.org/10.21468/SciPostPhysCodeb.4-r0.3} {\bibfield  {journal} {\bibinfo  {journal} {SciPost Phys. Codebases}\ ,\ \bibinfo {pages} {4}} (\bibinfo {year} {2022}{\natexlab{b}})}\BibitemShut {NoStop}%
\bibitem [{\citenamefont {Kahanamoku-Meyer}\ and\ \citenamefont {Wei}(2024)}]{gregjulia}%
  \BibitemOpen
  \bibfield  {author} {\bibinfo {author} {\bibfnamefont {G.~D.}\ \bibnamefont {Kahanamoku-Meyer}}\ and\ \bibinfo {author} {\bibfnamefont {J.}~\bibnamefont {Wei}},\ }\href {https://doi.org/10.5281/zenodo.10906046} {\bibinfo {title} {{GregDMeyer/dynamite: v0.4.0}}} (\bibinfo {year} {2024})\BibitemShut {NoStop}%
\bibitem [{\citenamefont {Szab{\'o}}\ and\ \citenamefont {Castelnovo}(2020)}]{szabo2020neural}%
  \BibitemOpen
  \bibfield  {author} {\bibinfo {author} {\bibfnamefont {A.}~\bibnamefont {Szab{\'o}}}\ and\ \bibinfo {author} {\bibfnamefont {C.}~\bibnamefont {Castelnovo}},\ }\bibfield  {title} {\bibinfo {title} {Neural network wave functions and the sign problem},\ }\href@noop {} {\bibfield  {journal} {\bibinfo  {journal} {Physical Review Research}\ }\textbf {\bibinfo {volume} {2}},\ \bibinfo {pages} {033075} (\bibinfo {year} {2020})}\BibitemShut {NoStop}%
\bibitem [{\citenamefont {Jing}\ \emph {et~al.}(2017)\citenamefont {Jing}, \citenamefont {Shen}, \citenamefont {Dubcek}, \citenamefont {Peurifoy}, \citenamefont {Skirlo}, \citenamefont {LeCun}, \citenamefont {Tegmark},\ and\ \citenamefont {Solja{\v{c}}i{\'c}}}]{jing2017tunable}%
  \BibitemOpen
  \bibfield  {author} {\bibinfo {author} {\bibfnamefont {L.}~\bibnamefont {Jing}}, \bibinfo {author} {\bibfnamefont {Y.}~\bibnamefont {Shen}}, \bibinfo {author} {\bibfnamefont {T.}~\bibnamefont {Dubcek}}, \bibinfo {author} {\bibfnamefont {J.}~\bibnamefont {Peurifoy}}, \bibinfo {author} {\bibfnamefont {S.}~\bibnamefont {Skirlo}}, \bibinfo {author} {\bibfnamefont {Y.}~\bibnamefont {LeCun}}, \bibinfo {author} {\bibfnamefont {M.}~\bibnamefont {Tegmark}},\ and\ \bibinfo {author} {\bibfnamefont {M.}~\bibnamefont {Solja{\v{c}}i{\'c}}},\ }\bibfield  {title} {\bibinfo {title} {Tunable efficient unitary neural networks (eunn) and their application to rnns},\ }in\ \href@noop {} {\emph {\bibinfo {booktitle} {International Conference on Machine Learning}}}\ (\bibinfo {organization} {PMLR},\ \bibinfo {year} {2017})\ pp.\ \bibinfo {pages} {1733--1741}\BibitemShut {NoStop}%
\bibitem [{\citenamefont {Astrakhantsev}\ \emph {et~al.}(2021)\citenamefont {Astrakhantsev}, \citenamefont {Westerhout}, \citenamefont {Tiwari}, \citenamefont {Choo}, \citenamefont {Chen}, \citenamefont {Fischer}, \citenamefont {Carleo},\ and\ \citenamefont {Neupert}}]{astrakhantsev2021broken}%
  \BibitemOpen
  \bibfield  {author} {\bibinfo {author} {\bibfnamefont {N.}~\bibnamefont {Astrakhantsev}}, \bibinfo {author} {\bibfnamefont {T.}~\bibnamefont {Westerhout}}, \bibinfo {author} {\bibfnamefont {A.}~\bibnamefont {Tiwari}}, \bibinfo {author} {\bibfnamefont {K.}~\bibnamefont {Choo}}, \bibinfo {author} {\bibfnamefont {A.}~\bibnamefont {Chen}}, \bibinfo {author} {\bibfnamefont {M.~H.}\ \bibnamefont {Fischer}}, \bibinfo {author} {\bibfnamefont {G.}~\bibnamefont {Carleo}},\ and\ \bibinfo {author} {\bibfnamefont {T.}~\bibnamefont {Neupert}},\ }\bibfield  {title} {\bibinfo {title} {Broken-symmetry ground states of the heisenberg model on the pyrochlore lattice},\ }\href@noop {} {\bibfield  {journal} {\bibinfo  {journal} {Physical Review X}\ }\textbf {\bibinfo {volume} {11}},\ \bibinfo {pages} {041021} (\bibinfo {year} {2021})}\BibitemShut {NoStop}%
\bibitem [{\citenamefont {Wu}\ \emph {et~al.}(2024)\citenamefont {Wu}, \citenamefont {Rossi}, \citenamefont {Vicentini}, \citenamefont {Astrakhantsev}, \citenamefont {Becca}, \citenamefont {Cao}, \citenamefont {Carrasquilla}, \citenamefont {Ferrari}, \citenamefont {Georges}, \citenamefont {Hibat-Allah} \emph {et~al.}}]{wu2024variational}%
  \BibitemOpen
  \bibfield  {author} {\bibinfo {author} {\bibfnamefont {D.}~\bibnamefont {Wu}}, \bibinfo {author} {\bibfnamefont {R.}~\bibnamefont {Rossi}}, \bibinfo {author} {\bibfnamefont {F.}~\bibnamefont {Vicentini}}, \bibinfo {author} {\bibfnamefont {N.}~\bibnamefont {Astrakhantsev}}, \bibinfo {author} {\bibfnamefont {F.}~\bibnamefont {Becca}}, \bibinfo {author} {\bibfnamefont {X.}~\bibnamefont {Cao}}, \bibinfo {author} {\bibfnamefont {J.}~\bibnamefont {Carrasquilla}}, \bibinfo {author} {\bibfnamefont {F.}~\bibnamefont {Ferrari}}, \bibinfo {author} {\bibfnamefont {A.}~\bibnamefont {Georges}}, \bibinfo {author} {\bibfnamefont {M.}~\bibnamefont {Hibat-Allah}}, \emph {et~al.},\ }\bibfield  {title} {\bibinfo {title} {Variational benchmarks for quantum many-body problems},\ }\href@noop {} {\bibfield  {journal} {\bibinfo  {journal} {Science}\ }\textbf {\bibinfo {volume} {386}},\ \bibinfo {pages} {296} (\bibinfo {year} {2024})}\BibitemShut {NoStop}%
\bibitem [{\citenamefont {Goodfellow}\ \emph {et~al.}(2016)\citenamefont {Goodfellow}, \citenamefont {Bengio},\ and\ \citenamefont {Courville}}]{goodfellow2016deep}%
  \BibitemOpen
  \bibfield  {author} {\bibinfo {author} {\bibfnamefont {I.}~\bibnamefont {Goodfellow}}, \bibinfo {author} {\bibfnamefont {Y.}~\bibnamefont {Bengio}},\ and\ \bibinfo {author} {\bibfnamefont {A.}~\bibnamefont {Courville}},\ }\href@noop {} {\emph {\bibinfo {title} {Deep learning}}}\ (\bibinfo  {publisher} {MIT press},\ \bibinfo {year} {2016})\BibitemShut {NoStop}%
\bibitem [{\citenamefont {Zen}\ \emph {et~al.}(2020)\citenamefont {Zen}, \citenamefont {My}, \citenamefont {Tan}, \citenamefont {H{\'e}bert}, \citenamefont {Gattobigio}, \citenamefont {Miniatura}, \citenamefont {Poletti},\ and\ \citenamefont {Bressan}}]{zen2020transfer}%
  \BibitemOpen
  \bibfield  {author} {\bibinfo {author} {\bibfnamefont {R.}~\bibnamefont {Zen}}, \bibinfo {author} {\bibfnamefont {L.}~\bibnamefont {My}}, \bibinfo {author} {\bibfnamefont {R.}~\bibnamefont {Tan}}, \bibinfo {author} {\bibfnamefont {F.}~\bibnamefont {H{\'e}bert}}, \bibinfo {author} {\bibfnamefont {M.}~\bibnamefont {Gattobigio}}, \bibinfo {author} {\bibfnamefont {C.}~\bibnamefont {Miniatura}}, \bibinfo {author} {\bibfnamefont {D.}~\bibnamefont {Poletti}},\ and\ \bibinfo {author} {\bibfnamefont {S.}~\bibnamefont {Bressan}},\ }\bibfield  {title} {\bibinfo {title} {Transfer learning for scalability of neural-network quantum states},\ }\href@noop {} {\bibfield  {journal} {\bibinfo  {journal} {Physical Review E}\ }\textbf {\bibinfo {volume} {101}},\ \bibinfo {pages} {053301} (\bibinfo {year} {2020})}\BibitemShut {NoStop}%
\bibitem [{\citenamefont {Bravyi}\ \emph {et~al.}(2022)\citenamefont {Bravyi}, \citenamefont {Gosset},\ and\ \citenamefont {Liu}}]{bravyi2022simulate}%
  \BibitemOpen
  \bibfield  {author} {\bibinfo {author} {\bibfnamefont {S.}~\bibnamefont {Bravyi}}, \bibinfo {author} {\bibfnamefont {D.}~\bibnamefont {Gosset}},\ and\ \bibinfo {author} {\bibfnamefont {Y.}~\bibnamefont {Liu}},\ }\bibfield  {title} {\bibinfo {title} {How to simulate quantum measurement without computing marginals},\ }\href@noop {} {\bibfield  {journal} {\bibinfo  {journal} {Physical Review Letters}\ }\textbf {\bibinfo {volume} {128}},\ \bibinfo {pages} {220503} (\bibinfo {year} {2022})}\BibitemShut {NoStop}%
\bibitem [{\citenamefont {Bravyi}\ \emph {et~al.}(2023)\citenamefont {Bravyi}, \citenamefont {Carleo}, \citenamefont {Gosset},\ and\ \citenamefont {Liu}}]{bravyi2023rapidly}%
  \BibitemOpen
  \bibfield  {author} {\bibinfo {author} {\bibfnamefont {S.}~\bibnamefont {Bravyi}}, \bibinfo {author} {\bibfnamefont {G.}~\bibnamefont {Carleo}}, \bibinfo {author} {\bibfnamefont {D.}~\bibnamefont {Gosset}},\ and\ \bibinfo {author} {\bibfnamefont {Y.}~\bibnamefont {Liu}},\ }\bibfield  {title} {\bibinfo {title} {A rapidly mixing markov chain from any gapped quantum many-body system},\ }\href@noop {} {\bibfield  {journal} {\bibinfo  {journal} {Quantum}\ }\textbf {\bibinfo {volume} {7}},\ \bibinfo {pages} {1173} (\bibinfo {year} {2023})}\BibitemShut {NoStop}%
\bibitem [{\citenamefont {Vehtari}\ \emph {et~al.}(2021)\citenamefont {Vehtari}, \citenamefont {Gelman}, \citenamefont {Simpson}, \citenamefont {Carpenter},\ and\ \citenamefont {B{\"u}rkner}}]{vehtari2021rank}%
  \BibitemOpen
  \bibfield  {author} {\bibinfo {author} {\bibfnamefont {A.}~\bibnamefont {Vehtari}}, \bibinfo {author} {\bibfnamefont {A.}~\bibnamefont {Gelman}}, \bibinfo {author} {\bibfnamefont {D.}~\bibnamefont {Simpson}}, \bibinfo {author} {\bibfnamefont {B.}~\bibnamefont {Carpenter}},\ and\ \bibinfo {author} {\bibfnamefont {P.-C.}\ \bibnamefont {B{\"u}rkner}},\ }\bibfield  {title} {\bibinfo {title} {Rank-normalization, folding, and localization: An improved r for assessing convergence of mcmc (with discussion)},\ }\href@noop {} {\bibfield  {journal} {\bibinfo  {journal} {Bayesian analysis}\ }\textbf {\bibinfo {volume} {16}},\ \bibinfo {pages} {667} (\bibinfo {year} {2021})}\BibitemShut {NoStop}%
\bibitem [{\citenamefont {Xu}\ \emph {et~al.}(2024)\citenamefont {Xu}, \citenamefont {Pollmann},\ and\ \citenamefont {Knap}}]{xu2024critical}%
  \BibitemOpen
  \bibfield  {author} {\bibinfo {author} {\bibfnamefont {W.-T.}\ \bibnamefont {Xu}}, \bibinfo {author} {\bibfnamefont {F.}~\bibnamefont {Pollmann}},\ and\ \bibinfo {author} {\bibfnamefont {M.}~\bibnamefont {Knap}},\ }\bibfield  {title} {\bibinfo {title} {Critical behavior of the fredenhagen-marcu order parameter at topological phase transitions},\ }\href@noop {} {\bibfield  {journal} {\bibinfo  {journal} {arXiv preprint arXiv:2402.00127}\ } (\bibinfo {year} {2024})}\BibitemShut {NoStop}%
\bibitem [{\citenamefont {Hastings}\ \emph {et~al.}(2010)\citenamefont {Hastings}, \citenamefont {Gonz{\'a}lez}, \citenamefont {Kallin},\ and\ \citenamefont {Melko}}]{hastings2010measuring}%
  \BibitemOpen
  \bibfield  {author} {\bibinfo {author} {\bibfnamefont {M.~B.}\ \bibnamefont {Hastings}}, \bibinfo {author} {\bibfnamefont {I.}~\bibnamefont {Gonz{\'a}lez}}, \bibinfo {author} {\bibfnamefont {A.~B.}\ \bibnamefont {Kallin}},\ and\ \bibinfo {author} {\bibfnamefont {R.~G.}\ \bibnamefont {Melko}},\ }\bibfield  {title} {\bibinfo {title} {Measuring renyi entanglement entropy in quantum monte carlo simulations},\ }\href@noop {} {\bibfield  {journal} {\bibinfo  {journal} {Physical review letters}\ }\textbf {\bibinfo {volume} {104}},\ \bibinfo {pages} {157201} (\bibinfo {year} {2010})}\BibitemShut {NoStop}%
\bibitem [{\citenamefont {Finzi}(2023)}]{marcfinzithesis}%
  \BibitemOpen
  \bibfield  {author} {\bibinfo {author} {\bibfnamefont {M.}~\bibnamefont {Finzi}},\ }\emph {\bibinfo {title} {Understanding and Incorporating Mathematical Inductive Biases in Neural Networks}},\ \href@noop {} {Ph.D. thesis},\ \bibinfo  {school} {New York University} (\bibinfo {year} {2023})\BibitemShut {NoStop}%
\end{thebibliography}
\end{document}